\documentclass[11pt]{cernrep}
\usepackage{graphicx}
\usepackage{here}
\begin{document}  

\title{{\small
\begin{flushright}
SI-HEP-2004-03 
\end{flushright}}
QUANTUM CHROMODYNAMICS AND HADRONS:\\ 
AN ELEMENTARY INTRODUCTION}
\author{Alexander~Khodjamirian }
\institute{Theoretische Physik 1, 
Fachbereich Physik, Universit\"at Siegen, D-57068, Siegen, Germany  }
\maketitle
\begin{abstract}
Notes of five lectures given at the  2003 European School
of High-Energy Physics,
Tsakhkadzor, Armenia, September 2003 
\end{abstract}

\section{QUARKS AND GLUONS }

\subsection{
Introduction} 

\begin{figure}
\begin{center}
\includegraphics*[width=14cm]{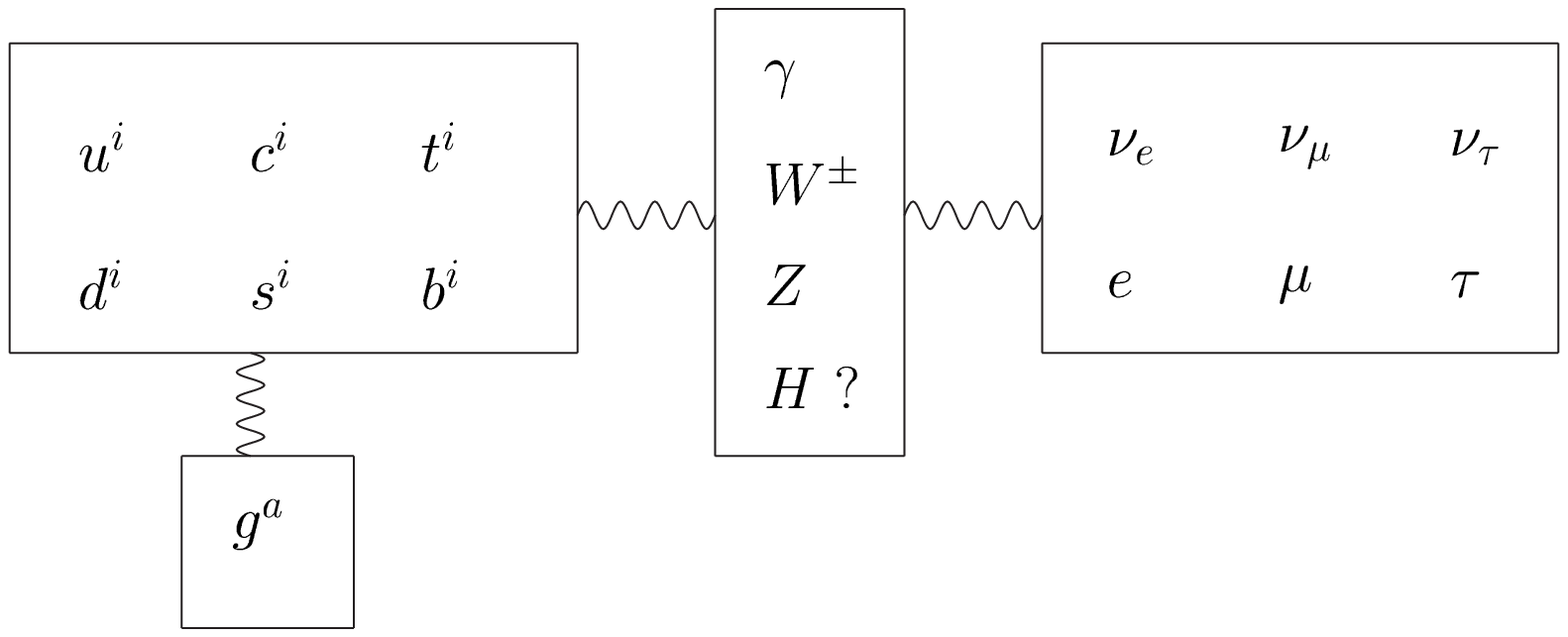}
\caption{{\em Particles of the Standard Model; 
$i=1,2,3$ and $a=1,...8$ are the color indices of quarks and gluons,
respectively}}
\label{fig:chart}
\end{center}
\end{figure}
In Standard Model the properties of quarks and leptons 
are remarkably similar, as far as the electroweak interactions
are concerned. The quarks with six different flavours 
are grouped into the three doublets
$(u,d)$,$(c,s)$,$(t,b)$, in one-to-one correspondence  
to the three lepton doublets 
$(\nu_e,e)$,$(\nu_\mu,\mu)$, $(\nu_\tau,\tau)$,
as shown in the schematic chart in Fig.~\ref{fig:chart}. 
Quarks and leptons interact in a similar way 
with the electroweak vector bosons $\gamma$,$~W^\pm$ and $Z$. 
Furthermore, it is anticipated that the universal 
Higgs mechanism is responsible for the generation of the quark 
and lepton masses (for more details see \cite{Aitchison}). 

In addition, quarks interact 
``on their own'', revealing their specific property,
the colour charge ({\em colour}), a conserved 
quantum number which is absent for leptons. 
A quark of a given flavour has three different colour
states with equal masses and electroweak charges.
Colour-induced interactions between quarks are  
mediated by {\em gluons}, the massless and 
electroweakly-neutral spin-1 particles, 

In these lectures I will discuss the quark-gluon ``corner''
of the Standard Model, introducing the basics of 
Quantum Chromodynamics (QCD), the theory of quark-gluon
interactions. Throughout this survey, the main emphasis 
will be put on the relation between QCD and {\em hadrons},
the observed bound states of quarks.  This first lecture   
is devoted to the basic properties of 
the quark-gluon interactions. I will frequently refer
to Quantum Electrodynamics (QED), the more familiar theory of 
electromagnetic (e.m.) interactions, which is a useful prototype of QCD.

\begin{figure}
\begin{center}
\includegraphics*[width=14cm]{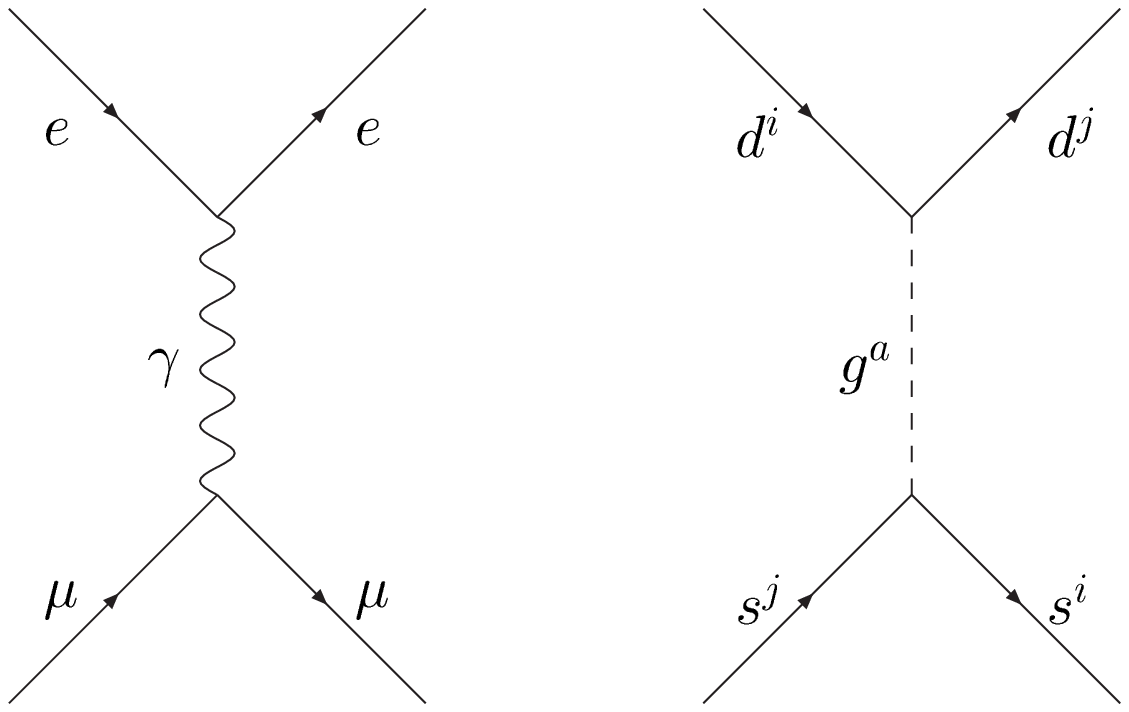}
\caption{{\em Diagrams of lepton-photon (left) and quark-gluon
(right) interactions. The dashed line is used to denote gluons
and wavy line photon.}}
\label{fig:diags}
\end{center}
\end{figure}

In Fig.~\ref{fig:diags} the Feynman 
diagram  of the electron-muon e.m. 
scattering is drawn together with the analogous diagram  of the 
quark-quark interaction. For definiteness, I specify the quarks as 
having $d$ and $s$
flavours, the counterparts of $e$ and $\mu$   in the Standard Model.
The two interactions have many important similarities: 
\begin{itemize}
\item[--]       
the colour-charged quarks emit and absorb gluons   
in the same way as the electrically charged leptons
emit and absorb photons;
\item[--]       
gluon and photon are massless; 

\item[--]  
both gluon and photon have spin 1. 
\end{itemize}

Being guided by this analogy, one would expect
that gluon exchanges generate a Coulomb-type 
interquark force, similar to the usual
attraction/repulsion between the electrically 
charged particles. In reality, quark-gluon interactions are
far more complicated. 
In particular, since the colour charge has three components, 
quarks can change their color states after emitting/absorbing gluons,
as indicated on the diagram in Fig.~\ref{fig:diags}.  Hence, due to the 
colour conservation, gluons also carry colour quantum numbers, 
and, as a result, interact with each other. 
In fact, it is the gluon self-interaction 
that makes QCD-dynamics so peculiar.
Yet QCD has one basic property in common with QED.
Both theories have specific ``internal'' symmetries, named
{\em local gauge symmetries}, to be discussed below.

\subsection{Local gauge symmetry in QED and QCD}  

The  lepton-photon dynamics is  described by  one 
compact formula  of the QED Lagrangian,  
\begin{equation}
L_{QED}(x)
=-\frac{1}{4}F_{\mu\nu}(x) F^{\mu\nu}(x)+
\sum\limits_{l=e,\mu, \tau} 
\bar{\psi}_l(x)(iD_\mu\gamma^\mu-m_l)\psi_l(x)\,,
\label{Lqed}
\end{equation}
where the physical degrees of freedom of the photon field 
(the electric and magnetic fields) are 
combined in the field strength tensor
$F_{\mu\nu}=\partial_\mu A_\nu-\partial_\nu A_\mu$, with  
$\partial_\nu A_\mu\equiv \frac{\partial A_\mu(x)}{\partial x_\nu}$.
In the above, $\psi_l(x)$ is the lepton 
Dirac field with spin 1/2  and mass $m_l$, and a compact notation for 
the covariant
derivative $D_\mu=\partial_\mu+ieA_\mu $ is introduced. 
Starting from $L_{QED}$ it is possible to derive Dirac equations for 
the leptons and Maxwell equations for the photon. 
Furthermore, Eq.~(\ref{Lqed}) yields 
the basics elements of the QED Feynman diagrams:
the photon and lepton propagators and the
lepton-photon interaction vertex.  Employing these elements, one 
obtains the amplitude of a given e.m. process  
in a form of a perturbative expansion in the numerically 
small coupling $\alpha_{em}=e^2/4\pi$.  

For us the most interesting property of $L_{QED}$ 
is the gauge symmetry which reveals itself  
if one locally 
changes the phase of the lepton fields: 
\begin{eqnarray}
\psi_l(x)& \to &\psi'_l(x) =exp\left[-i\chi(x)\right]\psi_l(x)\,,
\nonumber
\\
\bar{\psi}_l(x)& \to &\bar{\psi}'_l(x)=\bar{\psi}_l(x)
exp\left[i\chi(x)\right]\,,
\label{u1}
\end{eqnarray}
$\chi(x)$ being an arbitrary function of the 4-coordinate $x$.
If one simultaneously adds the derivative of the same function
to the photon field:
\begin{equation}
A_\mu \to A'_\mu(x)= A_\mu(x) + \frac{1}{e}\partial_\mu \chi(x)\,,
\label{u1A}
\end{equation}
the Lagrangian (\ref{Lqed}) remains invariant. 
It is easy to check that the transformations (\ref{u1}) and (\ref{u1A}) 
leave intact the physical observables, such as the e.m. current 
$\bar{\psi}_l\gamma_\mu\psi_l$  or $F_{\mu\nu}$.
Importantly, the 
local gauge symmetry implies  that the photon mass vanishes, 
$m_\gamma=0$. Indeed, adding to $L_{QED}$ 
a  mass term $m^2_\gamma A_\mu A^\mu $  for the photon field  
would evidently violate the symmetry
because the latter term changes under (\ref{u1A}). 
The particular case $\chi=const$  (when only the lepton fields
are transformed) corresponds to the 
{\em global} gauge symmetry which is  
responsible for the e.m. 
current conservation in QED.

From the mathematical point of view 
the QED gauge transformations form a {\em group}. Let me remind 
that a given set of elements $\{g_i\}$ can be qualified as a group if three
conditions are simultaneously fulfilled: 1) a multiplication rule 
can be defined $g_i*g_k=g_{l}$, that is, a correspondence is established 
between  
a given pair of elements $g_i$ and $g_k$ and a certain third element 
$g_l$ belonging to the same set; 2) the unit element 
$g_0$ exists, so that $g_0*g_i=g_i$ for each $g_i$ 
and 3) the inverse element $g_k^{-1}$ can be specified 
for  each $g_k$, so that $g_k*g_k^{-1}=g_0$.
All three conditions are  
valid for the (infinite and continuous) set of 
transformations (\ref{u1}) and (\ref{u1A}) generated by the set of 
the arbitrary functions $\chi(x)$. 
Indeed, performing two gauge transformations 
one after the other, with $\chi_1(x)$ and $\chi_2(x)$, 
is equivalent to the gauge transformation with 
$\chi_{12}(x)=\chi_1(x)+\chi_2(x)$.
The unit element of this ``multiplication '' rule 
is simply the transformation with $\chi_0(x)\equiv 0$ and the inverse element 
for each $\chi(x)$ is $-\chi(x)$ . Importantly, the group multiplication
is commutative (the group is Abelian) because the result 
of the overlap of two phase transformations is independent of their
order. 
The group we are discussing is called $U(1)$.
Mathematically, it is equivalent to the group of rotations of 
the Cartesian coordinate system around  one of its axes. 
The rotation angle plays the same role as the phase $\chi$.  
 
Gauge transformations in QCD have  a more rich geometry. 
They are somewhat similar 
to the general rotations in the 3-dimensional space,
involving  $3\times3$ matrices, which do not commute.
More specifically, a colour gauge transformation
of the quark field 
$$
\psi^i_q(x)=
\left(
\begin{array}{r@{}l}
\vspace{.1cm}
&\psi^1_q(x)\\
\vspace{.1cm}
&\psi^2_q(x)\\
\vspace{.1cm}
&\psi^3_q(x)\\
\end{array} \right)\,,
$$
with a given flavour $q=u,d,s,...$,  
involves transitions between different color 
components, a sort of ``rotations of colour coordinates'':
\begin{eqnarray}
\psi^i_q(x) \to  \psi'^i_q(x) = U^{i}_k(x)\psi^k_q(x)\,,\nonumber\\
\bar{\psi}_{q\,i}(x) \to  \bar{\psi}'_{q\,i}(x) 
= \bar{\psi}_k U^{\dagger\, k}_i(x)\,.
\label{su3transf}
\end{eqnarray}
The elements
of the $3\times 3$  matrix $U^{i}_k(x)$ depend arbitrarily on the 4-point $x$. 
Furthermore, this matrix is unitary: 
\begin{equation}
U^{\dagger i}_k U^{k}_j=\delta^{k}_j\,, 
\label{unit}
\end{equation}
or, in a symbolic form, $U^{\dagger}U=1$.  
To explain why unitarity is necessary, 
let me invoke the following physical argument.  
Since  photon does not distinguish quark colours, the only 
admittable form for the e.m. interaction of quarks is
\begin{equation}
L_{em}(x)=
eQ_q\sum\limits_{k=1,2,3}\bar{\psi}_{q\,k}(x) 
\gamma_\mu \psi_q^k(x) A^\mu(x)\,,
\label{lqed}
\end{equation}
where $Q_q=+2/3\,(-1/3)$ for $q=u,c,t\,(d,s,b)$ and 
the summation goes over the quark colour indices.
The gauge transformation (\ref{su3transf}) applied to the quark fields 
yields 
\begin{equation}
L_{em}(x)\to L'_{e.m.}(x) = \bar{\psi}_{q\,i}(x)U^{\dagger i}_k(x)\gamma_\mu 
U(x)^{k}_j\psi_q^j(x)A^\mu(x)\,.
\label{em2}
\end{equation}
Clearly, only if (\ref{unit}) is valid, 
$L_{em}$ remains invariant. 

The usual exponential representation of the gauge
transformation matrix is: 
\begin{equation}
U^{i}_k(x)= \exp\left[-i\sum\limits_{a=1}^{8}
\chi^a(x) \frac{\left( \lambda^a \right)^{i}_k}{2} 
\right]\,.
\label{matrix}
\end{equation}
It contains  eight independent and 
arbitrary functions  $\chi^a(x)$ multiplied by eight 
reference matrices $\lambda^a$ ($a=1,...8$). The latter 
have the form chosen 
by Gell-Mann: 
\begin{eqnarray}
\lambda^1=\left(\begin{array}{lll}
0&1&0\\
1&0&0\\
0&0&0\\
\end{array} \right),~
\lambda^2=\left(\begin{array}{lll}
0&-i&0\\
i&0&0\\
0&0&0\\
\end{array} \right),~
\lambda^3=\left(\begin{array}{lll}
1&0&0\\
0&-1&0\\
0&0&0\\
\end{array} \right),~
\nonumber
\\
\lambda^4=\left(\begin{array}{lll}
0&0&1\\
0&0&0\\
1&0&0\\
\end{array} \right),~
\lambda^5=\left(\begin{array}{lll}
0&0&-i\\
0&0&0\\
i&0&0\\
\end{array} \right),~
\nonumber
\\
\lambda^6=\left(\begin{array}{lll}
0&0&0\\
0&0&1\\
0&1&0\\
\end{array} \right),~
\lambda^7=\left(\begin{array}{lll}
0&0&0\\
0&0&-i\\
0&i&0\\
\end{array} \right),~
\lambda^8=\frac{1}{\sqrt{3}}\left(\begin{array}{lll}
1&0&0\\
0&1&0\\
0&0&-2\\
\end{array} \right).~
\end{eqnarray}
The $\lambda$-matrices have the following properties:
$\lambda^{a \dagger}=\lambda^a$ (hermiticity),
$Tr\lambda^a=0$, and  
$[\lambda^a,\lambda^b]= f^{abc}\lambda^c$ (noncommutativity),
where $f^{abc}$ are totally antisymmetric constants 
($f^{123}=-f^{213}$, etc.). It is easy to check that 
$U(x)$ defined in (\ref{matrix})
obeys unitarity and has a unit determinant $det~U=1$. 
The (infinite and continuous) set of noncommutative 
$U$ matrices forms a group, which is called $SU(3)$.
One may ask: why there are eight independent 
functions $\chi^a$ determining the rotations of the three color states? 
Isn't eight too many? The point is that the quark fields $\psi_q^i$  
are complex functions, hence the matrix $U(x)$  
is also a complex function of $x$
determined by  $2\times 9$ real functions.
Only 8 of them are independent because there are altogether 
10 constraints: nine provided by 
the unitarity relation (\ref{unit}) and  one 
by the unit determinant. With eight ``rotation angles'' 
the group $SU(3)$ is of course quite different from the 
group of rotations in three dimensions which has 
only 3 parameters.

\subsection{QCD Lagrangian} 

The QCD Lagrangian  has to 
be exactly symmetric with respect to the local gauge transformations
(\ref{su3transf}) with matrices (\ref{matrix}).
This property serves as a guiding principle for
constructing $L_{QCD}$. 

We start from the part of the Lagrangian 
which describes the propagation of free quarks:
\begin{equation}
L_{quark}(x)= 
\sum\limits_{q=u,d,s,...}\left(\sum\limits_{k=1,2,3}\bar{\psi}_{q\,k}(x)
(i\partial_\mu\gamma^\mu 
-m_q)\psi_q^k(x)\right)\,,
\label{quarkL}
\end{equation}
and yields the usual Dirac equation for 
spin 1/2 particle for each quark with a given flavour and 
colour. 
Similar to the case of QED, the  
expression (\ref{quarkL}) is not invariant with respect to the 
local gauge transformations (\ref{su3transf}). An additional term
with derivatives of $\chi^a(x)$ remains:
\begin{equation}
L_{quark}(x) \to L_{quark}(x) + \sum\limits_{q=u,d,s,...}
\bar{\psi}_{q\,i}(x)\left[iU^{\dagger\,i}_k(x)\partial_\mu U^k_j(x)
\right]\gamma^\mu\psi_{q}^{j}(x)\,.
\label{quarkp}
\end{equation} 
To restore gauge invariance one follows the same scenario 
as in QED  or in the electroweak theory \cite{Aitchison}. 
The idea (put forward long ago by Yang and Mills)  
is to introduce ``compensating''  spin-1 fields 
interacting with quark fields.  
There should be one separate spin-1 field for each 
of the eight degrees of freedom determining the gauge transformations
(\ref{su3transf}). In this way eight gluons enter the game,
with the following quark-gluon interaction term added to 
the Lagrangian: 
\begin{equation}
L_{int}(x)= 
g_s \sum\limits_{q=u,d,s,...}
\bar{\psi}_{q\,i}(x)\frac{(\lambda^a)^i_k}{2}\gamma^\mu \psi_q^k(x) 
A_\mu^a(x)\,.
\label{lintqcd}
\end{equation}
Here $g_s$ is the dimensionless coupling analogous to $e$ in 
$L_{QED}$. 
In contrast to e.m. interaction, where the photon field is electrically
neutral, the gluon fields also carry colour charge, so that the 
colour state of a quark changes after emitting/absorbing a gluon.
The  colour of the gluon distinguished by the index $a=1...8$  can be  
identified with a superposition of 
quark and antiquark colours. For example the gluon field 
$A_\mu^1(x)$ is in the same colour state  
as the quark-antiquark pair $\bar{\psi}_{q\,1}\psi^2_q 
+ \bar{\psi}_{q\,2}\psi^1_q$.   
The $A_\mu^a$-fields in (\ref{lintqcd})
have to  be gauge-transformed in the following way:
\begin{equation}
\frac{\lambda^a}{2}A_\mu^a(x)
\to U(x)\frac{\lambda^a}{2} A^a_\mu(x)U^{\dagger}(x) - 
\frac{i}{g_s}\partial_\mu  U(x) U^{\dagger}(x)\,,
\label{Atransf}
\end{equation} 
so that  the overall change of $L_{int}$ cancels 
the symmetry breaking term  on r.h.s. of (\ref{quarkp}).
It is a simple exercise to check that the combination of transformations
(\ref{su3transf}) and (\ref{Atransf}) leaves the sum  
$L_{quark}+L_{int}$ invariant. We see that in QCD the ``compensating'' 
transformation of gluon fields (\ref{Atransf}) is more complicated 
than its analog (\ref{u1A}) in QED:
the addition of the derivative over $\chi$-functions 
is accompanied by a ``color rotation''.

To complete the Lagrangian one has to 
add a gauge-invariant term describing the propagation 
of gluon fields :
\begin{equation}
L_{glue}(x)=-\frac{1}4 G^a_{\mu\nu}(x)G^{a\,\mu\nu}(x)\,,
\label{glueL}
\end{equation}
where
\begin{equation}
G^a_{\mu\nu}=\partial_\mu A^a_\nu-
\partial_\nu A^a_\mu+g_s f^{abc}A^b_\mu A^c_\nu\,,
\label{gmunu}
\end{equation}
is the gluon field-strength tensor.
The local gauge invariance of $L_{glue}$ 
implies that gluons are massless. At the same time,
$G^a_{\mu\nu}$ is a more complicated object than its 
QED analog $F_{\mu\nu}$.
Indeed, substituting (\ref{gmunu}) to  
(\ref{glueL}), we notice that not only the terms quadratic
in $A_\mu^a$ emerge (propagation
of gluons) but also the three- and four-gluon vertices 
(gluon self-interactions).  
Note that formally, both properties of gluons: colour 
quantum number and self-interactions 
are due to the noncommutativity
of the gauge-transformation group (the fact that $f_{abc}\neq 0$).
The final form of QCD Lagrangian 
is obtained by adding together the three pieces 
introduced above:
\begin{eqnarray}
L_{QCD}= L_{glue}+L_{quark}+ L_{int}
\nonumber
\\
=-\frac{1}{4}G^a_{\mu\nu} 
G^{a\mu\nu}+\sum\limits_q\bar{\psi}_q(iD_\mu\gamma^\mu-m_q)\psi_q\,,
\label{lqcd}
\end{eqnarray} 
where $D_\mu= \partial_\mu -ig_s \frac{\lambda^a}2 A^a_\mu$.
To summarize, $L_{QCD}$ describes not only 
quark-gluon interactions but also {\em gluodynamics},
the specific gluon self-interactions which have no analog
in QED \footnote{
In QCD "light emits light" at the level of the  
fundamental interactions entering Lagrangian, 
as opposed to QED where light-by-light interaction appears 
only as $O(\alpha_{em}^2)$ quantum correction 
(when photons exchange virtual electrons via box diagrams).
For a classical Maxwellian electrodynamics, 
self-interacting e.m. fields would mean  
``new physics beyond standard theory''. I am not aware 
of any discussions of photon self-interactions 
in the times before quantum field theory.  
Interestingly, the light emitting
light was mentioned in poetry. I found the following sentence 
written in XIII century by  Armenian poet  Kostandin Erznkazi \cite{KE}:
``And so the light was born from the light, the great light of Sun...''
(in translation from Armenian)}.

The quark and gluon propagators and vertices  derived
from $L_{QCD}$  are collected in Table 1. 
The formula of the gluon propagator
has a certain degree of freedom related to the fact that 
the physical massless gluon  has only two polarization/spin states
whereas the field $A_\mu^a$ has four components. To make
things working, one follows the same recipe as in QED. 
An additional constraint on the gluon field is introduced, 
the so called gauge-fixing condition. The gluon 
propagator given in Table~\ref{tab:FR} corresponds to the usual Feynman gauge.
Note that in QCD purely gluonic loop diagrams are possible,
in which case one has to take care of subtracting 
the contributions of unphysical components of $A_\mu^a$ also
in these loops. It is usually done by adding specially designed 
fictitious particles (the so called Fadeev-Popov ghosts) 
which only appear in the loops, and are not shown in Table 1.
\begin{table}
\begin{center}
\begin{tabular}{| c | c |}
\hline
&\\
Quark propagator & 
$
\langle 0|T\{\psi_{q}^i(x)\bar{\psi}_{qk}(0)\}|0\rangle=
i\delta^i_k
\int d^4p \left(\frac{p_\alpha\gamma^\alpha+m_q}{p^2-m_q^2}\right)e^{-ipx}
$\\
&\\
\hline
&\\
Gluon propagator &
$
\langle 0|T\{A^a_\mu(x)A_\nu^b(0)\}|0\rangle=-i\delta^{ab}\int d^4k 
\frac{g_{\mu\nu}}{k^2} e^{ikx}
$\\
&\\
\hline
&\\
Quark-gluon vertex &
$g_s\bar{\psi}_q(x)\gamma_\mu \frac{\lambda^a}{2}\psi_q(x)A^{a\,\mu}(x)$
\\
&\\
\hline
&\\
3-gluon vertex &
$
-\frac{g_s}2 f^{abc}[\partial_\mu A^a_\nu(x)-
\partial_\nu A^a_\mu(x)]A^{b\mu}(x)A^{c\nu}(x)
$\\
&\\
\hline
&\\
4-gluon vertex&
$
-\frac{g_s^2}4 f^{abc}f_{ade} A^b_\mu(x) A^c_\nu(x) A^{d \mu}(x) 
A^{e\nu}(x)
$\\
&\\
\hline
\end{tabular}
\caption{{\it Propagators and vertices in QCD.}}
\label{tab:FR}
\end{center}
\end{table}

Feynman diagrams in QCD are obtained 
employing the quark-gluon propagators and vertices as building blocks.
However, the use of diagrams makes sense
only if the perturbative expansion in $g_s$ is meaningful. 
To obey this condition, the {\em quark-gluon coupling} 
\begin{equation}
\alpha_s = \frac{g_s^2}{4\pi}\,,
\label{alphas}
\end{equation}
the QCD analog of the e.m. coupling $\alpha_{em}=e^2/4\pi$,
has to be sufficiently small, $\alpha_s \ll 1 $.
If this condition is fulfilled, then, for example, 
the $O(\alpha_s)$  diagram of the quark-quark
interaction in Fig.~\ref{fig:diags} dominates over the higher-order 
diagrams with additional gluon exchanges. One has then 
a tractable situation,  
with quarks and gluons propagating quasi-freely.
Thus, there is an important question to be addressed: 
how large is $\alpha_s$ ?

\subsection{Running of the quark-gluon coupling} 

To answer the above question,
one has to investigate quantum effects in QCD, i.e.,
creation and annihilation of virtual 
gluons and quarks described by loop diagrams. 
In QED, which we use as a prototype, quantum loops 
generate very small effects, such as      
Lamb shift (the correction to the Coulomb force due 
to the virtual electron-positron  pairs) or the $O(\alpha_{em})$ 
one-loop correction to the muon magnetic moment. 
These effects, being accessible in precision experiments,
play  a minor role in the bulk of electromagnetic processes. 

\begin{figure}
\begin{center}
\includegraphics*[width=14cm]{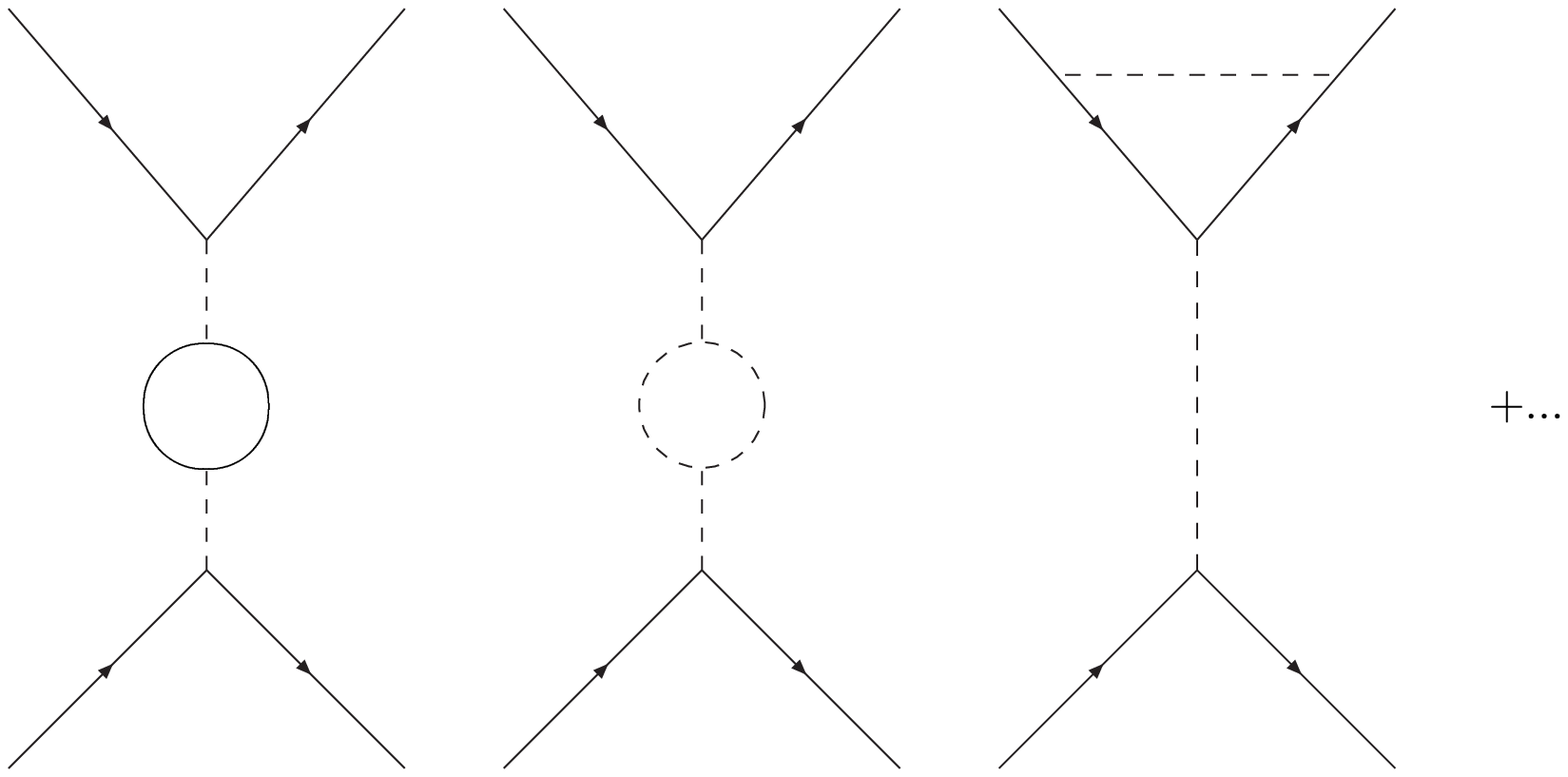}
\caption{{\em Some of the diagrams corresponding to 
quantum loop corrections  to the quark-quark scattering}}
\label{fig:loops}
\end{center}
\end{figure}

In QCD, quark-gluon  loops are far more 
pronounced  and play a crucial role in determining $\alpha_s$.
To have a closer look, let us consider 
the quark-quark scattering amplitude.
In the lowest-order (at the tree level) 
the amplitude is given by the one-gluon exchange diagram 
shown in Fig.~\ref{fig:diags}
for the particular choice of quark flavours. 
Having at hand the Feynman rules in QCD it is  
easy to write down the amplitude:
\begin{equation}
 {\cal A}= \frac{4\pi \alpha_s}{q^2}
\left(\bar{\psi_s}_i \gamma_\mu 
\frac{(\lambda^a)^i_{k}}2
\psi_s^k\right) 
\left(\bar{\psi}_{d\,j} \gamma^\mu 
\frac{(\lambda^a)^j_{l}}2
\psi_d^l\right)\,, 
\label{zeroth}
\end{equation}
where $q^2<0$ is the momentum transfer squared. Below
I will also use the notation 
for the momentum scale: $Q\equiv \sqrt{Q^2}$, where 
$Q^2=-q^2>0$.  
Considering $O(\alpha_s)$ corrections to the amplitude
(\ref{zeroth}) one encounters diagrams shown in Fig.~\ref{fig:loops}.
They contain gluon or quark loops 
inserted within the exchanged gluon line or in the vertices. 
These loop effects turn out to be extremely important for 
evaluating $\alpha_s$.

Let me first explain the loop diagram calculation in more detail.
After substituting propagators of virtual particles, one arrives at    
four-dimensional Feynman integrals over the 4-momentum
$k$ flowing inside the loop, typically: 
\begin{equation}
I_{loop}(q,m_q)=\int \frac{d^4k}{(2\pi)^4}\frac{...}{(k^2-m_q^2)((k+q)^2-m_q^2)}\,,
\label{loopint}
\end{equation}
where the explicit expression in the numerator (depending on the spin  
of the loop particles) does not play role in our discussion.
For simplicity, I will also put to zero the quark masses, 
a reasonable approximation if $Q\gg m_q$.
To calculate loop integrals, one usually
employs the method of dimensional regularization. The idea is to lower 
the number of dimensions in the integral (\ref{loopint}) making it convergent. 
One replaces 4 by a generic integer number $D$, then calculates 
the integral as a function of $D$ and after that 
considers the result at an arbitrary noninteger $D= 4-\epsilon$, 
schematically:
\begin{equation}
I_{loop}(q,0)=\int d^4k f(k,q)\to \mu^{(4-D)} \int d^D k f(k,q)= I(q,\mu,D)
\to I(q,\mu,4-\epsilon)\,.
\label{intI}
\end{equation}
The auxiliary mass scale $\mu$ is introduced to
keep unchanged the physical dimension of the integral.
The major advantage of dimensional regularization 
is  in preserving  gauge symmetry of the amplitudes
at each step of the calculation. 
The  divergent part of the integral in this scheme 
is represented in a form of terms proportional to $1/\epsilon$. 
A generic expression for the loop integral has a form:
\begin{equation}
I_{loop}(q,0) \rightarrow\mu^{4-D}\int
\frac{d^Dk}{(2\pi)^D}\frac{...}{k^2(k+q)^2} =
I_1\log(-q^2/\mu^2)+\frac{1}{\epsilon} +I_2\,.  
\label{loop}
\end{equation}
where $I_{1,2}$ are calculable finite coefficients.

QCD is a {\em renormalizable} 
theory (similar to QED),  which means one can absorb 
all divergent $1/\epsilon$ terms into the so called 
$Z$ -factors. Multiplying by $Z^{-1}_g$ the coupling $g_s$, 
and by the corresponding factors the quark masses, quark and gluon
fields, one defines the finite (renormalized) quantities; 
e.g., the renormalized coupling is $g_s^{ren}= Z_g^{-1} g_s$. 
Having in mind the validity of the renormalization procedure 
I will simply ignore divergent terms appearing in (\ref{loop})
and in other loop integrals.

Adding the $O(\alpha_s)$ diagrams in Fig.~\ref{fig:loops} 
to the tree-level amplitude, results in the same 
expression (\ref{zeroth}), but with $\alpha_s$ replaced by 
an {\em effective coupling} depending on the momentum scale:
\begin{equation}
\alpha_s^{eff}(Q)\equiv \alpha_s\left[1-\frac{\alpha_s}{4\pi}
\left(\beta_0 \log\frac{Q^2}{\mu^2}+const \right)\right]\,,
\label{alphaseff}
\end{equation}
where a shorthand notation
\begin{equation}
\beta_0= 11-\frac23 n_f
\label{eff1}
\end{equation}
is introduced. Here $n_f$ is the number of ``active'' quark flavours
in the loop diagrams. Only those quarks which have $m_q\ll Q$ contribute 
to $\alpha_s^{eff}(Q^2)$. 
Importantly, $\beta_0$ is positive, because the term 11 
originating from the gluon loops
exceeds the quark-loop contribution $2n_f/3$ (since $n_f<6$ in any case).

Taking $\alpha_s^{eff}$ at a different scale $Q_0$,
\begin{equation}
\alpha_s^{eff}(Q_0)\equiv \alpha_s\left[1-\frac{\alpha_s}{4\pi}
\left(\beta_0 \log\frac{Q_0^2}{\mu^2}+const \right)\right]\,,
\label{eff2}
\end{equation}
and dividing (\ref{alphaseff}) by (\ref{eff2})
one obtains, with an accuracy of $O(\alpha_s^2)$: 
\begin{equation}
\alpha_s^{eff}(Q) = \alpha_s^{eff}(Q_0)\left[1-\frac{\alpha_s^{eff}(Q_0)}{4\pi}
\beta_0 \log \frac{Q^2}{Q_0^2}\right ]\,,
\label{alphas1}
\end{equation}
the relation between effective couplings at two different scales.
The approximation (\ref{alphas1}) is valid only 
if $\alpha_s^{eff}(Q_0)$ 
is sufficiently small and the higher-order corrections 
are negligible. Remarkably, (\ref{alphas1}) predicts
that $\alpha_s^{eff}(Q)$ decreases 
when the momentum scale $Q$ increases. Thus, 
if the perturbative expansion 
in  $\alpha_s$ is applicable at certain $Q_0$, it should  
behave even  better   at $Q>Q_0$.

Note that (\ref{alphas1}) still has to be improved.
At $Q\to \infty$ the logarithm becomes 
very large driving the combination $\alpha_s \log Q^2$ to rather big
values, so that the $O\left([\alpha_slog\left(Q^2/Q_0^2\right]^2\right)$ 
correction 
originating from the two-loop diagrams shown in Fig.~\ref{fig:multiloops}
becomes important, and the whole perturbative construction is again in danger. 
Fortunately, a systematic resummation of all 
$O\left([\alpha_s log\left(Q^2/Q_0^2\right]^n \right)$ corrections 
is possible. In practice, one does not need to calculate 
all multiloop diagrams, which would be a tremendous work. 
Instead, the renormalization-group method is used, 
which is however beyond our scope. 
The resulting expression for the {\em running coupling} is:
\begin{equation}
\alpha_s(Q)=\frac{\alpha_s(Q_0)}{1+\frac{\alpha_s(Q_0)}{4\pi}\beta_0\,.
\log\frac{Q^2}{Q_0^2}}\,.
\label{alphasrun}
\end{equation}
(hereafter the superscript '$eff$' is omitted).
Expanding the denominator in (\ref{alphasrun})
and retaining only the first two
terms we return to the relation (\ref{alphas1}).

\subsection{Asymptotic freedom}
A consistent use of the running QCD coupling
is possible if one can find a range 
of $Q$ where $\alpha_s(Q)$ is numerically small. 
The first indications that $\alpha_s$  is indeed 
small at large momentum transfers were obtained in the studies 
of deep-inelastic lepton-nucleon scattering (to be
discussed in Lecture 3). This remarkable discovery paved the way 
for using QCD perturbation theory with the running coupling 
in many other processes. 

To trace the numerical behavior of $\alpha_s(Q)$ 
one has to fix the coupling at a certain large scale 
using an experimental input. 
One possibility is the decay of $Z$-boson
to a quark-antiquark pair. Quarks originating in this decay 
inevitably  build hadrons in the final state (see the next subsection).
To avoid hadronic complexity, one measures the total inclusive 
width $\Gamma(Z\to hadrons)$, so that the probabilities 
of all possible quarks $\to$ hadrons transitions add up to a unit. 
The majority of $Z\to hadrons$ events observed at LEP 
has a spectacular structure of two hadronic jets 
originating from the initial, very energetic quark and antiquark
($E_q=E_{\bar q}=m_Z/2$ in the $Z$ rest-frame), see, e.g. \cite{jets}. 
On the other hand, the  share of $\geq 3$-jet  events in $Z\to hadrons$, with additional jets originating from gluons and/or from 
secondary quark-antiquark pairs, is small.
This observation clearly indicates that the quark-gluon 
coupling at the scale $m_Z$ is small, or in other words 
the initial quark pair interacts weakly during the short time after
its creation. 

\begin{figure}
\begin{center}
\includegraphics*[width=12cm]{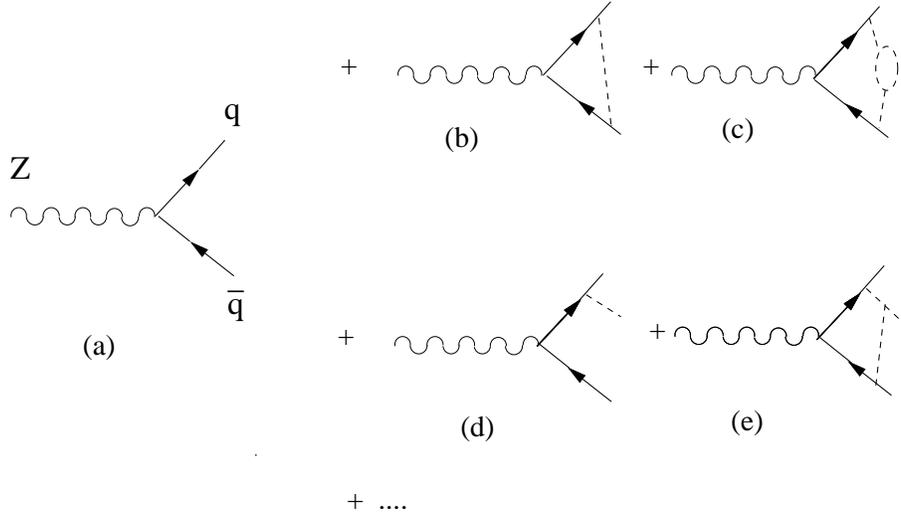}
\caption{{\em  The lowest-order diagram (a) and some of the 
higher-order in $\alpha_s$ diagrams (b-d) determining the
total width $Z\to hadrons$.}}
\label{fig:Zhad}
\end{center}
\end{figure}

The perturbative diagrams of $Z\to q\bar{q}$ ( $q=u,d,s,c,b$)
including the gluon emission $Z\to \bar{q}q G$ 
are shown  in Fig.~\ref{fig:Zhad}. 
\begin{figure}
\begin{center}
\includegraphics*[width=8cm]{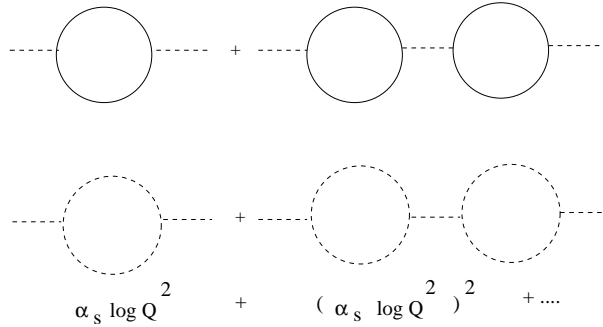}
\caption{{\em  One- and two-loop diagrams
contributing to the running of $\alpha_s$.}}
\label{fig:multiloops}
\end{center}
\end{figure}
Evaluating these diagrams  one gets the perturbative expansion
for the total hadronic width, schematically: 
\begin{eqnarray}
\Gamma(Z\to hadrons)= 
\sum\limits_{q=u,d,s,..}\Bigg[
\Gamma (Z\to \bar{q}q)\left(1+C^q_1\alpha_s(m_Z)+C^q_2\alpha_s^2(m_Z)
+ ...\right)
\nonumber
\\
+\Gamma (Z\to \bar{q}q G)\left(1+C^{qG}_1\alpha_s(m_Z)+...
\right)+...\Bigg]\,,
\label{zhadrons}
\end{eqnarray}   
where $\Gamma (Z\to \bar{q}q)$ and 
$\Gamma (Z\to \bar{q}q G)$ (the latter starting from $O(\alpha_s)$)
are the corresponding perturbative widths, 
and $C^{q}_{1,2},C^{qG}_1,..$ are 
the calculable coefficients. Smallness of $\alpha_s$ allows
to neglect all higher-order corrections starting, say from $O(\alpha_s^3)$.
In the above expression $\alpha_s$ is taken at the scale $m_Z$, 
the characteristic scale in this process.
One can trace how the running coupling  
builds up in (\ref{zhadrons}), by summing up all logarithmic 
corrections generated by the loop insertions similar to the one
shown in Fig.~\ref{fig:multiloops}c.

Comparing the result (\ref{zhadrons}) of 
the theoretical calculation with the experimental data 
yields \cite{PDG}: 
\begin{equation}
\alpha_s(m_Z)\simeq 0.12.
\label{alphasZ}
\end{equation}     
\begin{figure}
\begin{center}
\includegraphics*[width=14cm]{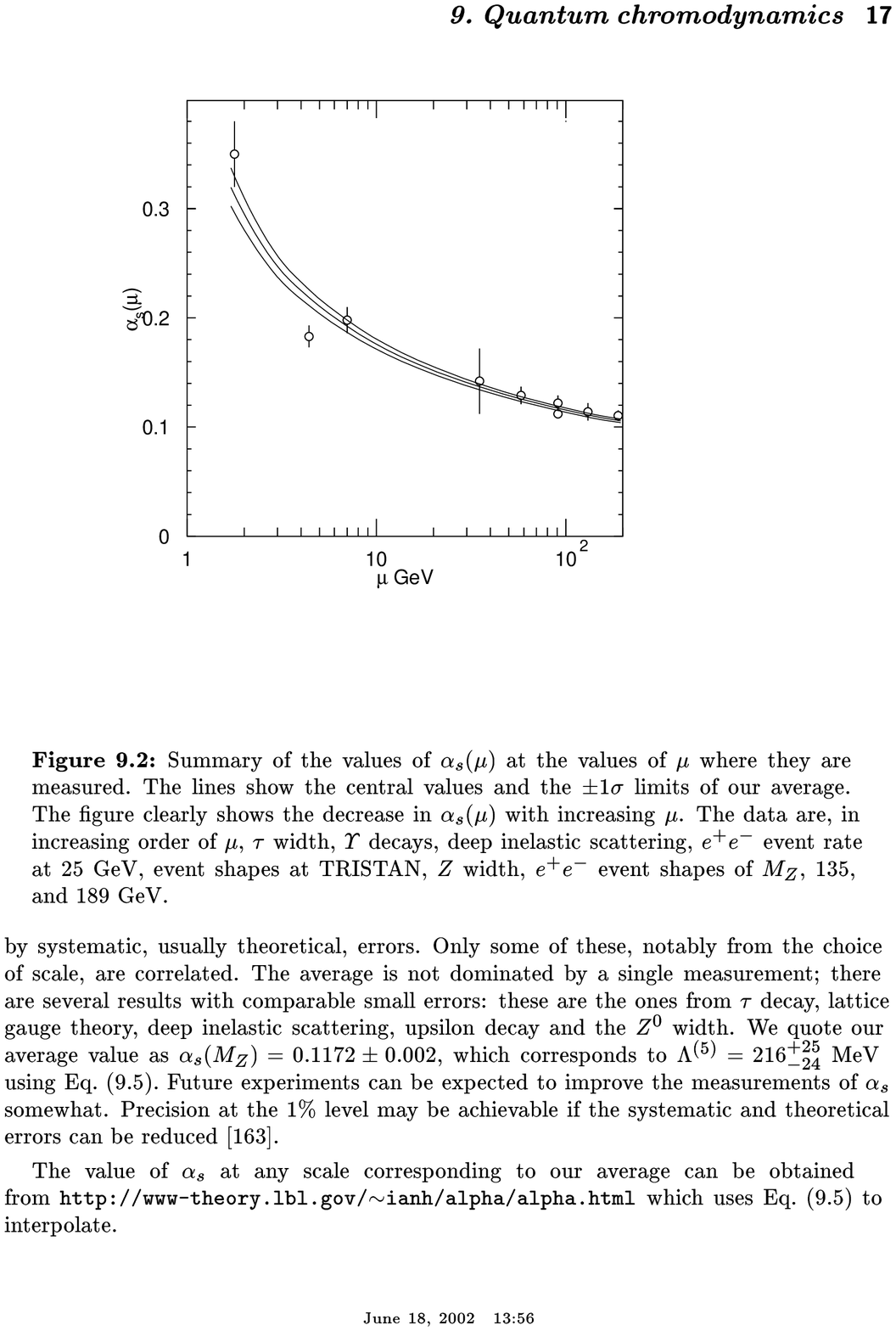}
\caption{{\em  The dependence of $\alpha_s(\mu)$ on the 
running scale $\mu$ \cite{PDG}: 
the upper and lower curves  indicate the theoretical uncertainty,
the points with errors are the values of 
$\alpha_s$ extracted at different scales  using various methods. }}
\label{fig:alphasPDG}
\end{center}
\end{figure}
As expected, it is a rather small number, 
so that the whole perturbative treatment turns out to be quite consistent. 
Using (\ref{alphasrun}) and (\ref{alphasZ}),
one  predicts  $\alpha_s(Q)$ at lower scales.
The curve plotted in Fig.~\ref{fig:alphasPDG} is taken from \cite{PDG} 
and reflects the current status of the running coupling,
including all known (and reasonably small) 
higher-order corrections to (\ref{alphasrun}).
As can be seen in this figure, there is a wide 
region spreading up to $Q\sim $ 1 GeV, where $\alpha_s$ is 
small and perturbative QCD is applicable. 
Furthermore, $\alpha_s$ extracted from various processes
at different scales agrees with the running 
behavior predicted in QCD.
The most spectacular consequence of (\ref{alphasrun})
is the vanishing of the running quark-gluon coupling
at $Q\to \infty$, revealing that QCD is asymptotically 
free.

\subsection{Confinement and hadrons}

\begin{figure}
\begin{center}
\includegraphics*[width=8cm]{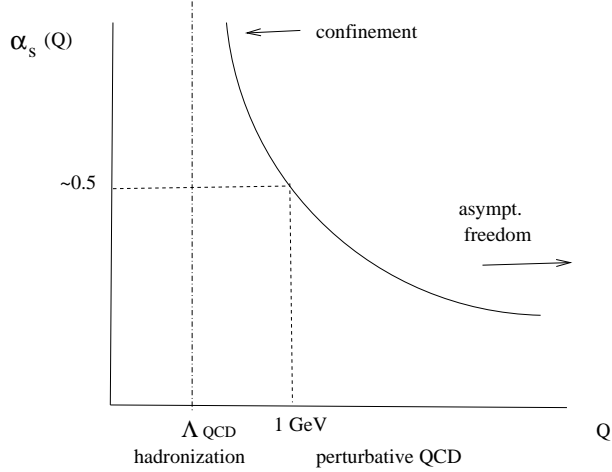}
\caption{{\em  Schematic view of the  $\alpha_s$ behavior at 
different scales.}}
\label{fig:alphaslow}
\end{center}
\end{figure}

Quite an opposite situation takes place in the quark-gluon
interactions at small momentum transfers (at long distances).  
According to (\ref{alphasrun}), if one starts from $Q\gg 1$~GeV 
and  drifts towards smaller scales, 
$\alpha_s$ grows to $O(1)$ at 
$Q < 1 ~\mbox{GeV}$ (see Fig.~\ref{fig:alphaslow}).  
Perturbation theory becomes useless in this region,
because e.g. in the quark-quark scattering, 
an infinite amount of  higher-order quark-gluon diagrams   
has to be taken into account. 
Moreover, at a certain momentum scale 
denoted by $\Lambda_{QCD}$ the denominator in (\ref{alphasrun}) 
vanishes and $\alpha_s(Q)$ diverges. 
The relation between $\alpha_s(Q)$ and $\Lambda_{QCD}$
following from (\ref{alphasrun}) is:
\begin{equation}
\alpha_s(Q)= \frac{2\pi}{\beta_0\log\left(\frac{Q}{\Lambda_{QCD}}
\right)}\,.
\label{lambda}
\end{equation}
The experimentally measured value (\ref{alphasZ}) corresponds, roughly, 
to 
\begin{equation}
\Lambda_{QCD}=200-300~\mbox{MeV}. 
\label{lambdaQCD}
\end{equation}
I skip some details concerning the dependence
of $\Lambda_{QCD}$ on the number of active quark flavours, and
on the theoretical scheme of the loop-diagram evaluation,
relevant for the higher-order corrections to (\ref{lambda}).
The important fact is that (\ref{lambda}) and (\ref{lambdaQCD})
were derived neglecting the quark masses. We could have put
to zero the quark masses in $L_{QCD}$ from the very beginning, 
starting with a  theory without dimensionful parameters.
The emergence of an intrinsic energy scale in a theory with dimensionless 
coupling $g_s$ ({\em dimensional transmutation})  is 
a specific property of QCD, due entirely to 
quantum effects.

The breakdown of perturbation theory and the 
exploding behavior of $\alpha_s(Q)$ at $Q\to \Lambda_{QCD}$
are actually anticipated.
Long before QCD was invented it was known that 
at long distances quarks and antiquarks strongly  interact
and form {\em hadrons}, the quark-antiquark ({\em meson})
and 3-quark ({\em baryon}) bound
states. The properties of hadrons 
will be considered in a more
systematic way in the next Lecture. 
For the present discussion it is important
that the characteristic energy scale of  
hadronic interactions is of
$O(\Lambda_{QCD})$. Hence, it is quite natural that 
the formation of hadrons is due to the  
strong, nonperturbative quark-gluon force emerging 
in QCD at momenta $\sim \Lambda_{QCD}$.

Moreover, hadronic matter is the only observable 
form of quarks and gluons at long distances. In any process
of quark and antiquark production,  
independent of the energy/momenta involved, 
quarks form hadrons in the final state 
\footnote{There is one exception:
t-quark, decaying via weak interactions, is too short-lived 
to be bound by quark-gluon forces.}. 
Note that in QED the situation is quite different: 
isolated leptons and other electrically charged particles 
are observed, and the e.m. bound states 
(e.g., hydrogen atom, positronium or muonium)
can always be split into constituents if a sufficient 
energy is supplied.

In QCD, the non-observation of free colour-charged particles  
(quarks, antiquarks and gluons) is 
arranged in a form  of  the {\em colour confinement}
principle, postulating that all observable states, 
i.e., all hadrons, have to be colour-neutral. 
The confinement principle 
was never rigorously proved, because of our limited
ability to work with QCD beyond perturbation theory.
Nevertheless, all experimental results concerning hadrons,
as well as lattice simulations of QCD at 
long distances, unambiguously support colour confinement.

Before the era of QCD the search for free quarks 
was very popular among experimentalists. 
The fractional electric charge was a 
smoking-gun signal to be observed.
The hunt for quarks was a part of many accelerator
experiments. Not surprisingly,
free quarks have never been found at accelerators or in other places 
(in  cosmic rays, water, ice, meteorites etc.).
Nowadays, one would hardly invest efforts in the search for free quarks.
We are confident that QCD obeys confinement.

\subsection{Quark masses}

The current intervals of  quark masses presented
in the Particle Data Tables \cite{PDG} are:  
\begin{eqnarray}
&m_u= 1.5 \div 4.5~\mbox{MeV},& m_c= 1.0 \div 1.4~\mbox{GeV},~~m_t= 174.3\pm 5.1~\mbox{GeV};
\nonumber \\ 
&m_d= 5 \div 8.5~\mbox{MeV},&m_s= 80\div 155~\mbox{MeV},
~~m_b= 4.0 \div 4.5~\mbox{GeV}\,. 
\label{qmasses}
\end{eqnarray}
The fact that the masses are spread over five orders of magnitude, 
is a reflection of some fundamental flavour physics not related to QCD,
e.g. the Higgs mechanism of the Standard Model.
Hence, the masses $m_q$  
entering QCD Lagrangian (\ref{lqcd}) are  
``external'' parameters. At the same time quark masses are evidently
changed in the presence of quark-gluon interactions.  
At long distances, within hadrons, each 
quark acquires, roughly speaking,  an extra addition 
of $O(\Lambda_{QCD})$ to its ``bare'' mass. It is however very
difficult, if not impossible at
all, to define this {\em constituent} quark mass in a 
model-independent way. Representing
the hadron mass as a sum of the constituent quark masses
plus some interaction energy, e.g. for a meson:
\begin{equation} 
M_{meson}= m_q^{constit}+ m_{\bar{q}}^{constit}+E_{int}\,,
\end{equation}
one can always redistribute the part of 
$E_{int}$ between the quark and antiquark masses.
Since free, on-shell quarks are not observed,
the usual definition  of the particle mass  
(the minimal possible  energy of the one-particle state) 
does not work.

\begin{figure}
\begin{center}
\includegraphics*[width=8cm]{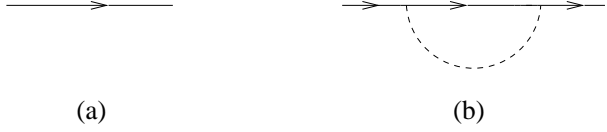}
\caption{{\em The free quark propagator (a) and 
the gluon correction to it (b). }}
\label{fig:prop}
\end{center}
\end{figure}

The mass values presented in (\ref{qmasses}) have 
nevertheless quite a definite meaning.  These are 
``short-distance'' masses  of the virtual quarks.
As we already know, at large virtualities  quark propagation 
is quasi-free and a  consistent 
use of perturbation theory derived from  QCD Lagranigian
is possible. In particular, the free quark propagator 
\begin{equation}
S(p)=\frac{p_\alpha\gamma^\alpha+m_q}{p^2-m_q^2}\,
\label{prop}
\end{equation}
is applicable at $|p^2|\gg \Lambda_{QCD}^2,m_q^2$, 
with the bare mass  $m_q$ from $L_{QCD}$. 
Furthermore, at short distances the 
quark-gluon interactions are calculable in terms of series in $\alpha_s$.
The quark propagating at short distances can emit and absorb
a gluon (see diagram  (b) in Fig.~\ref{fig:prop}). From the experience 
with the running coupling one expects such loops
to be important.    
Adding the loop diagram to the free propagator 
yields the same expression as (\ref{prop})
with $m_q$ replaced by  an effective mass 
depending on the momentum scale 
\begin{equation}
m^{eff}_q(Q)=m_q \left [1-\frac{\alpha_s}{4\pi}\left(\gamma_0 \log \frac{Q^2}{\mu^2} 
+const \right)\right]\,,
\label{meff}
\end{equation}
where I omit the divergent part, knowing that it will 
be absorbed by renormalization. In this case the scale 
$Q\sim \sqrt{|p^2|}$ is determined by the virtuality of the quark
and $\gamma_0=4$ is the result of the explicit calculation.
Again, as in the case of $\alpha_s$, 
we can relate the effective masses at two 
different scales. Writing down the above equation for another 
scale $Q_0$ and using instead of $m^{eff}_q$ a conventional notation
$\overline{m}_q$ we obtain:  
\begin{equation}
\overline{m}_q(Q)=\overline{m}_q (Q_0)\left (1-
\left( \frac{\gamma_0}{\beta_0}\right)
\frac{\alpha_s(Q_0)}{4\pi}\beta_0 \log \frac{Q^2}{Q_0^2}\right)\,,
\label{meff2}
\end{equation}
where I have multiplied and divided the logarithmic term by $\beta_0$
and used $\alpha_s=\alpha_s(Q_0)$ which is correct 
with $O(\alpha_s)$ accuracy. With the same accuracy 
the expression in parentheses can be transformed further:
\begin{equation}
1-\left( \frac{\gamma_0}{\beta_0}\right)
\frac{\alpha_s(Q_0)}{4\pi}\beta_0 \log \frac{Q^2}{Q_0^2}\simeq
\left( 1-\frac{\alpha_s(Q_0)}{4\pi}\beta_0 \log \frac{Q^2}{Q_0^2}
\right)^{\gamma_0/\beta_0}\,.
\label{meff3}
\end{equation}
Using (\ref{alphas1}) we obtain 
\begin{equation}
\bar{m}_q(Q)=\bar{m}_q(Q_0)
\left(\frac{\alpha_s(Q)}{\alpha_s(Q_0)} \right)^{\gamma_0/\beta_0}\,,
\label{runmass}
\end{equation}
the formula for the {\em running mass}. A more rigorous 
derivation is possible using the renormalization group method. 
Also in recent years, the higher-order corrections to 
(\ref{runmass}) have been calculated.

The masses presented in \cite{PDG} are the running masses
(also called $\overline{MS}$ masses if one specifies the 
particular renormalization procedure) normalized at some large scale.
The light $u,d,s$ quark masses are traditionally 
taken at $Q=2$ GeV, whereas a more appropriate scale for the 
heavy $c$ and $b$ quark masses
is the quark mass itself, $Q=m_{c}$ and $Q=m_b$, respectively, 
(which means, e.g., the virtuality of 
the $c$ quark is $p^2=-m_c^2$).  
The fact that quark masses run with the scale, 
is in accordance with the absence of  
isolated quarks among observable states.

\subsection{Two branches of QCD}

To summarize, QCD yields two qualitatively 
different pictures of quark-gluon interactions:

1) at high momentum-transfers i.e., at short average distances, 
perturbative expansions in $\alpha_s$  
are applicable in terms of Feynman diagrams with 
quark and gluon propagators and vertices.
In this region the scale-dependence (running) of the coupling
and quark masses should be properly taken into account.

2) at low scales, that is, at long distances,  
one loses control over perturbative interactions 
between individual quarks and gluons; the latter strongly interact
and form hadrons.  

Accordingly, QCD is being developed in two different 
directions. The first one deals with short-distance physics accessible 
at high energies. One studies specific processes/observables
calculable (at least partly)  in a form of a perturbative expansion in
$\alpha_s$.  A typical short-distance process is the jet production
in $Z$ decays considered above, other examples will be presented
in Lecture 3. 

The second direction deals with  
nonperturbative quark-gluon interactions at long distances
and with hadron dynamics.  
A complete analytical evaluation of 
hadronic masses and other parameters directly from $L_{QCD}$  
is not yet accessible. Instead, a powerful numerical 
method of simulating QCD on the space-time lattice has been developed. 
Lattice QCD has become a separate field, which 
is beyond the scope of these lectures
(for a pedagogical introduction see e.g.,\cite{lattice}). 
Still there is a lot of interesting
advances in the long-distance 'branch'' of QCD, so that
I will only be able to cover a part of them.  
As demonstrated in Lecture 2, many important features
of hadron spectroscopy follow from QCD  
at the qualitative level. The relation 
of long-distance dynamics to the nontrivial structure of 
the QCD vacuum  will be discussed in  Lecture 4. 
An approximate analytical method of QCD sum rules 
based on this relation  and used to calculate hadronic parameters 
will be overviewed in Lecture 5.

One might think  that physics of hadrons  
plays a secondary role, because  
the most important direct tests of QCD  in terms of 
quarks and gluons are done at short distances.
Let me emphasize the fundamental importance of hadron dynamics 
by mentioning two topical problems:

1) {\it The origin of the nucleon mass}
 
Proton  and neutron are the lowest and most stable baryons, 
with the quark content $uud$ and $udd$, 
respectively. Their masses 
\begin{equation}
m_p\simeq m_n\simeq 940~\mbox{MeV}\,,
\label{mqproton}
\end{equation}
are substantially larger than the tripled  quark mass 
$m_{u,d}= O(\mbox{ few ~MeV})$.
We conclude that $\sim 99\%$ of the baryon matter in the Universe 
is not related  to the  ``fundamental'' quark masses 
generated by  the Higgs mechanism or from some other flavour dynamics.
The  bulk of the baryon mass is  
due to the long-distance quark-gluon interactions. 
Indeed, adding a ``constituent''
mass of $O(\Lambda_{QCD})$ to each quark, one gets 
the order-of-magnitude value of $m_{p,n}$. Certainly, the 
problem of the nucleon mass is quite fundamental and has 
to be solved within long-distance QCD.

2) {\it Extracting electroweak parameters
from $B$ decays} 

\begin{figure}
\begin{center}
\includegraphics*[width=8cm]{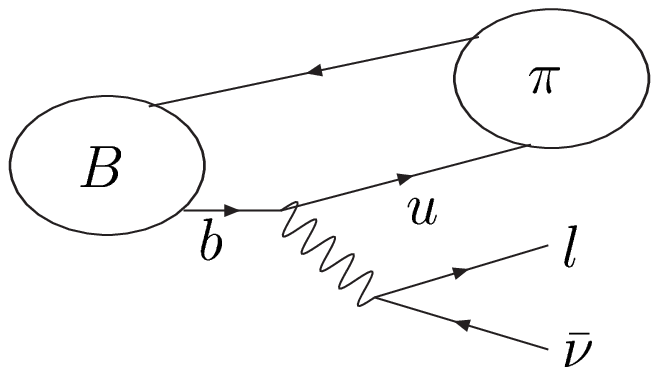}
\caption{{\em $B\to \pi l \nu_l$ decay involves a 
hadronic transition between $B$ and $\pi$ states.}}
\label{fig:Bpi}
\end{center}
\end{figure}

The weak decays of $B$ mesons (the bound states 
of $b$ quark and light antiquark) represent a valuable source
of information on fundamental aspects
of electroweak interactions, such as  
the quark-mixing CKM matrix and the origin of CP-violation 
(see \cite{Fleischer}). 
One topical example is the $B \to \pi l \nu_l$ decay 
observed at $B$ factories \cite{PDG}. This decay 
(see Fig.~\ref{fig:Bpi}) is 
driven by the $b\to u$ weak transition, proportional 
to $V_{ub}$, one of the poorly known  CKM matrix elements. In order
to extract this fundamental parameter from
the experimentally measured partial width, one needs
to divide out the hadronic $B\to \pi$ 
transition amplitude (form factor). The latter 
is essentially determined  by the long-distance quark-gluon
interactions. 

Below, in Lecture 5 we will discuss the (approximate) solutions
of the two abovementioned problems.


\section{FROM QUARKS TO HADRONS } 

\subsection{Mesons and baryons}
Let us now have a closer look at the properties of
hadrons \footnote{ The word 'hadron' was coined  
by Okun in \cite{okun}
where he wrote: ''It is reasonable to call strongly
interacting particles hadrons, and the corresponding decays - hadronic.
In Greek the word hadros means ``large'',''massive'', in contrast to 
the word ``leptos'', which means ``small'',''light''. The term hadron 
refers to long-lived mesons and baryons, as well as to resonances.``}, 
the bound states of quarks. 
There are lots of  
experimental data on hadrons 
accumulated in the Review of Particle Physics \cite{PDG}.       
As we shall see, employing various 
symmetries of QCD, it is possible to predict, at least qualitatively, 
many observable regularities of hadronic spectra 
and interactions.

The pre-QCD quark model of hadrons 
failed to explain why only mesons ($\bar{q}q$), baryons ($qqq$)
and  antibaryons $(\bar{q}\bar{q}\bar{q})$
are observed. Why, for example, the diquark ($qq$) or 
four-quark ($qqqq$) bound states are absent? 
In QCD, one immediately gets an explanation
based on the colour confinement principle. The quark-antiquark
and three-quark combinations can form  
colour-singlet states, whereas diquark and four-quark 
compounds are always colour-charged. 
The colour-singlet meson state is obtained by simply 
summing over the colour indices of the quark and antiquark. 
For example, the $\pi^+$-meson has the following 
flavour/colour structure:
\begin{equation}
|\pi^+ \rangle =\frac1{\sqrt{3} }\sum\limits_{i=1}^3 |u^i \bar{d}_i\rangle\,. 
\label{pion1}
\end{equation}
The way  baryons are built  
is less trivial. Two quarks are arranged in a coloured 
{\em diquark} state $\epsilon_{ijk}|q^jq^k\rangle $,  
where $\epsilon_{ijk}$ is totally antisymmetric 
with $\epsilon_{123}=1$. This state has the  
colour SU(3)-transformation properties of  
an antiquark (with a colour index $i$).
Hence, we can obtain a colour-neutral state 
combining diquark with the third quark 
and summing over colours in the same way as 
in mesons. One important example of a three-quark baryon state   
\footnote{
Note that if there were  four colours of quarks,
with the corresponding SU(4) symmetry,  
baryons would have been built from four quarks,
with profound consequences for the 
physical world, e.g., atoms with fractional electric charge.} 
is $\Omega^-$ baryon with spin 3/2: 
\begin{equation}
|\Omega\rangle = \frac1{\sqrt{6} }\sum\limits_{i,j,k=1}^3 
\epsilon_{ijk}|s^i s^j s^k\rangle. 
\label{baryon}
\end{equation}
With respect to 
the flavour and spin, this state is totally symmetric: 
all quark have the same flavour and parallel spins. 
Therefore, colour degrees of freedom provide
the antisymmetry demanded by 
the Pauli  principle for any bound state of fermions.

\subsection{Quark model of hadrons}

The hadron decompositions (\ref{pion1}) and (\ref{baryon})
resemble chemical formulae, displaying the content of a 
composite state. 
In reality hadrons are far more complicated than atoms 
and molecules, because the masses of 
the $u,d,s$ quarks are smaller than the QCD long-distance scale: 
$m_{u,d}\ll \Lambda_{QCD}$ and $m_s < \Lambda_{QCD}$.
Hence, light quarks are purely relativistic and 
the number of quarks, antiquarks and gluons within a hadron 
cannot be fixed. Since QCD is a quantum field theory, 
additional quark-antiquark pairs
or gluons are created and annihilated
inside the bound states {\em virtually}, 
i.e. within the time/distance intervals of $O(1/\Lambda_{QCD})$.
As a result, the general decomposition of the 
physical pion state is not simply (\ref{pion1}) but rather 
\begin{equation}
\mid \pi^+\rangle = ~\mid \!u^i \bar{d}_i  \rangle +
\sum_{q=u,d,s,..}\!\!\mid u^i \bar{d}_i q^j \bar{q}_j \rangle +
\mid \!u^k (\lambda^a)^i_k \bar{d}_i G^a\rangle +
...\,,
\label{decomp}
\end{equation}
where $G$  denotes a gluon, the Lorentz indices are not specified 
and a summation over colour indices is implied. In the above sum, 
the first term represents a state with the minimal 
particle content  (the so called valence-quark content), and 
ellipses indicate all other multiparticle fluctuations. 
Naturally, all components   
of the pion state have to be colourless, with the same 
$u \bar{d}$ overall flavour, to obey the colour-neutrality
and flavour conservation.
We conclude that hadrons are, in general, many-body systems with 
relativistic constituents. Therefore, 
simple quantum-mechanical models with an interquark potential
cannot fully describe pion or other light-quark hadrons. 

Before attempting to solve the QCD dynamics, it is useful
to apply the symmetries of QCD Lagrangian.
The space-time (Lorentz-Poincare) invariance 
implies that the total angular momentum (or total spin)
$J$ of a hadron is a well defined and conserved quantum number. In addition 
$P$- and $C$- parities are conserved in QCD 
(as opposed to the electroweak theory). Therefore, for a given
hadron, the spin-parity 
combination $J^P$  ($J^{PC}$ for flavour-neutral hadrons) is 
the next important signature after the mass. Spin-parities 
are indicated for each observed hadron in its entry in \cite{PDG}.

To proceed in hadron spectroscopy, 
let us have a more detailed look at 
the mesons having the same flavour content 
$u\bar{d}$ as $\pi^+$.
Each meson state is a complicated coherent decomposition similar 
to (\ref{decomp}). Nevertheless, since $J^P$ is conserved, 
it is sufficient to consider the valence-quark component
to count all possible combinations of $J^P$ starting
from the lowest possible spin. The total angular momentum 
of the valence quark-antiquark state is a sum of 
the quark and antiquark spins plus the orbital angular momentum:
\begin{equation}
\vec{J}= \vec{S}+ \vec{L}\,,
\label{spin}
\end{equation} 
where $ \vec{S}=\vec{s}_q+\vec{s}_{\bar{q}}$ is the total
quark spin;  $S= 1(0)$ if the 
individual spins are parallel (antiparallel). Accordingly, there are
two possible states with $L=0$ and $J=S$. One
is with $J^P=0^-$ (pion) and the other with 
$J^P=1^-$ ($\rho$ meson).  The negative $P$-parity  
attributed to these states is obtained from 
the following rule: $P=-(-1)^L$. Notice an additional minus
which has to be added to account for the so called 
``intrinsic'' $P$-parity of the relativistic quark-antiquark system.
\begin{table}
\begin{center}
\vskip0.2cm
\begin{tabular}{|c|c|c|}
\hline
      & S=0                     & S=1\\ 
\hline
&&\\
L& $J^P$~~~  meson      & $J^P$~~~ meson \\
&&\\
\hline
$0$ &   $0^-$~~~  $\pi(140)$ &  $1^-$~~~ $\rho(770)$ \\
&&\\
\hline
&&\\
$1$ &   $1^+$~~~  $b_1(1235)$ &  $0^+$~~~ $a_0(980)$ \\
 &&\\
&    &  $1^+$~~~ $a_1(1260)$ \\
 &&\\
&    &  $2^+$~~~ $a_2(1320)$ \\
\hline
\hline
\end{tabular}
\caption{{\it Spectrum of the lowest $u\bar{d}$ states.}}
\label{tab:ud}
\end{center}
\end{table}
Turning to the states with  $L=1$, one encounters 
three mesons with $S=1$ ($J^P=0^+,1^+,2^+$) and 
one with $S=0$ ($J^P=1^+$). They are listed in Table \ref{tab:ud}
 according to 
the classification of \cite{PDG}. A similar counting
can be done for $L=2,3,...$, predicting $J >2$ mesons.
Some of them can be found in \cite{PDG}.  
Generally, it is rather difficult 
to observe hadrons with higher spins.
Having larger masses, these states have many decay channels
and, therefore, a large total width, complicating 
their experimental identification in a form of a resonance. 
Baryons from $u,d,s$ quarks with different $J^P$ listed in \cite{PDG}  
can also be interpreted, at least qualitatively, in terms
of three-quark valence states with the orbital momentum
$L$ between diquark and the third quark.

The angular momentum is not the only source of excited hadron
resonances. There are
mesons which  have the same $J^P$ as  $\pi$ or $\rho$ 
but a larger mass. These states 
are somewhat similar to the radially excited levels in a potential. 
For the pion a natural candidate of such excitation 
is the $\pi'(1300)$ state with $J^P=0^-$, 
whereas $\rho$ meson has at least
two experimentally established radially-excited partners with $J^P=1^-$:
$\rho'(1450)$ and $\rho''(1700)$ \cite{PDG}. 
Note that because the orbital momentum $L$ is not
conserved in relativistic theory,  the $L=2$ state with $J^P=1^-$ 
cannot be simply distinguished from the ``radial excitation''
of the $L=0$ state with the same spin-parity, - another difficulty
for the potential models of light-quark hadrons. 
Ultimately, one has to think in terms 
of purely relativistic extended objects, some kind of 
{\em quark-gluon strings} having a spectrum of ``radial'' and 
$J$ excitations. However, attempts to derive a string picture for hadrons 
directly from $L_{QCD}$ were not successful so far.
\begin{table}[t]
\begin{center}
\begin{tabular}{| c | c | c | c |c |c|}
\hline
& $u$& $d$ & $s$& $c$& b\\
\hline
&&&&&\\
&  $\pi^0,\eta,\eta'$ & $\pi^-$& $K^-$ &$D^0$& $\bar{B}^-$\\
$\bar{u}$ &&&&&\\
&  $\rho^0,\omega$ & $\rho^-$& $K^{*-}$ &
$D^{*0}$& $\bar{B}^{*-}$\\
&&&&&\\
\hline
&&&&&\\
&  $\pi^+$ & $\pi^0,\eta,\eta'$& $\bar{K}^0$ &$D^+$& $\bar{B}^0$\\
$\bar{d}$&&&&&\\
&  $\rho^+$ & $\rho^0,\omega$ & $\bar{K}^{*0}$ &
$D^{*+}$& $\bar{B}^{*0}$\\
&&&&&\\
\hline
&&&&&\\
& $K^+$ & $K^0$& $\eta,\eta'$& $D_s$ & $\bar{B}_s$\\
$\bar{s}$&&&&&\\
& $K^{*+}$ & $\bar{K}^{*0}$& 
$\phi$& $D^*_s$ & $\bar{B}^*_s$\\
&&&&&\\
\hline
&&&&&\\
& $\bar{D}^0$ & $D^-$& $\bar{D}_s$ &$\eta_c$& $\bar{B}_c$\\
$\bar{c}$ &&&&&\\
&  $\bar{D}^{*0}$ & $D^{*-}$& $\bar{D}^*_s$ &$J/\psi$& $\bar{B}^*_c$\\
&&&&&\\
\hline
&&&&&\\
& $B^+$ & $B^0$& $B_s$ & $B_c$& $\eta_b$\\
$\bar{b}$ &&&&&\\
&  $B^{*+}$ & $B^{*0}$& $B^*_s$ & $B^*_c$ &$\Upsilon$\\ 
&&&&&\\
\hline
\end{tabular}
\caption{{\it Pseudoscalar ($J^P=0^-$) (upper lines) 
and vector ($J^P=0^-$) (lower lines) mesons with different flavour content.
}}
\label{tab:mesons}
\end{center}
\end{table}

Quark-gluon interaction 
is flavour-independent.
Therefore, given that a $u\bar{d}$ meson with 
a certain $J^P$ exists,  the mesons 
with the same $J^P$ containing all possible 
quark-antiquark flavour combinations should also be observed. 
The  flavour partners of the pion 
($\rho$-meson) with $J^P=0^-$ ($J^P=1^-$) are listed in 
Table~\ref{tab:mesons}.  
Almost all of them have been observed;
the masses and other characteristics 
are given in \cite{PDG}. The only temporary 
exceptions are the pseudoscalar $b\bar{b}$ 
meson ($\eta_b$) and the vector $\bar{b}c$ meson 
($B_c^*$). These two states are not yet in \cite{PDG}, 
due mainly to experimental reasons.

The {\em heavy quarkonia}, i.e., the mesons consisting 
of heavy quark and antiquark ($\bar{c}c$ or $\bar{b}b$) are of a special
interest. Here the masses of interacting quarks are 
large enough compared to their 
characteristic energies within  hadrons: $ m_{b,c} \gg \Lambda_{QCD}$.  
In other words, heavy quarks
are nonrelativistic objects with respect to QCD interactions.
It is therefore  possible to approximate the quark-gluon interactions
with a nonrelativistic potential, putting the hadronic calculus
on the safe ground of quantum mechanics.
The Coulomb quark-antiquark potential $V(r) =\alpha_s/r$, derived 
from the one-gluon exchange, is valid 
at small distances. To provide
quark confinement, a certain long-distance part 
of the potential, infinitely growing at $r\to \infty$
should also exist (e.g., oscillator or linear potential). 
This part of the potential cannot be directly calculated
from $L_{QCD}$ and is usually modeled and fitted to 
the observed quarkonium spectra.
Importantly, numerical studies of QCD on the lattice
confirm the existence of the confining potential force
between heavy quark and antiquark.

\subsection{
Isospin}
In addition to the exact colour- and space-time symmetries, 
QCD Lagrangian possesses 
approximate flavour symmetries
originating from the pattern of quark masses. Since the 
latter are generated by some external mechanism, 
the flavour symmetries do not have fundamental roots in QCD. Nevertheless, 
they provide very important relations for hadron masses
and hadronic amplitudes.

Let us start with the $u$ and $d$  quarks and 
rewrite the QCD Lagrangian, isolating these  two flavours: 
\begin{eqnarray}
L_{QCD}= 
\bar{\psi}_u(iD_\mu\gamma^\mu-m_u)\psi_u+
\bar{\psi}_d(iD_\mu\gamma^\mu-m_d)\psi_d
+L_{glue}+ L_{s,c,b,t}\,.
\label{lqcdisosp}
\end{eqnarray} 
The smallness of the $u$ and $d$ masses,
$m_{u}\sim m_{d}\ll  \Lambda_{QCD}$, implies that also their difference 
is small:
\begin{equation}
m_{d}- m_{u}\ll  \Lambda_{QCD}\,.
\label{udmass2}
\end{equation}
Neglecting this difference we use
a new notation for the common  $u,d$ quark mass:
\begin{equation}
m_u\simeq m_d\simeq \tilde{m}\,.
\label{tildem}
\end{equation} 
In this approximation, 
\begin{equation}
L_{QCD} \simeq  L_{QCD}^{(u=d)}=
\overline{\Psi}(D_\mu\gamma^\mu-\tilde{m})\Psi + L_{glue}+.. \,,
\label{qcdisosp}
\end{equation}
where a new, two-component 
spinor field (doublet) is introduced: 
$$
\Psi=\left(\begin{array}{l}\psi_u\\\psi_d\end{array} \right)\,, ~~
\bar{\Psi}= (\bar{\psi}_u,\bar{\psi}_d)\,.
$$
The theory described by the r.h.s. of (\ref{qcdisosp}) 
is not exactly QCD, but is very close to it.
The new Lagrangian $L_{QCD}^{(u=d)}$ contains two degenerate 
quark flavours and has a symmetry with respect to the 
general phase rotations in the ``two-flavour space'':
\begin{equation}
\Psi \to \Psi'=\exp\left(-i\sum\limits_{a=1}^3
\omega^a\frac{\sigma^a}2\right)\Psi,~~~
\overline{\Psi} \to \overline{\Psi}'=
\overline{\Psi}\exp\left(i\sum\limits_{a=1}^3
\omega^a\frac{\sigma^a}2\right)\,,
\label{su2transf}
\end{equation}
where $\omega^a$ are arbitrary ($x$-independent) parameters 
and $\sigma^1,\sigma^2,\sigma^3$ are the 
$2\times2$ Pauli matrices. The symmetry transformations (\ref{su2transf}) form
a group SU(2) \footnote{ The number of independent parameters 
for SU(2) is determined in the same way as we did for SU(3) in Lecture 1: one
counts the number of independent elements in the unitary 2$\times$ 2
matrix with the unit determinant.}.

One has to emphasize that the approximate  
SU(2)-flavour symmetry  emerges ``by chance'', simply because 
the $u$ and $d$ quark masses turn out to be almost 
degenerate. Note also that e.m. interaction violates this symmetry,
due to different electric charges of $u$ and $d$ quarks.
But this $O(\alpha_{em})$ effect is again small, 
at the same level of $\sim 1 \%$, 
as the $O(\frac{m_d-m_u}{\Lambda_{QCD}})$ violation due to the quark
mass difference.

The approximate degeneracy of $u$ and $d$ flavours
manifests itself in hadrons.
Replacing $u$ quarks by $d$ quarks or vice versa 
in a given  hadron, yields a different hadron 
which has a very close mass 
and other properties.
This qualitative prediction is nicely confirmed 
by the measured mass differences between proton ($uud$) and neutron ($udd$),  
$\pi^+ (\bar{u}d)$ and $ \pi^0 ([\bar{u}u -\bar{d}d]/\sqrt{2})$, 
$K^+(u\bar{s})$ and $K^0 (d\bar{s})$, etc.. The typical 
mass splittings for the $u\leftrightarrow d $ hadronic partners  
are at the level of few MeV.  
Thus, QCD nicely explains the origin of {\em isospin} symmetry 
introduced by Heisenberg in 30's in nuclear physics, to describe  
the similarities between the ``mirror'' isotopes, obtained from each 
other by interchanging protons and neutrons. 
The second part in the word 'isospin'
reflects the analogy with   
the electron spin symmetry, the degeneracy of the spin-up 
and spin-down electron states in quantum mechanics.

In fact, one introduces a similar formalism in QCD,  
attributing isospin $I=1/2$~
to the doublet of $u$ and $d$ quarks and treating these two
flavours as ``up'' and ``down'' components with 
$I_3=+1/2$ and $I_3=-1/2$,
respectively. The hadrons containing $u$ and $d$ quarks 
in different combinations form isomultiplets, with
the isospin counting similar to  the spin algebra in quantum mechanics.
In the case of proton and neutron,
the diquark $ud$ has isospin 0, therefore, adding $u$ or $d$ quark 
to the diquark, we get the nucleon isodoublet ($I=1/2$) consisting 
of proton with $I_3=+1/2$ and neutron with $I_3=-1/2$.
Another isodoublet is formed by  $K^+$ and $K^0$, 
where $\bar{s}$ quark, which has no isospin, is combined
with $u$ and $d$, respectively .
In the same way, $D^0(u\bar{c})$ and $D^-(d\bar{c})$, or 
$B^+(u\bar{b})$ and $B^0(d\bar{b})$ build isodoublets. 

Note that antiquarks have the opposite signs of $I_3$:
$\bar{u}$ ($\bar{d}$) has $I_3=-1/2$ ($I_3=+1/2$).
Combining $u$ and $d$ quark with their antiquarks, 
one gets four states. Three of them belong to   
isotriplet ($I=1$): 
\begin{equation}
u\bar{d}~(I_3=+1),~~
\frac{u\bar{u}-d\bar{d}}{\sqrt{2}} ~(I_3=0), ~~
d\bar{u}~(I_3=-1)\,.
\label{isotriplet}
\end{equation}
For example, pions ($\pi^+$, $\pi^0$ and $\pi^-$),
as well as $\rho$ mesons ($\rho^+$, $\rho^0$ and $\rho^-$)
form isotriplets.  
The fourth quark-antiquark state is an isosinglet ($I=0$):
\begin{equation}
\frac{u\bar{u}+d\bar{d}}{\sqrt{2}}\,,
\label{iso0}
\end{equation}
which deserves a separate discussion.
In general, the  $u\bar{u}$ and $d\bar{d}$ states 
transform into each other  via intermediate gluons. In mesons this transition 
takes place at long distances, due to some 
nonperturbative mechanism,
not necessarily described by diagrams with a fixed number of gluons.
In any case, the transition amplitude has a characteristic
scale of $O(\Lambda_{QCD})$, much larger than the mass
difference $\simeq 2(m_u-m_d)$ between the $u\bar{u}$ and $d\bar{d}$ 
states. The $\bar{u}u-\bar{d}d$ degeneracy    
yields two orthogonal physical states, the $I_3=0$ component
of the isotriplet (\ref{isotriplet}) and 
the isosinglet (\ref{iso0}). Turning to strange quarks 
one encounters the second $I=0$ state $\bar{s}s$.  
The same gluonic transition mechanism provides 
a mixing between (\ref{iso0}) and the $\bar{s}s$-state,  
Now, the difference between the masses is not small, 
being of $O(2m_s-2\tilde{m}) \sim\Lambda_{QCD}$.  Hence, the amount of mixing  
depends on the magnitude of the 
$s\bar{s} \leftrightarrow (u\bar{u} +d\bar{d})$ 
transition amplitude.
The latter is quite sensitive to
the spin-parity of the meson state.  
For example, 
the two isosinglet mesons with $J^P=0^-$ 
shown in  Table~\ref{tab:mesons} have (up to small deviations)  
the following quark content:
\begin{eqnarray}
\eta(547)\simeq \frac1{\sqrt{6}}(u\bar{u}+d\bar{d}-2\bar{s}s)\,,
\label{eta}
\\
\eta'(958)\simeq \frac1{\sqrt{3}}(u\bar{u}+d\bar{d}+\bar{s}s)\,,
\label{etaprime}
\end{eqnarray}
indicating that mixing in the $O^-$ channel
is large. For the $J^P=1^-$ mesons the situation is 
completely different. From the quark content of 
the isosinglet mesons:
\begin{equation}
\omega(782) \simeq \frac{u\bar{u}+d\bar{d}}{\sqrt{2}},~~
\phi(1020)\simeq \bar{s}s \,,   
\label{omega}
\end{equation}
one concludes that the transition from strange to nonstrange
quark pairs in the $1^-$ state is suppressed.

\begin{figure}
\begin{center}
\includegraphics*[width=8cm]{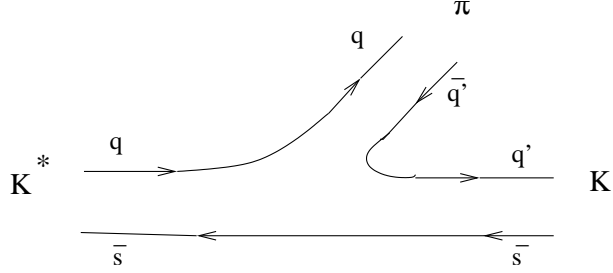}
\caption{{\em Quark diagram for $K\to K\pi$ 
hadronic decays; the four possible combinations of $q=u,d$ and $q'=u,d$ 
correspond to the four decay modes related by isospin symmetry.}}
\label{fig:KstKpi}
\end{center}
\end{figure}

Returning to the isospin symmetry, it is worth to mention that   
it yields useful relations between hadronic amplitudes. 
To give just one simple example,
let us consider the four observable hadronic decays 
of $K^*$ -meson.
The amplitudes of these decays are related via isospin symmetry, so that  
only one amplitude is independent: 
\begin{equation}
A(K^{*0}\to K^+\pi^-) = -\frac1{\sqrt{2}}A(K^{*0}\to K^0\pi^0)=
A(K^{*+}\to K^0\pi^+) = \frac1{\sqrt{2}}A(K^{*+}\to K^+\pi^0)\,. 
\label{su2rel}
\end{equation}
To obtain these relations, one does not
necessarily need to apply formulae for SU(2) group.
In the isospin limit,
all four decays are described by a single quark diagram, 
shown in Fig.~\ref{fig:KstKpi}, 
where the initial and final mesons 
are taken in the valence-quark state.
Each individual decay mode has its own
combination of $u$ and $d$ quarks to be substituted in this diagram. 
Furthermore, the coefficients $1/\sqrt{2}$ originate from the quark 
content of $\pi^0$.  Naturally, the hadronic amplitude attributed 
to the quark diagram in Fig.~\ref{fig:KstKpi}  
is a nonperturbative, long-distance object and cannot be 
directly calculated in QCD. However all we need are the 
amplitude relations between 
individual decay modes and not the value of the amplitude itself.
Squaring the relations (\ref{su2rel}) and taking into account
the phase space, one predicts the ratios of branching 
fractions confirmed by the experimental values given in \cite{PDG}.

\subsection{SU(3)-flavour symmetry}

Another flavour symmetry, widely used
in hadron phenomenology, is $SU(3)_{fl}$ corresponding 
to the limit of QCD with all three quarks, $u$,$d$ and $s$, 
having equal masses. Since in reality $m_s$ is only inessentially
smaller than $\Lambda_{QCD}$, the magnitude of $SU(3)_{fl}$-violation  
in hadrons is not universal, depending on their quark content and 
quantum numbers.

Neglecting the mass differences between $u$,$d$,$s$ quarks 
and introducing a common mass $\tilde{m}_3$ one 
approximates $L_{QCD}$  as 
\begin{equation}
L_{QCD} \simeq  L_{QCD}^{(u=d=s)}=
\overline{\Psi}_3(D_\mu\gamma^\mu-\tilde{m}_3)\Psi_3 + L_{glue}+.. \,,
\label{su3}
\end{equation}
where $\Psi_3$ is a triplet:
$$
\Psi_3=\left(\begin{array}{l}\psi_u\\\psi_d \\ \psi_s\end{array} \right)\,, ~~
\bar{\Psi}_3= (\bar{\psi}_u,\bar{\psi}_d,\bar{\psi}_s)\,.
$$
The modified QCD with $L_{QCD}^{(u=d=s)}$ 
has a symmetry with respect to the transitions between 3 flavour states:
\begin{equation}
\Psi_3\to \Psi'_3 =\exp\left[-i\sum\limits_{a=1}^8\omega^a\frac{
\lambda^a}2\right]\Psi_3\,,~~~
\bar{\Psi}_3\to \bar{\Psi'}_3 =\bar{\Psi}_3\exp
\left[i\sum\limits_{a=1}^8\omega^a\frac{
\lambda^a}2\right]\,.
\label{su3flav}
\end{equation}  
Although physically, $SU(3)_{fl}$ 
and the fundamental $SU(3)$-colour have completely different origin,
the group-theoretical formalism of both symmetries is the same. 
In particular, the eight $\lambda^a$-matrices entering (\ref{su3flav})
are already given in  (\ref{matrix}).

The $SU(3)_{fl}$ symmetry is very helpful in ``organizing'' the spectra 
of strange and nonstrange hadrons in multiplets.
The light-quark meson multiplets are obtained by combining 
quark flavour-triplets and antiquark flavour-antitriplets.  
Without using the specific rules of $SU(3)$-algebra,
which can be found in many textbooks, it is easy to figure  out 
that  the nine quark-antiquark states split into a singlet
$\bar{\Psi}_3\Psi_3$ and octet $\bar{\Psi}_3\lambda^a\Psi_3$. 
The octet has its own isospin substructure \footnote{
In mathematical terms $SU(2)_{isospin}$ is a subgroup of $SU(3)_{fl}$.}. 
The singlet-octet pattern provides a reasonable  
description for pseudoscalar mesons, in particular, 
$\eta'$ meson is close to the $SU(3)_{fl}$ singlet state (\ref{etaprime}). 
The isotriplet of pions ($\pi^+,\pi^0,\pi^-$), 
two isodoublets of kaons ($K^+,K^0$ and $\bar{K}^0,K^-$) and the 
isosinglet $\eta$, given by (\ref{eta}), together form an octet.
However, this pattern is not universal.
E.g., in the case of vector mesons, $\omega$ and $\phi$ states in
(\ref{omega}) are neither octets, nor singlets. 
To complete the counting of meson $SU(3)_{fl}$ multiplets, 
one has to mention also triplets and antitriplets
of heavy-light mesons. For example, in the case of $c$ quark, $\bar{D}^0,D^-,\bar{D}_s$ 
($D^0,D^-,D_s$) form  a triplet (antitriplet). 

Generally, $SU(3)_{fl}$ works quite well for baryons,
because their characteristic mass scale is a few  
times larger than $\Lambda_{QCD}$. Importantly, the spin-parity of the lowest 
baryon $SU(3)_{fl}$ multiplets is fixed, due to the total antisymmetry 
of the baryon ``wave-function'' in the  $SU(3)_{fl}$/spin/colour
coordinates required by Fermi-statistics. There is an octet with $J=1/2$ (including proton and neutron) 
and decouplet with $J=3/2$. Let us, for instance, have a look at the latter. It contains 
the isoquadruplet ($I=3/2$) of $\Delta$-resonances
($\Delta^{++}(uuu)$, $\Delta^+(uud)$, $\Delta^0(udd)$, $\Delta^-(ddd)$), 
the isotriplet of $\Sigma$ resonances ($\Sigma^+(uus)$, $\Sigma^0(uds)$,
$\Sigma^-(dds)$), the isodoublet of $\Xi$-resonances ($\Xi^0(uss)$,
$\Xi^-(dss)$) and  the isosinglet $\Omega(sss)$. 
Consulting \cite{PDG} for the masses of these baryons, one notices a
distinct hierarchy: each constituent $s$ quark adds an amount of $O(m_s)$ to the baryon mass.  

Returning to the quark diagram in Fig.~\ref{fig:KstKpi}, we may now
replace the $s$ quark
by $u$ or $d$ quark. In $SU(3)_{fl}$ limit the new 
diagram obtained after this replacement is equal to the one with 
$s$-quark, yielding relations between the 
$K^*\to K\pi$ and $\rho\to \pi\pi$ hadronic amplitudes, e.g.
\begin{equation}
A(K^{*+}\to K^0\pi^+) \simeq -\frac{1}{\sqrt{2}}A(\rho^{+}\to \pi^0\pi^+)\,. 
\label{}
\end{equation} 
Such relations are typically violated at the level of 20-30 \%, but are 
still useful from the phenomenological point of view. 

\subsection{
Heavy quark symmetry}
To complete our survey of flavour symmeties, we 
now turn to the $\{c, b \}$ quark sector
of QCD Lagrangian. The fact 
that $m_{c,b} \gg \Lambda_{QCD}$, together with the 
flavour-independence of quark-gluon interactions, allows
one to consider an interesting limit 
of $L_{QCD}$ where both $c$ and $b$ quarks have infinitely
heavy mass: 
\begin{equation}
m_{c}\sim m_b\sim m_Q \to \infty\,. 
\label{hlimit}
\end{equation}
At first sight, the limit is not justified, because 
in reality $m_b$ is substantially larger than $m_c$. 
As we shall see, the fact that both masses are large 
turns out to be more important for QCD dynamics.
Formally, in the limit (\ref{hlimit}) one can introduce a doublet 
of heavy-flavour fields 
$$
\Psi=\left(\begin{array}{l}\psi_c\\ \psi_b \end{array} \right)\,, ~~
\bar{\Psi}= (\bar{\psi}_c,\bar{\psi}_b)\,,
$$ 
and rewrite the Lagrangian in a form
invariant with respect to SU(2)-rotations in the $c,b$ flavour space:
\begin{equation}
L_{QCD}= \sum\limits_{Q=c,b}\overline{\Psi}_Q(iD_\mu\gamma^\mu-m_Q)\Psi_Q
+L_{glue}+L_{u,d,s}\,.
\label{hqet}
\end{equation}
This particular form of the heavy-quark limit for $L_{QCD}$  
is however not convenient, because the heavy mass scale $m_Q$ 
is still present explicitly. To understand why it is desirable 
to effectively remove that scale, let us 
consider the heavy-quark limit (\ref{hlimit}) for $D$ or $B$ 
meson. Since $m_c$ and $m_b$ have  no direct relation to QCD,
it makes sense to discuss a hypothetical 
heavy-light meson $H$ with a mass $m_H$ containing a heavy quark
with an arbitrarily large mass $m_Q$.
In the rest frame of $H$   the constituent heavy
quark stays almost at rest,  providing a static source
of color charge which emits and absorbs gluon fields. 
The meson mass, to a good approximation is
\begin{equation}
m_H= m_Q + \bar{\Lambda}\,,
\label{massH}
\end{equation}
where $\bar{\Lambda}$ is the energy of the light quark-gluon ``cloud''
surrounding the heavy quark.
The situation very much resembles the hydrogen atom where
the total mass of the atom is a 
sum of an extremely large ($\sim $ GeV) proton mass
and a small energy of the electron cloud ($\sim$ MeV). The essential
point is that the electron itself is nonrelativistic.  
One can isolate the electron mass from the rest of the energy,
introduce the Coulomb potential and kinetic energy,
and eventually solve the equations of motion, 
determining the electron energy levels. In heavy hadrons 
the light-quark cloud is purely relativistic ($m_{u,d,s}< \Lambda_{QCD}$)
and has a complicated long-distance nature. 

Nevertheless, one essential feature is common
for both bound states. In the atom 
the energy of the electron cloud does not depend on  
the proton mass. Likewise, in the heavy-light meson 
$\bar{\Lambda}$  in (\ref{massH})
is (up to $1/m_Q$ corrections) independent of $m_Q$. 
From atomic physics we know that the electron energy levels
in hydrogen and deuterium coincide to a great precision.
The fact that deuteron is twice more massive
than proton, does not play a role for the energy levels,
because in both cases the atomic nuclei are static. Important is that 
the electric charge does not change by switching from proton
to deuteron. Similarly, in the $H$ meson $\bar{\Lambda}$ changes 
very little if one replaces $m_Q$ by $m_b$  or by $m_c$,
because the colour charge of the heavy quark 
does not change. Thus, the heavy-flavour symmetry is 
in reality the symmetry 
between the light-quark remnants of the heavy hadrons, so that
the heavy-quark mass scale indeed plays a secondary role.  

To achieve a  quantitative level, a special formalism 
of {\em heavy quark effective theory} (HQET)  was developed
for applications of QCD to heavy-light hadrons. 
One starts from $L_{QCD}$ and introduces 
transformations which decouple  the $\sim m_Q$ part 
of the heavy-quark field from 
the part which has the remnant momentum $\sim \Lambda_{QCD}$. Only the
latter part strongly interacting with the light quark-gluon cloud 
is relevant for QCD dynamics.  In HQET 
one integrates out the heavy degrees of freedom and
works with the Lagrangian containing a new effective quark field
carrying the flavour of $Q$  but no mass. Not only the static limit
(\ref{hlimit}) but also an expansion in powers of $1/m_Q$  can be
systematically treated.  
The field-theoretical aspects 
of heavy-mass expansion are nicely explained in the literature
(see e.g., \cite{ShifmanHQ},\cite{tm}); I will only focus 
on some important  phenomenological consequences of heavy-flavour symmetry.

One famous example is the $B\to D l \nu_l $ decay involving weak 
$b\to c $ transition in Standard Model. The unknown part of the decay
amplitude is the hadronic matrix element 
\begin{equation}
\langle D(p_D)| \bar{c} \gamma_\mu b |B(p_B\rangle 
\label{BD}
\end{equation}
determined by the long-distance interactions involving 
the initial and final heavy quarks as well as the surrounding
light quark-gluon ``cloud''. One chooses a special
kinematical configuration, the ``zero recoil point''
where the momentum transfer to the lepton pair is equal
to 
\begin{equation}
q^2=(p_B-p_D)^2=(m_B-m_D)^2\,.
\label{zerorec}
\end{equation} 
In the $B$ meson rest system $p_B=(m_B,0,0,0)$ 
this point correspond to the final $D$ meson at rest.
In the heavy-quark limit the replacement of 
$b$ quark by $c$-quark does not change the hadronic state:
$$
\langle D(p_D)|= \langle B(p_B)| 
$$
and the matrix element (\ref{zerorec})
reduces to a trivial normalization factor. 
One can therefore predict the decay amplitude
in the zero recoil point up to $1/m_Q$ corrections.
In fact there is a theorem stating that 
these corrections are even smaller and 
start from $O(1/m_Q^2)$, but to derive this and other
important details one needs a full-scale HQET framework.

Without resorting to the effective theory, 
it is possible to understand the origin of another
important symmetry emerging in the heavy-quark limit.
In the hydrogen atom,  the electron and proton have magnetic moments 
related to their spins and yielding interactions 
with the external magnetic fields or with each other
(spin-spin interactions). 
The magnetic moments are inversely proportional to the masses,
so that the proton magnetic moment plays no role for the electron 
energy levels. Each level is degenerate with respect to the proton spin
direction. Since QED and QCD have very similar vector boson interactions
with spin 1/2 particles,   
the spin $1/2$ quarks also have {\em chromomagnetic} moments.
and  interact with the ``magnetic'' parts of gluonic 
fields and with other quarks.
For the heavy nonrelativistic quark the chromomagnetic moment is 
inversely proportional to the heavy quark mass $m_Q$.
In the infinite mass limit the interaction vanishes 
and, hence the light-cloud energy $\bar{\Lambda}$ is independent 
of the spin orientation of the heavy quark.

One arrives at a new classification of heavy-light states
based on this {\em heavy-quark spin symmetry}.
Instead of adding together the orbital momentum and the total
spin of quarks as we did in (\ref{spin}) it is more appropriate to introduce,
for a $Q\bar{q}$ meson ($Q=c,b$; $q=u,d,s$), the total angular momentum  
of light degrees of freedom:
\begin{equation}
\vec{J}_{light}=\vec{L} +\vec{s}_q
\label{totl}
\end{equation}
Adding the heavy quark spin $s_Q=1/2$ to $J_{light}$ one gets 
degenerate doublets of heavy-light mesons with total angular
momentum $J=J_{light}\pm 1/2$.
At $L=0$ one simply has $J_{light}=1/2$ and therefore 
a doublet of mesons with $J^P=1^-$ and $J^P=0^-$ consisting of $B$ and $B^*$  
($D$ and $D^*$) in the $b$ quark (charm)
sector. The mass differences within doublets are indeed very small \cite{PDG}:
\begin{equation}
\delta_B=m_{B^*}-m_B= 47~\mbox{MeV}, ~~ \delta_D=m_{D^*}-m_D= 142~\mbox{MeV},
\label{diffM}
\end{equation}
indicating that the heavy-quark spin symmetry works
quite well, especially for the heavier $b$ quark.
Taking  into account that the mass differences  are $\sim 1/m_Q$
effects, one expects that $\delta_B/\delta_D\simeq (m_c/m_b)$ 
which is also in accordance with (\ref{diffM})
and (\ref{qmasses}).
I leave as an exercise to show that at $L=1$ there
are two degenerate doublets: one with $J^P=0^+,1^+$  
and another one with  $J^P=2^+,1^+$.

\subsection{ 
Exotic hadrons }

The colour confinement principle does not exclude
hadronic states with an ``exotic''  valence quark content, 
different from $q\bar{q}$ or $qqq$.  
Quarks, antiquarks and gluons can be added together in any combination, 
e.g., $q\bar{q}G$ , $GG$ , $q\bar{q} q \bar{q}$ or $q\bar{q}qqq$, provided
they are in a colour-neutral state. 
Since one cannot calculate the 
spectrum of hadrons in QCD with a good precision,  
predictions of exotic states are generally model-dependent.
It is always problematic to distinguish an exotic hadron 
from the  excitation of an ordinary hadron with the same $J^P$ and flavour quantum numbers. 
Moreover, in this case one expects mixing
between ordinary and exotic hadrons.
For example, if there is a $J^P=0^-$ state composed of two gluons 
$GG$ ({\em glueball}), it should be mixed with $\eta'$ to a certain degree,
so that $\eta'$ acquires a glueball component. 

Therefore,  the most interesting, ``smoking gun'' signatures
are the hadrons with exotic quantum numbers
(flavour content and/or $J^{PC}$), forbidden 
for quark-antiquark mesons or three-quark baryons. 
For example it is impossible to arrange 
a flavour-neutral quark-antiquark state with 
$J^{PC}=1^{-+}$. The $P$ and $C$ ( charge-conjugation) parities of a 
fermion-antifermion state are determined by the rules:
$P=-(-1)^{L}$ and $C=(-1)^{L+S}$, so that $P=-1$ means
$L=0,2,...$. Hence,  the only possibility to have
$C=+1$ is $S=0,2,..$. However, adding together even values 
of $L$ and $S$  one cannot get $J=1$. 
On the other hand, adding one constituent 
gluon to a $q\bar{q}$ pair, one easily makes a  ``hybrid'' meson 
with $q \bar{q}G$ content and $J^{PC}=1^{-+}$ quantum numbers.
Searches for hybrid mesons are currently being carried out, but the
experimental situation is not yet settled.

The recently observed narrow baryon resonance $\Theta(1540)$ decaying to $K^+n$\cite{penta}
is another promising candidate for hadron exotics,
a state $\bar{s}uudd$ with five valence constituents ({\em pentaquark}). 
Flavour symmetries are important model-independent tools
to confirm/reject the experimental candidates for exotic resonances. 
In particular, an important task is to find the symmetry partners
of these hadrons
and to fill the relevant isospin and $SU(3)_{fl}$  multiplets 
(in case of the pentaquark it is actually the $SU(3)_{fl}$ antidecouplet).

\section{QCD AT SHORT DISTANCES}

\subsection{ Probing short distances with electroweak quark currents}

\begin{figure}
\begin{center}
\includegraphics*[width=8cm]{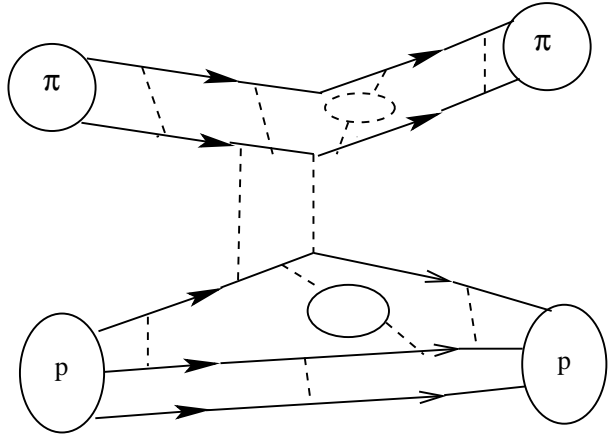}
\caption{{\em A schematic view of pion-proton  elastic scattering at 
large momentum transfers.}}
\label{fig:piN}
\end{center}
\end{figure}

In this lecture we return to the quark-gluon interactions
at large momentum transfers (short distances).
In this region, practically at $Q \geq 1$ GeV,   
$\alpha_s$ is small, allowing one to apply the perturbative 
expansion. It is then possible to test QCD 
quantitatively, calculating various quark-gluon interaction 
processes at large $Q$ 
and comparing the results with the available experimental data. 
Note however, that the traditional way to study interactions by
scattering one object on the other is not  applicable 
to quarks and gluons. They simply are not available in free-particle states.
One needs to trace quarks inside hadrons, where the long-distance
forces are important. Take as an example 
the elastic pion-proton scattering at large 
momentum transfers (see Fig.~\ref{fig:piN}). Here one 
has to combine the perturbative quark-quark scattering amplitudes 
(two-gluon exchange) with the ``wave functions'' of quarks inside the initial and final hadrons.
To obtain these functions one needs to go beyond perturbative QCD. 
Therefore, an unambiguous extraction of the perturbative
amplitude from the data on the scattering cross section is  
not a realistic task. 

The situation is not so hopeless, actually, 
since we have at our disposal electroweak bosons
($\gamma,W,Z$) interacting with quarks in a pointlike way. 
Electroweak interactions at large momentum transfer
serve as external probes of short-distance dynamics. 
In Lecture 1, we already discussed one example: 
the quark-antiquark pair production
in $Z$ decay. To list all possible    
electroweak sources of quarks in a more systematic way,
I start with the photon. Its interaction with the quark e.m. current
was already given in  (\ref{lqed}), let me write
it down again:
\begin{equation}
L_{em}(x)= -ej_\mu^{em} A^\mu\,,
\label{em1}
\end{equation}
introducing a compact notation for the quark e.m. current:
\begin{equation}
j^{em}_\mu= \sum\limits_{q=u,d,s,c,..}Q_q\bar{\psi}_q\gamma_\mu\psi_q\,,
\label{jem}
\end{equation}
where the summation over colour indices is not shown for brevity.
The quark weak current entering the quark-$W$ flavour-changing
interaction:
\begin{equation}
L_{W}(x)= -\frac{g}{2\sqrt{2}}j^{W}_\mu W^\mu + c.c.\,
\label{weak}
\end{equation}
is more complicated and includes
the CKM mixing matrix:
\begin{equation}
j^{W}_\mu(x)=(\bar{u},\bar{c},\bar{t})\gamma_\mu(1-\gamma_5)
\left(\begin{array}{lll}
V_{ud}&V_{us}&V_{ub}\\
V_{cd}&V_{cs}&V_{cb}\\
V_{td}&V_{ts}&V_{tb}\\
\end{array} \right)
\left(\begin{array}{l}
d\\
s\\
b\\
\end{array} \right)\,.
\label{jweak}
\end{equation}
Finally, the quark-Z interaction is
\begin{equation}
L^Z=-\frac{g}{2cos\theta_W}j_\mu^Z Z_\mu\,,
\label{qZ}
\end{equation}
where the quark electroweak neutral current
is a mixture of vector and axial-vector parts.

\subsection{
Perturbative QCD 
in $e^+e^- \to hadrons$} 
\begin{figure}
\begin{center}
\includegraphics*[width=8cm]{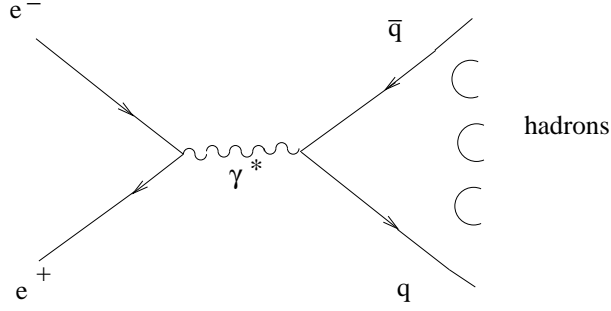}
\caption{{\em The virtual-photon exchange diagram for $e^+e^- \to hadrons $}}
\label{fig:eehad}
\end{center}
\end{figure}
In $e^+e^-$ annihilation at high energies the virtual photon 
provides a short-distance source of quark-antiquark pairs.
This process  is  hadron-free in the initial state. 
The photon-exchange diagram 
is depicted in Fig.~\ref{fig:eehad} 
(for simplicity, I ignore the additional $Z$-exchange diagram).
The experimentally measured total cross section
$\sigma_{tot}(e^+e^-\to h)$ depends
on one kinematical variable $s=(p_{e-}+p_{e^+})^2$. The virtual
timelike photon transfers its energy $\sqrt{s}$ 
to the hadronic state. At very large $\sqrt{s}\gg  \Lambda_{QCD}$
and $\sqrt{s}\gg m_q$, 
the initial pair of quarks is produced at   
an average distance of $O(1/\sqrt{s})$, much smaller than the typical 
hadronic distance scale $1/\Lambda_{QCD}$. 
Due to asymptotic freedom of QCD,  
gluonic interactions  of the  produced quark pair 
are suppressed by small $\alpha_s$.
Hence, $e^+e^-$ annihilation at high energies 
provides an almost pointlike source of quasi-free
quark pairs. At long distances the created quarks and antiquarks 
are inevitably converted into some hadronic state. Since in the total cross
section the summation is done over 
all hadronic states produced at a given energy,  
the total probability of hadronization sums up to a unit.  
Hence, at $\sqrt{s}\to \infty$  
the hadronic cross section is well approximated by 
the cross section of the free quark-antiquark pair production:
\begin{equation}
\sigma_{tot}^{(e^+e^-\to h)}(s)\simeq 
\sum_{q=u,d,s,..}\sigma^{(e^+e^-\to q\bar{q})}(s)\,,
\label{sigmatot}
\end{equation}   
summed over all quark flavours with $m_q \ll \sqrt{s}$. 
This, so called {\em parton model} approximation for $e^+e^-\to h$ 
is confirmed by experimental data. Moreover,
the majority of events saturating the cross section at high $\sqrt{s}$ 
consists of two distinct hadronic jets originating from
the initial quark pair. 

The way we obtained (\ref{sigmatot}) may seem too qualitative
and a bit ``hand-waving''. In the following, we shall 
derive the asymptotic 
cross section (\ref{sigmatot})
in a more rigorous way. In this derivation several important concepts 
will be introduced, to be used in discussing 
further topics covered by these lectures.  

\begin{figure}
\begin{center}
\includegraphics*[width=12cm]{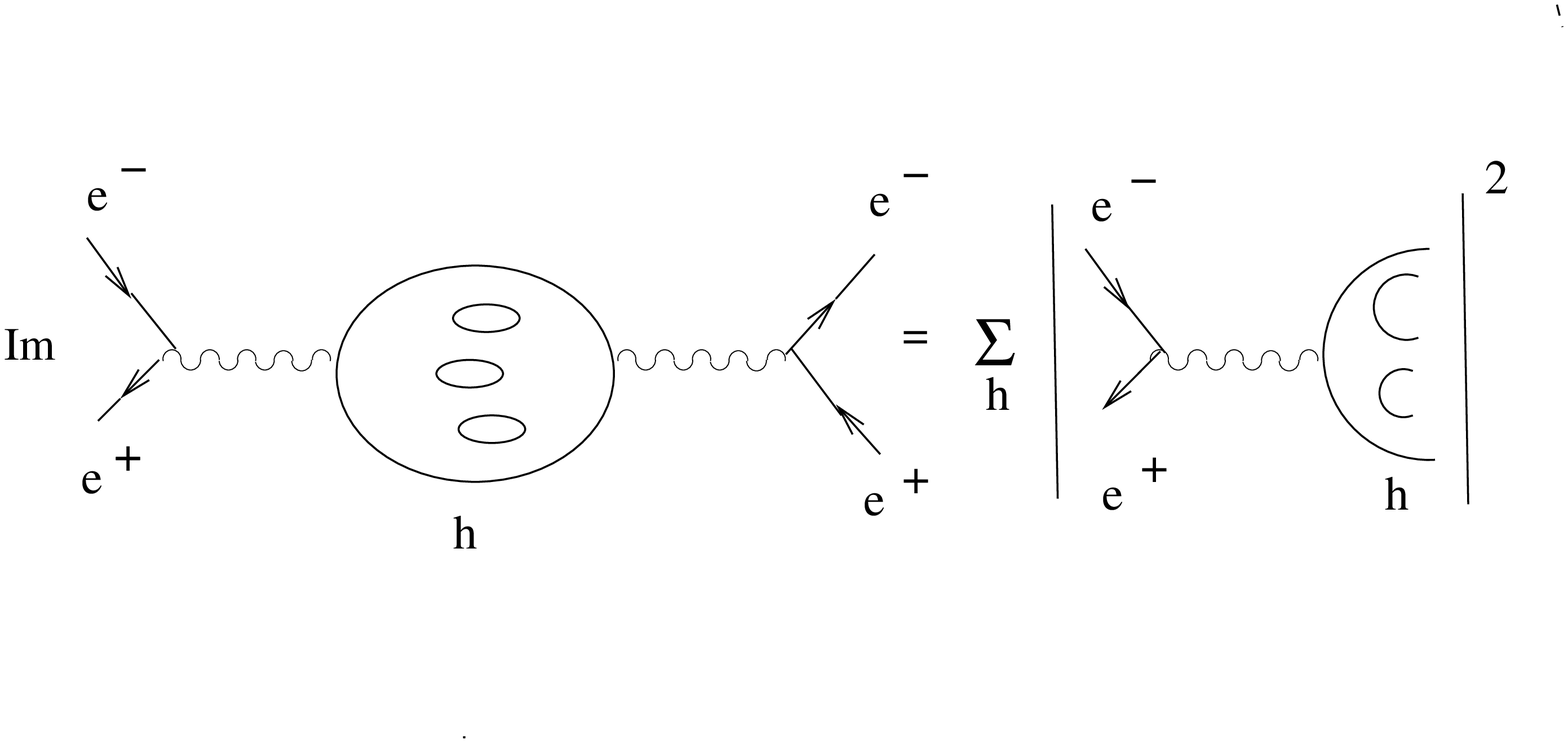}
\caption{{\em  Unitarity relation for $e^+e^- \to hadrons $}}
\label{fig:unit}
\end{center}
\end{figure}

We start with the formal definition of the total cross section:
\begin{equation}
\sigma_{tot}^{(e^+e^-\to h)}(s)=\frac1{2s}\sum\limits_{h_n} \left|
\langle h_n \mid T\mid e^+e^-\rangle\right|^2\,, 
\label{sum}
\end{equation}
where the sum over the final hadronic states $h_n$
includes phase-space integration 
and implies summation over spins (polarizations). 
The matrix elements 
$\langle f |T|i\rangle \equiv T_{fi}$  
(in (\ref{sum})
 $|i\rangle = |e^+e^-\rangle$ and $\langle f |=\langle h_n|$)  
determine the general $S$-matrix of the theory:
\begin{equation}
S_{fi}\equiv \langle f | S | i \rangle= \delta_{fi}+iT_{fi}\,.
\label{s}
\end{equation}
The usual representation of $S$-matrix in terms of 
Lagrangian  has the time-ordered exponential form: 
\begin{equation}
S=T\left \{ \exp \left [ i\int d^4x\left(L_{QCD}(x)+
L_{QED}(x)\right)\right] \right\}\,,
\label{smatr}
\end{equation}
where $L_{QED}$ includes e.m. interactions of quarks and leptons.
Furthermore, the unitarity of $S$ matrix is used: 
\begin{equation}
SS^\dagger=1\,,
\label{unitS}
\end{equation}
or 
\begin{equation}
\sum\limits_n\langle f|S|n\rangle \langle n | S^\dagger|i\rangle=\delta_{fi}\,.
\label{unit1}
\end{equation} 
From now on we consider the forward scattering $f=i$.
Replacing $\langle n | S^\dagger|i\rangle = 
\langle i | S|n\rangle^*$
and substituting (\ref{s}) in (\ref{unit1}), 
one obtains the unitarity relation for $T_{ii}$ 
(the optical theorem):
\begin{equation}
2\mbox{Im} T_{ii}= \sum\limits_n|T_{ni}|^2\,.
\label{tunit}
\end{equation}
To apply this universal relation to $e^+e^-$- scattering, 
we take $|i \rangle = |e^+e^-\rangle$ with  four-momentum 
$q=p_{e^+}+p_{e^-}$, so that $q^2=s>0$. Furthermore, we choose 
$|n\rangle=|h_n\rangle$ restricting
the set of intermediate states by hadronic states.
As a result we obtain a rigorous  
relation between the amplitude of the forward $e^+e^-\to h\to e^+e^-$ 
scattering via hadronic intermediate states, 
$T_{ii}={\cal A}^{(e^+e^-\to h \to e^+e^-)}$ , and the sum over 
the squared $e^+e^-\to h_n$  amplitudes. 
The latter sum, according to (\ref{sum})
is proportional to the total hadronic cross section.
The optical theorem (\ref{tunit}) takes the form:
\begin{equation}
2~\mbox{Im}~{\cal A}^{(e^+e^-\to h \to e^+e^-)}(s)=
\sum\limits_{h_n} |\langle h_n|T|e^+e^-\rangle|^2
= 2s\sigma_{tot}^{(e^+e^-\to h)}(s)\,.
\label{unit2}
\end{equation}
Diagrammatically, this relation is represented 
in Fig.~\ref{fig:unit}. The amplitude 
\begin{equation}
{\cal A}^{(e^+e^-\to h \to e^+e^-)}(q^2)= \frac{e^4}{(q^2)^2}
(\bar{\psi}_e\gamma^\mu\psi_e)(\bar{\psi}_e\gamma^\nu \psi_e)\Pi_{\mu\nu}(q)
\label{Aee}
\end{equation}
contains the photon propagators
and the products of electron and positron spinors
in both initial and final states, written according 
to  the QED Feynman rules. 
The  nontrivial part of this 
amplitude, denoted as $\Pi_{\mu\nu}$, describes the 
$j_\mu^{em} \to h \to  j_\nu^{em}$ transition,
and is called the  {\em correlation function}
(or correlator) of quark currents. The formal expression
for this object reads:  
\begin{equation}
\Pi_{\mu\nu}(q)=i\int d^4x ~e^{iqx}\langle 0 | 
T\{j^{em}_\mu(x)j^{em}_\nu(x)\}|0\rangle\,.
\label{pimunu}
\end{equation}
Due to the
conservation of e.m. current ($\partial^\mu j_\mu^{em}=0$),
the correlation function depends 
on one invariant amplitude:
\begin{equation}
\Pi_{\mu\nu}(q)=(-g_{\mu\nu}q^2+q_\mu q_\nu)\Pi(q^2)\,.
\label{pimunu2}
\end{equation}

Substituting (\ref{pimunu2}) in (\ref{Aee}) and taking 
imaginary part from both sides, we obtain
\begin{equation}
\mbox{Im}~{\cal A}^{(e^+e^-\to h \to e^+e^-)}(s)= -\frac{e^4}{s}
(\bar{\psi}_e\gamma^\mu\psi_e)(\bar{\psi}_e\gamma_\mu \psi_e)
\mbox{Im}\Pi(s)\,.
\label{impart}
\end{equation}

It is convenient to normalize the hadronic cross section to the 
$e^+e^- \to \mu^+\mu^-$ cross section known from QED. One can literally
repeat the derivation done above, taking  
instead of hadronic states the $ \mu^+\mu^-$ states: $|n\rangle =|\mu^+\mu^-\rangle$. 
The resulting relations  
are quite similar to (\ref{unit2}), (\ref{Aee}) and (\ref{impart}): 
\begin{equation}
2~\mbox{Im}~{\cal A}^{(e^+e^-\to \mu^+\mu^- \to e^+e^-)}(s)=
\sum\limits_{\mu^+\mu^-} |\langle \mu^+\mu^-|T|e^+e^-\rangle|^2
= 2s\sigma^{(e^+e^-\to \mu^+\mu^-)}(s)\,,
\label{unit_mu}
\end{equation}
\begin{equation}
{\cal A}^{(e^+e^-\to \mu^+\mu^- \to e^+e^-)}(q^2)= \frac{e^4}{(q^2)^2}
(\bar{\psi}_e\gamma^\rho\psi_e)(\bar{\psi}_e\gamma^\lambda \psi_e)\Pi^{(\mu)}
_{\rho\lambda}(q)\,,
\label{Amu}
\end{equation}
and 
\begin{equation}
\mbox{Im}~{\cal A}^{(e^+e^-\to \mu^+\mu^- \to e^+e^-)}(s)= -\frac{e^4}{s}
(\bar{\psi}_e\gamma^\mu\psi_e)(\bar{\psi}_e\gamma_\mu \psi_e)
\mbox{Im}\Pi^{(\mu)}(s)\,,
\label{impartmu}
\end{equation}
where the muonic correlation function $\Pi^{(\mu)}$ is 
nothing but a 2-point muon-loop diagram.
Furthermore, the cross section in (\ref{unit_mu}) taken from QED 
textbooks, reads:  
\begin{equation}
\sigma^{(e^+e^- \to \mu^+\mu^-)}(s)=\frac{4\pi\alpha_{em}^2}{3 s}\,.
\label{mumu}
\end{equation}
Dividing the hadronic unitarity relation 
(\ref{unit2}) by the muonic one (\ref{unit_mu})
and using (\ref{impart}) and (\ref{impartmu}),
we obtain a useful ratio:
\begin{equation}
\frac{\mbox{Im}\Pi(s)}{\mbox{Im} \Pi^{(\mu)}(s)}= 
\frac{\sigma^{(e^+e^- \to h)}_{tot}(s)}{
\sigma^{(e^+e^- \to \mu^+\mu^-)}(s)}
\equiv R(s)\,.
\label{R}
\end{equation}
\begin{figure}
\begin{center}
\includegraphics*[width=12cm]{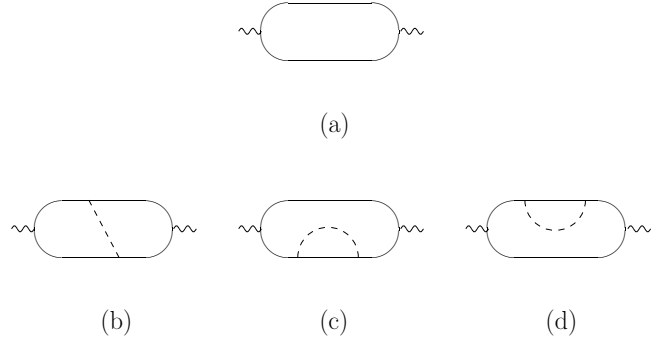}
\caption{{\em  Diagrams corresponding to the 
perturbative contributions to the correlation function of two quark currents: 
(a) the leading-order loop and (b-d) the $O(\alpha_s)$ corrections.
Wavy lines denote external currents with 4-momentum $q$, solid lines
quarks and dashed lines gluons. }}
\label{fig:2loop}
\end{center}
\end{figure}

The next key point in our derivation
is the analysis of the correlation
function $\Pi(q^2)$ at spacelike $q^2<0$.
At large $|q^2|=Q^2\gg \Lambda_{QCD}^2$, 
the long-distance domain  
in the space-time integral in (\ref{pimunu}) is  
suppressed by the strongly oscillating exponent 
and the short distances/times $|\vec{x}|\sim x_0\sim 1/Q$
dominate. This justifies using 
QCD perturbation theory with 
$\alpha_s(Q)\to 0$ at $Q\to \infty$. The 
leading-order asymptotically free result is 
given  by the 2-point quark-loop diagram (Fig.~\ref{fig:2loop}a),
and the next-to-leading corrections are determined 
by $O(\alpha_s)$ two-loop diagrams (Fig.~\ref{fig:2loop}b,c,d).
Calculation of these diagrams yields (at $m_q\ll Q$):
\begin{equation}
\Pi(q^2)=-\frac{1}{4\pi^2}\left(\sum\limits_qQ_q^2\right)
\log\left(\frac{-q^2}{\mu^2}\right)
\left(1+\frac{\alpha_s}{\pi}\right)+ O\left(\frac{1}{\epsilon}\right)+const\,, 
\label{pimunuloop}
\end{equation}
where each flavour $q$ contributes with the same expression and a
coefficient $Q_q^2$. The natural scale for $\alpha_s$ is $q^2$. 
In recent years, due to tremendous calculational efforts,
the $O(\alpha_s^2)$ and even $O(\alpha_s^3)$ corrections
to $\Pi$ have been calculated; also the loop diagrams for the massive
quark are known with a high accuracy.

\begin{figure}
\begin{center}
\includegraphics*[width=8cm]{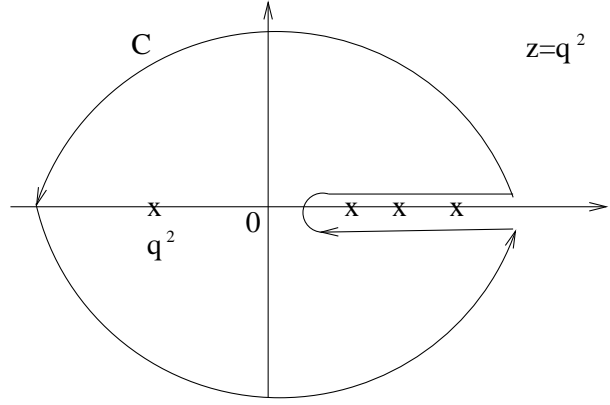}
\caption{{\em The integration contour in (\ref{Cauchy}). The crosses
on the positive real axis indicate singularities of $\Pi(q^2)$ 
}}
\label{fig:cont}
\end{center}
\end{figure}

In the final stage of our derivation 
the calculated
function  $\Pi(q^2)$ at $q^2 <0$  is related to 
$\mbox{Im}\Pi(s)$  at positive $s$.
One employs Cauchy theorem for the 
function $\Pi(z)$ obtained from  $\Pi(q^2)$ by 
analytically continuing the real variable  $q^2$ to 
the complex values, $q^2\to z$:
\begin{equation}
\Pi(q^2)=\frac{1}{2\pi i}\int\limits_C dz\frac{\Pi(z)}{z-q^2}\,.
\label{Cauchy}
\end{equation}
The integration contour C is shown in Fig.~\ref{fig:cont}.
It circumvents the singularities of the function $\Pi(z)$, i.e.
the points/regions where $\mbox{Im}\Pi(z)\neq 0$. According to (\ref{unit2}) and (\ref{impart})  
the location of singularities at real $q^2>0$ 
is determined by the masses of resonances and/or 
the thresholds of multiparticle hadronic states 
produced in $e^+e^-\to h$; the lowest one is at $s_{min}=4m_\pi^2$ 
corresponding  to the threshold of the lightest two-pion state.
Subdividing the contour $C$ into: 1) a large circle with the 
radius $R$, 2) an infinitely  small semicircle
$\tilde{C}$ surrounding $s_{min}$ 
and 3) two straight lines from $s_{min}$ to $R$, we can rewrite 
the integral in terms of three separate contributions:   
\begin{equation}
\Pi(q^2)=\frac{1}{2\pi i}\int\limits_{|z|=R} dz\frac{\Pi(z)}{z-q^2}
+ \frac{1}{2\pi i}\int\limits_{s_{min}}^{R} dz
\frac{\Pi(z+i\delta)-\Pi(z-i\delta)}{z-q^2}\label{disp1}+
\frac{1}{2\pi}\int\limits_{\tilde{C}} dz\frac{\Pi(z)}{z-q^2}\,.
\end{equation}
Suppose the function decreases at $|q^2|\to \infty$:
$\Pi(q^2)\sim 1/|q^2|^\lambda$, where $\lambda>0$.
Then the first integral vanishes at $R\to \infty$.
Taking infinitely small semicircle, one makes the third
integral also vanishing. Furthermore, since there are no singularities of 
$\Pi(z)$ at $\mbox{Re} z<s_{min}$, the integrand in the second integral reduces
to the imaginary part: $\Pi(q^2+i\delta)-\Pi(q^2-i\delta)=2i\mbox{Im}\Pi(q^2)$
(due to Schwartz reflection principle).
Finally, we obtain the desired {\em dispersion relation} 
\begin{equation}
\Pi(q^2)=\frac{1}{\pi}\int\limits_{s_{min}}^\infty ds 
\frac{\mbox{Im}\Pi(s)}{s-q^2-i\delta}\,
\label{disprel}
\end{equation}
with l.h.s. calculated 
in QCD in a form of perturbative expansion
(\ref{pimunuloop}) and r.h.s. related to the cross section via (\ref{R}):
\begin{equation}
\mbox{Im} \Pi(s)= R(s)\mbox{Im}\Pi^{(\mu)}(s)\,.
\label{impi}
\end{equation}
The quantity $R(s)$ is directly measurable in $e^+e^-$ experiments. 
It remains to determine  $\mbox{Im}\Pi^{(\mu)}(s)$.
Knowing the answer for the quark loop diagram in
Fig.~\ref{fig:2loop}a it is very easy
to write down the expression for the muon loop.
Since we neglect masses in both diagrams, 
the only difference is the factor 3, 
from summing up the colour states in the quark loop.
This factor is naturally absent for the muon loop.
From (\ref{pimunuloop}), taking imaginary part, one obtains 
\begin{equation}
\mbox{Im}\Pi^{(\mu)}(s)=\frac{1}{12\pi} \,.
\label{Immuloop}
\end{equation} 
Finally, to guarantee the convergence of the dispersion integral 
(\ref{disprel}), \footnote{For brevity, 
I avoid a longer derivation  which includes
some special mathematical construction (subtractions).}
let us differentiate both parts in $q^2$:
\begin{equation}
\frac{d \Pi(q^2)}{d q^2}=\frac{1}{\pi}\int\limits_{s_{min}}^\infty ds 
\frac{\mbox{Im}\Pi(s)}{(s-q^2)^2}\,.
\label{dispreldiff}
\end{equation}
where (\ref{pimunuloop}) gives for l.h.s. 
\begin{equation}
\frac{d\Pi(q^2)}{dq^2}=\left(\sum\limits_q^{n_f}Q_q^2 \right)
\left(-\frac{1}{4\pi^2 q^2}\right)\left(1+\frac{\alpha_s}{\pi}\right)\,.
\label{PiQCD1}
\end{equation}
Note that divergent and constant terms disappeared
after differentiation and play no role in our derivation.
The final form of the dispersion relation is 
\begin{equation}\frac{3\left(\sum\limits_q^{n_f}Q_q^2 \right)}{-q^2}\left(1+
\frac{\alpha_s}{\pi}+ ...\right)=
\int\limits_{s_{min}}^\infty ds\frac{R(s)}{
(s-q^2)^2}\label{rel2}\,,
\end{equation} 
where ellipses denote higher-order in $\alpha_s$ corrections.
The fact that (\ref{rel2}) is valid at $(-q^2)\to \infty$ unambiguously
fixes the constant limit of $R(s)$ at large $s$:
\begin{equation}
R(s) \to 3\sum\limits_q^{n_f}Q_q^2\,, 
\label{Rasympt}
\end{equation}    
where $n_f$ indicates that $R$ includes 
all ``active'' quark flavours, for which the condition $\sqrt{s}\gg m_q$ 
is fulfilled. Finally, we notice that 
(\ref{Rasympt}) coincides with the parton model prediction
(\ref{sigmatot}), taking into account that the free-quark- and muon-pair
cross sections differ only by the colour factor times the quark charge squared:  
$$
\sigma^{(e^+e^- \to q\bar{q})}(s)= 
3Q_q^2 \sigma^{(e^+e^- \to \mu^+\mu^-)}(s)\,.
$$ 
Importantly, QCD not only reproduces the parton model prediction for  $R(s)$ 
but also provides perturbative corrections,
as well as predicts the integral (\ref{rel2}) over $R(s)$. 
\begin{figure}
\begin{center}
\includegraphics*[width=14cm]{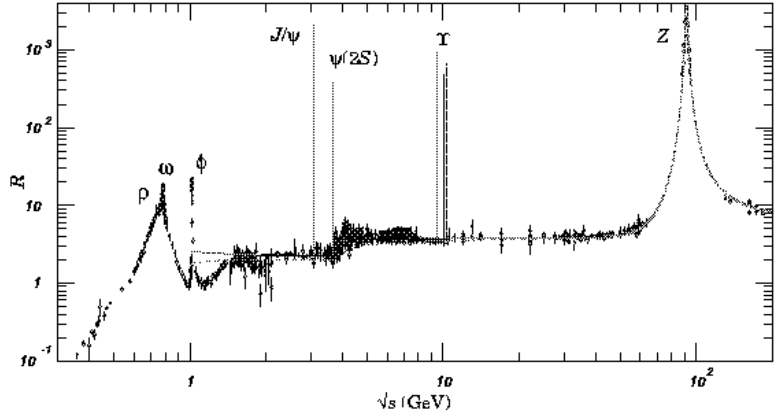}
\caption{{\em Ratio $R$ from \cite{PDG}}}
\label{fig:Ree}
\end{center}
\end{figure}

The experimental data collected in various regions
of $\sqrt{s}$ nicely confirm (\ref{Rasympt}).
According to Fig.\ref{fig:Ree} taken from \cite{PDG},
the ratio $R(s)$ approaches first the constant value $R\simeq 2$
at energies $\sim 2-3$ GeV, above the region
of vector meson resonances $\rho$,$\omega$,$\phi$
(and below the charm threshold). That is exactly the value anticipated from (\ref{Rasympt})
for $n_f=3$. Well above charmonium resonances, a new  
constant level is achieved: $R=2+ 3Q_c^2=10/3$. And 
finally, $R= 10/3+3Q_b^2=11/3$ 
is settled at energies above  
$\Upsilon$ resonances, where all five quark flavours are 
in their asymptotic regime. Actually, the current data on $R(s)$ 
are so precise that one should also include 
small $\alpha_s$-corrections to $R_{QCD}$.

There are other similar inclusive observables 
calculable in QCD, among them the total 
widths $\Gamma_{tot}(Z\to hadrons)$ and
$\Gamma_{tot}(W\to hadrons)$. They have the same status 
as $R(s)$, but a fixed scale $m_Z$ or $m_W$ instead of $\sqrt{s}$. 
One has also to mention an interesting and well developed sub-field 
of perturbative QCD related to the jet and/or heavy-quark production 
in $e^+e^-$ and hadron collisions at high energies.  The underlying 
short-distance quark-gluon processes are successfully traced in the 
experimentally observed multijet structure of the final state. 
Naturally, hadrons cannot be completely avoided, 
because, after all, quarks and gluons hadronize. 
In fact, hadronization in jet physics is nowadays a 
somewhat routine procedure described by QCD-oriented 
models (e.g. the Lund model integrated within PYTHIA \cite{Lund}).
At lower scales, $Q\sim 1-2 $ GeV, inclusive decays of $\tau$-lepton 
are among useful tools to study perturbative QCD
(see, e.g. \cite{tau}).

\subsection{Deep inelastic scattering and operator-product expansion}

The processes of lepton-nucleon {\em deep inelastic scattering} (DIS),
$ l N \to l h$  and $\nu_l N \to l h $ ($l=e,\mu$),
are among the best testing grounds of QCD. 
The long-distance structure of the initial nucleon  
makes these processes  more complicated than $e^+e^- \to hadrons$. 
For definiteness, let us consider the electron-nucleon 
scattering mediated by the virtual photon $\gamma^*$ with 
the 4-momentum  $q$, whereas $p$ is the nucleon 4-momentum. 
To measure the DIS cross section, 
one only has to detect the final electron.
In the nucleon rest frame 
the invariant variables $Q^2=-q^2$ and $\nu=q\cdot p$  
are related to the initial and final energies
and the scattering angle of the electron: 
$ Q^2= 4EE' sin^2\theta$, $\nu=(E-E')m_N$.

The specific kinematical region 
$Q^2 \sim \nu \gg \Lambda_{QCD}^2$ has to be chosen
to reveal the spectacular effect of asymptotic
freedom.    
In this region the experimentally measured DIS cross section,  normalized to
the cross section of the electron scattering on a 
free pointlike quark $\sigma_{point}$,  
depends essentially on the ratio $x=Q^2/2\nu$. 
This effect ({\em Bjorken scaling}),
was first interpreted in terms of a  beautiful phenomenological
model suggested by Feynman.
One considers the reference frame with large nucleon momentum $p$,
so that $-\vec{q}$ gives the longitudinal direction.
Neglecting  long-distance binding forces between quarks, 
the initial nucleon is represented 
as a bunch of free constituents ({\em partons}): 
quarks, antiquarks and gluons moving in the longitudinal direction 
\footnote{We refer here to a 
generic picture of the nucleon, 
where all possible multiparticle components are coherently 
added to the valence three-quark state, similar to (\ref{decomp})
for the pion.}. 
For simplicity the quark masses and transverse 
momenta as well as the nucleon mass are neglected, in comparison 
with $Q^2$ and $\nu$. The electron scatters  on one of the nucleon
constituents (excluding gluons, of course, because they are
electrically neutral) which has the momentum fraction $\xi$ , so that 
after the pointlike collision the quark 4-momentum is $p\xi+q$. 
The massless quark has to remain on-shell, $(\xi p+ q)^2=0$,
therefore $2\xi (p\cdot q) +q^2=0$ and 
\begin{equation}
\xi= Q^2/2\nu=x\,.
\label{condi}
\end{equation}
The  cross section is then represented as a sum of all possible 
``elementary'' processes integrated over $\xi$ 
\begin{equation}
d\sigma^{(eN\to eh)}(Q^2,\nu)= \int\limits_0^1 \sum\limits 
_{i=u,\bar{u},d,\bar{d}...} f_i(\xi)d\sigma^i_{point}(Q^2,\nu)
\delta(\xi-x)d\xi = \sum\limits_{i}d\sigma^i_{point}(Q^2,\nu) f_i(x)\,,
\label{sigmaDIS}
\end{equation} 
where $\delta$-function takes into account (\ref{condi}). 
The sum goes over all quark and antiquark species
inside proton, and $f_i(\xi)$ are the
{\em parton distributions} defined as the probabilities
to find the $i$-th constituent with the momentum fraction $\xi$.

In QCD, the scaling behavoiur (\ref{sigmaDIS}) corresponds 
to the asymptotic-freedom 
approximation ($\alpha_s=0$). Switching on the perturbative 
quark-gluon interactions one predicts  logarithmic corrections $\sim log (Q^2)$ 
to this formula which are nicely reproduced by experiment.
There are many excellent reviews and lectures 
where DIS in perturbative QCD are discussed (see e.g.\cite{dis}). 
Here we shall concentrate 
on one essential aspect: separation of long- and short-distance
effects.

Let us approach DIS from the  
quantum-field theory side, as we did above for $e^+e^-\to hadrons$.
Omitting for simplicity the initial and final electrons, one can 
represent the hadronic part of the DIS cross section in a form 
of $\gamma^*N\to h$ cross section.  
Employing unitarity relation (\ref{tunit}) 
with $|i\rangle = | \gamma^*N\rangle$ 
one is able to relate 
the DIS cross section to the $\gamma^*(q) N(p)\to \gamma^*(q)N(p)$ 
forward-scattering amplitude (see Fig.~\ref{fig:DIS}): 
\begin{figure}
\begin{center}
\includegraphics*[width=10cm]{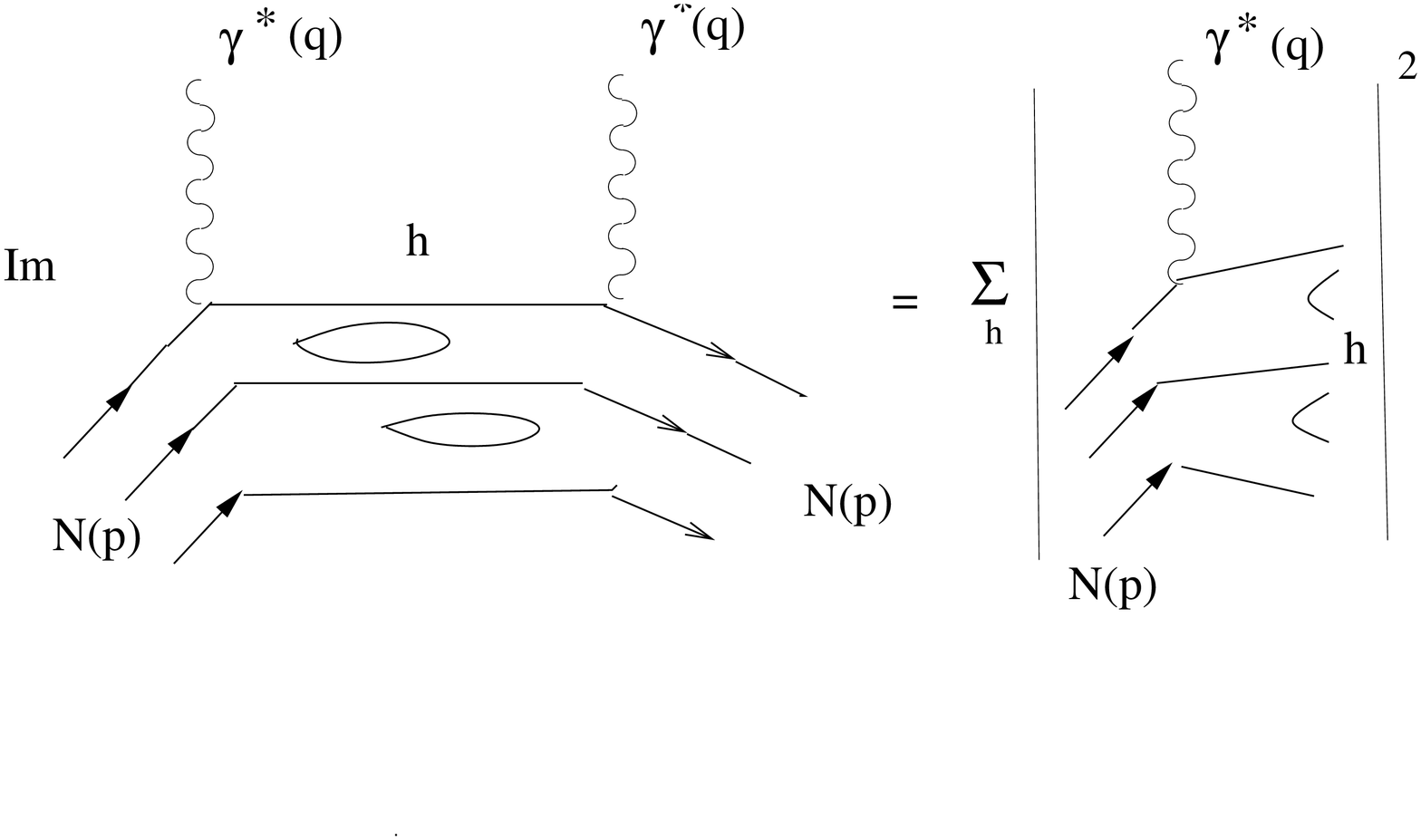}
\caption{{\em Unitarity relation for DIS cross section}}
\label{fig:DIS}
\end{center}
\end{figure}
\begin{equation}
2~\mbox{Im}~{\cal A}^{(\gamma^*N\to h \to \gamma^*N)}(q,p)=
\sum\limits_{h_n} |\langle h_n|T|\gamma^*N\rangle|^2
\sim d\sigma^{(\gamma^*N\to h)}(Q^2,\nu)\,.
\label{unitDIS}
\end{equation}
The amplitude 
\begin{equation}
{\cal A}^{(\gamma^*N\to h \to \gamma^*N)}(q,p)=\epsilon^\mu(q)\epsilon^\nu(q) 
T_{\mu\nu}(q,p)
\label{amplit}
\end{equation}
contains the photon polarization vectors 
multiplied by a new purely hadronic object 
\begin{equation}
T_{\mu\nu}(p,q)=i\int d^4x e^{iqx}\langle N(p) |T\{j^{em}_\mu(x)j^{em}_\nu(0)\}|N(p)\rangle\label{Tmunu}\,,
\end{equation}
which 
resembles the correlation function of two currents we introduced 
above, but, instead of vacuum, has nucleons 
in the initial and final states.

One can prove that at large $Q^2$ and $\nu$ 
the dominant contribution to the above integral stems from small space-time intervals 
$x^2\sim 1/Q^2\sim 1/\nu~$
\footnote{Note that the light-cone dominance $x^2=x_0^2-\vec{x}^2\sim 0$ 
is a more general condition than the small-distance/time 
dominance $x_0\sim\vec{x}\sim 0$ which takes place in $e^+e^-\to hadrons$.}. 
In other words, the points of absorption and emission of the virtual 
photon are located close to the light-cone $x^2=0$. 
The process takes place in the asymptotic freedom regime, that is, 
a single quark absorbs the photon, and penetrates quasi-freely at
small $x^2$ before emitting the photon. All other contributions, 
for example,  with different quarks emitting and absorbing 
initial and final photons are suppressed by powers of $1/Q^2,1/\nu$.   
To describe the free-quark propagation 
from $x$ to 0 we use the quark propagator 
\begin{equation}
\langle 0|T\{\psi_{q}(x)\bar{\psi}_{q}(0)\}|0\rangle=
\int d^4p \left(\frac{p_\alpha\gamma^\alpha}{p^2}\right)e^{-ipx}
=\frac{ix_\alpha\gamma^\alpha}{2\pi^2(x^2)^2}\,.
\label{prop1}
\end{equation}
neglecting the quark mass.
Substituting the e.m. currents in (\ref{amplit}) 
in terms of quark fields, contracting the fields of the propagating 
quark and using (\ref{prop1}) we obtain
\begin{eqnarray}
T_{\mu\nu}(p,q)=i\int d^4x e^{iqx}\sum\limits_{q=u,d,s,..}\langle N(p) 
|T\{\bar{\psi}_q(x)\gamma_\mu \psi_q(x)\bar{\psi}_q(0)\gamma_\nu\psi_q(0) 
\}|N(p)\rangle \nonumber\\
=i\int d^4x e^{iqx}\frac{ix_\alpha}{2\pi^2(x^2)^2}
\sum\limits_{q=u,d,s,..}\langle N(p) 
|\bar{\psi}_q(x)\gamma_\mu\gamma_\alpha\gamma_\nu\bar{\psi}_q(0)| N(p)\rangle 
+\ldots
\label{Tmunu1}
\end{eqnarray}   
In this expression, where only the leading term is shown, 
the calculable short-distance part (the quark propagator)
is separated from  the long-distance part which is 
represented by the quark-antiquark
matrix element taken between nucleon states. This hadronic 
matrix element is a complicated object which has to be either 
determined  
from experiment or calculated using methods beyond perturbative QCD
which will be discussed in the next two lectures.  
I skip the derivation of the 
cross section formula from the imaginary part of $T_{\mu\nu}$ 
which allows to relate the matrix element introduced
in (\ref{Tmunu1}) with the parton distributions. 
Also $\alpha_s$ corrections can be systematically calculated;
they contain important and observable $logQ^2$ effects.
Important is that the long-distance matrix element (or 
parton distributions) are universal characteristics of the nucleon
and they do not change if the short-distance part of the 
process changes (e.g., by switching to neutrino-nucleon scattering 
$W^*N\to h_c$ where the charmed quark is produced).
To summarize, in DIS it is possible to separate the short-distance 
domain by choosing appropriate kinematical region 
and defining convenient physical observables. The short-distance 
quark-gluon interactions  
are calculable in a form of perturbation theory in $\alpha_s$,
whereas the long-distance part is parameterized in terms 
of universal hadronic matrix elements. The procedure
of approximating the product of currents in (\ref{Tmunu1})
by a quark-antiquark operator and separating short and long distances
is called {\em operator-product expansion}
(OPE). Implicitly or explicitly, OPE is used in almost any perturbative QCD 
treatment of hadronic processes.

\section{LONG-DISTANCE DYNAMICS 
AND QCD VACUUM }

\subsection{Vacuum condensates} 

QCD at short distances does not essentially help in 
understanding the long-distance dynamics of quarks and gluons. 
From the short-distance side we only know that the running 
coupling $\alpha_s(Q)$ increases at low-momentum 
scales and eventually diverges at $Q\sim \Lambda_{QCD}$
(see Fig.~\ref{fig:alphaslow}). Is the growth of $\alpha_s$  
the only dynamical reason for confinement?
It is not possible to answer this question
remaining within the perturbative QCD framework, 
because the language of Feynman
diagrams with propagators and vertices is not applicable 
already at $\alpha_s \sim 1$. QCD in nonperturbative 
regime is currently being studied using other methods, 
first of all, the lattice simulations.
From these studies,  there is a growing confidence 
that long-distance dynamics 
is closely related to the nontrivial properties of the 
physical vacuum in QCD. 

For a given dynamical system, 
vacuum is a state with the minimal 
possible energy. Evidently, in QCD the vacuum state contains 
no hadrons, because creating any hadron always costs a certain 
amount of energy. In that sense, QCD vacuum has to be identified
with the $\langle 0 \!\mid$  state in  the 
correlation function  (\ref{pimunu}).
Given that the vacuum state contains no hadrons
does not yet mean that it is completely empty.
There could be quantum fluctuations
of quark and gluon fields with nonvanishing densities. 
The existence of vacuum fields is manifested, e.g. by 
{\em instantons},  special solutions of QCD equations
of motion having a form of localized dense gluonic fields
(for an introductionary review on instantons see, e.g., \cite{inst}).
Lattice QCD provides an independent 
evidence for quark/gluon fields in the vacuum. 
Without going into further theoretical details, 
I will rather concentrate on the phenomenology 
of vacuum fields in QCD. We shall see that 
properties of hadrons are influenced 
by the existence of quark and gluon vacuum
fluctuations with nonvanishing average densities,
the so called {\em vacuum condensates}.  

Formally, in the presence of vacuum fields,   
the matrix elements of quark and gluon field operators
between the initial $\mid \!\!0 \rangle $ 
and final $\langle 0 \!\!\mid$ states are different from zero.  
The combinations of fields have to obey Lorentz-invariance, colour gauge symmetry and
flavour conservation, so that the  
simplest allowed composite operators are  
\begin{eqnarray}
O_3=\bar{\psi}_{q\,i} \psi^i_q,~~    
O_4=G^a_{\mu\nu}G^{a\mu\nu},~~ 
O_5=\bar{\psi}_{q\,i} (\lambda^a)^i_k\sigma_{\mu\nu}G^{a \mu\nu}\psi_q^k,~~
\nonumber
\\
O_6=\left[\bar{\psi}_{q\,i} (\Gamma^a)^i_k\psi^k_q \right]
\left[\bar{\psi}_{q'\,j}(\Gamma^a)^j_l \psi^l_{q'}\right]\,,
\label{vacuum}
\end{eqnarray}
where $\sigma_{\mu\nu}=(i/2)[\gamma_\mu,\gamma_\nu]$ and $\Gamma^a$ are 
various combinations of Lorentz- and colour matrices. The indices at
$ O_d$ reflect their dimension $d$ in GeV units. 
Furthermore, if the operators are taken at different 4-points, 
care should be taken of the  local gauge invariance.
For instance, the quark-antiquark nonlocal matrix element has  
the following form:
\begin{equation}
\langle 0 \mid \bar{\psi}_q(x) [x,0]\psi_q(0)\mid 0\rangle\,, 
\label{qqbarvac}
\end{equation}
where $[x,0]=\exp\left[ig_s\int_0^1 dv x^\mu A^a_\mu(vx)(\lambda^a/2)\right ]$ is
the so-called gauge factor.
Only the matrix elements with the light quarks 
$q=u,d,s$ are relevant for the nonperturbative long-distance dynamics. 
A pair of heavy $c$ ($b$) quarks can be created in vacuum 
only at short distances/times of $O(1/2m_c)$ ($O(1/2m_b)$), that is, 
perturbatively.

Without fully solving QCD, very 
little could be said about vacuum fields, in particular
about their fluctuations at long distances which have 
typical ``wavelengths'' of $O(1/\Lambda_{QCD})$. 
Thus, we are not able to calculate  
the matrix element (\ref{qqbarvac})  
explicitly, as a function of  $x$. 
It is still possible to investigate the vacuum phenomena 
in QCD applying certain approximations. One possibility is to 
study  the average local densities.
The vacuum average of the product of quark and antiquark fields,
\begin{equation}
\langle 0 \mid \bar\psi_{q\,k} \psi_q^k \mid 0\rangle \equiv
\langle \bar{q}q\rangle\,, 
\label{qqbar}
\end{equation}
corresponds to the $x\to 0$ limit of the matrix element
(\ref{qqbarvac}). The simplest vacuum density of gluon fields 
is
\begin{equation}
\langle 0 \mid G_{\mu\nu}^a G^{a \mu\nu}\mid 0 \rangle \equiv
\langle GG\rangle\,.
\label{glucond}
\end{equation}
Due to translational invariance, 
both $\langle \bar{q}q\rangle$ and $\langle GG\rangle$ are 
independent of the 4-coordinate.
These universal parameters are
usually called the densities of quark and gluon condensates, 
respectively. As we shall see in the following subsection, 
the nonvanishing quark condensate 
drastically influences the symmetry properties of QCD. 

\subsection{
Chiral symmetry and its violation in QCD }

Let us return to the isospin 
symmetry limit (\ref{qcdisosp}) of the QCD Lagrangian:
\begin{equation}
L_{QCD}^{(u=d)}= \overline{\Psi}(iD_\mu\gamma^\mu-\tilde{m})\Psi + L_{glue}+... \,.
\label{qcdchiral}
\end{equation}
Since $\tilde{m}\simeq m_u\simeq m_d \ll \Lambda_{QCD}$,  
a reasonable approximation is to put $\tilde{m}\to 0$, 
so that $u$- and $d$-quark components of the $\Psi$-doublet
become massless.

Each Dirac spinor can be decomposed into the left- and right-handed
components,
\begin{equation}
\psi_q=\frac{1+\gamma_5}{2}\psi_q+\frac{1-\gamma_5}2\psi_q \equiv
\psi_{q\,R} +\psi_{q\,L}\,,
\label{psiRL}
\end{equation}
where, by definition, the left-handed (right-handed) quark
has an antiparallel (parallel) spin projection on its $3$-momentum. 
Similarly, for the conjugated fields one has:
\begin{equation}
\bar{\psi}_q = \bar{\psi}_q \frac{1-\gamma_5}{2}+\bar{ \psi}_q\frac{1+
\gamma_5}2 \equiv \bar{\psi}_{q\,R} +\bar{\psi}_{q\,L}\,.
\label{barpsiRL}
\end{equation}
Rewriting $\Psi$ in terms of the 
left- and right-handed components
we obtain the following decomposition of the Lagrangian
(\ref{qcdchiral})
in the massless limit:
\begin{equation}
L_{QCD}^{(u=d)}= \overline{\Psi}_R\,iD_\mu\gamma^\mu\Psi_R+
\overline{\Psi}_L\,iD_\mu\gamma^\mu \Psi_L + L_{glue}+...\,.
\label{Lud}
\end{equation}
The quark-gluon interaction term 
in $L_{QCD}^{(u=d)}$ is now split into two parts,  
$g_s\overline{\Psi}_R\gamma_\mu A^{\mu a} (\lambda^a/2) \Psi_R$ 
and $g_s\overline{\Psi}_L\gamma_\mu A^{\mu a} (\lambda^a/2) \Psi_L$, 
so that quarks conserve their {\em chirality} (left- or right-handedness)
after emitting /absorbing an arbitrary number of gluons.
Importantly, also the interaction of quarks with photons 
has the same property (take the e.m. interaction (\ref{em1})
and decompose it into $L$ and $R$ parts in the same way as above).
In the massless ({\em chiral}) limit, quarks of left- and right 
chiralities propagate and interact independently from each other. 
In fact, it is possible to introduce two independent 
isospin  $SU(2)$ transformations, separately for $L$ and $R$ fields:
\begin{eqnarray}
\Psi_L&\to& \Psi_L'=\exp\left(-i\frac{\sigma^a}{2}\omega^a_L\right)\Psi_L\,,
\nonumber
\\
\Psi_R&\to& \Psi_R'=\exp\left(-i\frac{\sigma^a}{2}\omega^a_L\right)\Psi_R\,.
\label{chiraltransf}
\end{eqnarray}
Restoring the mass in $L_{QCD}^{(u=d)}$ leads to the violation 
of chiral symmetry. The Lagrangian mass term 
can be represented as an effective   
transition between left- and right-handed quarks:  
\begin{equation}
\tilde{m}\overline{\Psi}\Psi=\tilde{m}\left(\overline{\Psi}_L\Psi_R+
\overline{\Psi}_R\Psi_L\right)\,.
\label{mterm}
\end{equation}
Having in mind the smallness of the $u,d$-quark masses, 
one naturally expects that an approximate chiral symmetry 
manifests itself  
in the observable properties of hadrons. In reality, 
the symmetry is violated quite substantially,
as the two following examples demonstrate.
\begin{figure}
\begin{center}
\includegraphics*[width=8cm]{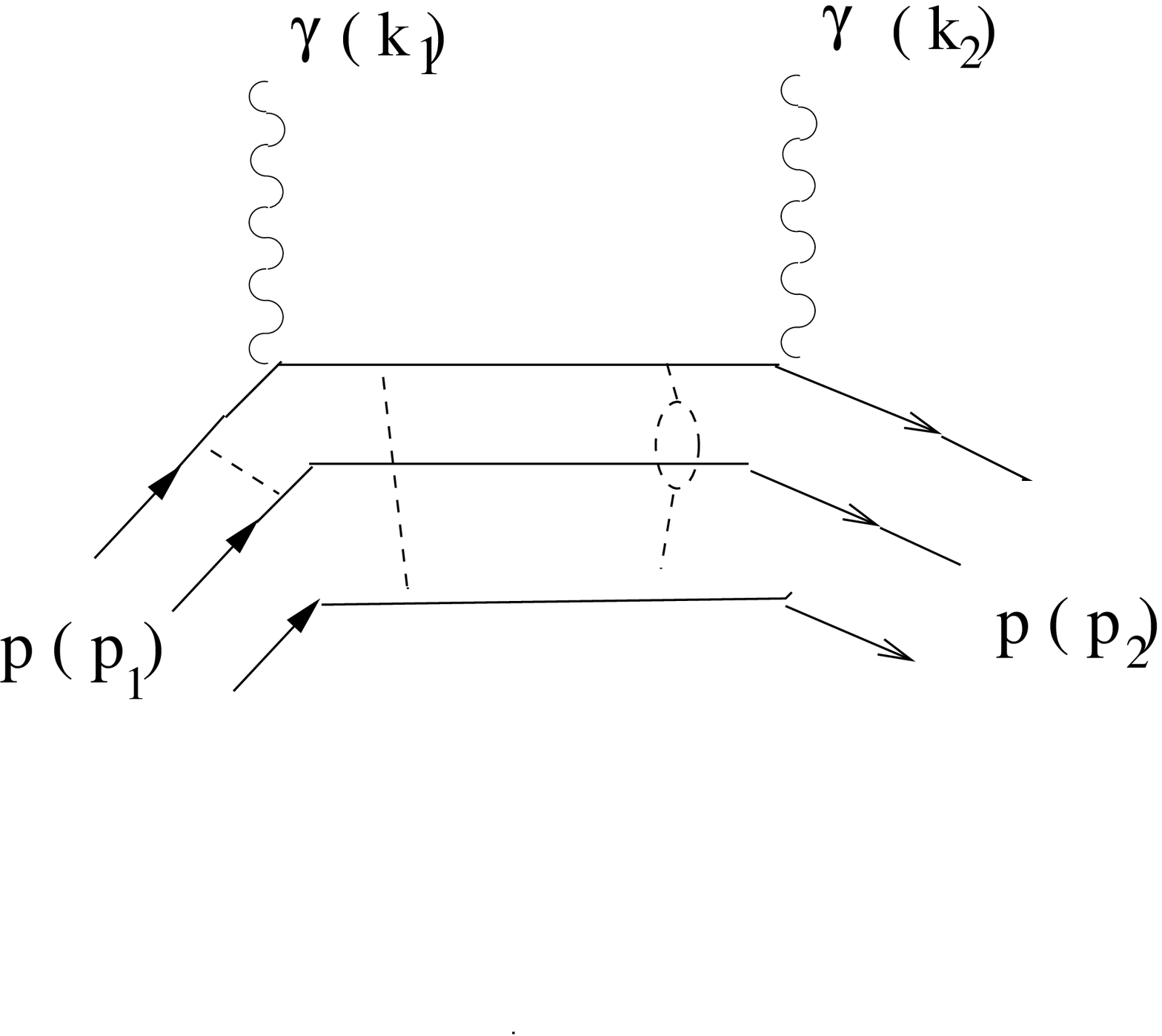}
\caption{{\em One of the diagrams describing the photon-proton scattering.
The second diagram is obtained by interchanging the photon lines.}}
\label{fig:compt}
\end{center}
\end{figure}

The first one is a real hadronic process: the 
photon scattering on a polarized 
(e.g. left-handed) proton.
One of the diagrams is shown in Fig.~(\ref{fig:compt}). This process is a very complicated
mixture of quark-photon and quark-gluon interactions
at long distances, determined by the quark structure of 
the proton. Importantly, all these interactions obey 
an approximate chiral symmetry at the Lagrangian level.
Hence, the amplitude of the proton-chirality flip is expected 
to be very small, $O(m_{u,d}/\Lambda_{QCD})\sim 1\%$ 
of the total  scattering amplitude.
To check this conjecture, let me consider 
the case when the initial photon energy is much smaller
than the proton mass. In this case, it is possible 
to approximate the proton with a point-like particle.
The $\gamma p\to \gamma p$ amplitude is then simply obtained from
the $\gamma e\to \gamma e$ amplitude in QED (Compton effect), replacing  
the electron spinors and propagators by the proton ones:
\begin{eqnarray}
A(\gamma p \to \gamma p) \simeq e^2\bar{u}(p_2)\left[
\gamma_\alpha\frac{(p_1+k_1)_\mu\gamma^\mu +m_p}
{(p_1+k_1)^2-m_p^2}\gamma_\beta
+\gamma_\beta\frac{(p_1-k_2)_\mu\gamma^\mu +m_p}
{(p_1-k_2)^2-m_p^2}\gamma_\alpha\right] 
u_L(p_1)\epsilon_2^\alpha\epsilon^\beta_1,
\label{compt}
\end{eqnarray}     
where $\epsilon_{1,2}$ are the polarization vectors of the 
initial and final photons with the 4-momenta $k_1$ and $k_2$, 
respectively.
Without even completing the calculation, we 
notice that the $\sim m_p$  terms in the amplitude 
flip the initial proton chirality $L\to R$, 
whereas the $\sim\gamma_\mu$ terms preserve chirality.
Importantly, the contributions of both types are 
of the same order, determined by the scale $m_p$,
indicating that chiral symmetry for the photon-proton
scattering is broken at 100\% level.

To present the second example of the chiral symmetry violation,
I start from the correlation function
\begin{equation}
\Pi_{\mu\nu}=i\int d^4x e^{iqx}\langle 0 | T\{j_\mu(x)j_\nu(0)\}|0\rangle
= (-g_{\mu\nu}q^2+q_\mu q_\nu)\Pi(q^2)\,,
\label{corrV}
\end{equation}
very similar to the one introduced in Lecture 3, 
but containing a slightly different quark current:
\begin{equation}
j_\mu=\frac{1}{2}(\bar{\psi}_u\gamma_\mu\psi_u-\bar{\psi}_d\gamma_\mu
\psi_d)\,,
\label{vectcurr}
\end{equation} 
which produces the $I=1$ and $J^P=1^-$ 
quark-antiquark states. Note that this current is conserved, 
$\partial _\mu j^\mu=0$, even if $m_{u,d}\neq 0$. 
We then follow the same derivation as in Lecture 3 and obtain
the dispersion relation (\ref{disprel})
for $\Pi(q^2)$.
The only change is in the imaginary part 
(\ref{impi}), where now only the states with $I=1$ contribute
to the total cross section
($\rho$ meson and its radial excitations, the two-pion state and 
other states with an even number of pions), so that the 
ratio $R(s)$ has to be replaced by 
\begin{equation}
R^{(I=1)}(s)=\frac{\sigma_{tot}(e^+e^-\to h(I=1))}{
\sigma(e^+e^-\to \mu^+\mu^-)}\,.
\label{sigmaI1}
\end{equation}
In parallel, we consider the correlation function
\begin{equation}
\Pi^5_{\mu\nu}=i\int d^4x e^{iqx}\langle 0 | T\{j_{\mu 5}(x)j_{\nu 5}(0)\}|0\rangle
\label{corrA}
\end{equation}
of two axial-vector currents with $I=1$: 
\begin{equation}
j_{\mu 5}=\frac{1}{\sqrt{2}}(\bar{\psi}_u\gamma_\mu\gamma_5\psi_u-
\bar{\psi}_d\gamma_\mu\gamma_5\psi_d)\,.
\label{axialcurr}
\end{equation} 
This current is conserved only 
in the chiral symmetry ($m_{u,d}=0$) limit:
\begin{equation}
\partial ^\mu j_{\mu 5}=\frac{1}{\sqrt{2}}(2m_u\bar{\psi}_u\gamma_5\psi_u-
2m_d\bar{\psi}_d\gamma_5\psi_d)\,.
\label{axialcurr1}
\end{equation} 
Decomposing the correlation function (\ref{corrA})
in two tensor structures:
\begin{equation}
\Pi^5_{\mu\nu}(q)= -g_{\mu\nu}\Pi'_5(q^2)+ q_\mu q_\nu\Pi_5(q^2)\,,
\label{struct}
\end{equation}
we notice that in the chiral limit  
there is only one independent invariant amplitude:
\begin{equation}
\Pi_5(q^2)=\frac{\Pi'_5(q^2)}{q^2}\,.
\label{strAx}
\end{equation}
The dispersion relation 
for $\Pi_5$ has the same form as (\ref{disprel}):
\begin{equation}
\Pi_5(q^2)=\frac1\pi\int\limits_{s^5_{min}}^\infty ds\frac{
\mbox{Im} \Pi_5(s)}{s-q^2-i\delta}\,,
\label{dispa}
\end{equation}
where $s^5_{min}$ is the corresponding threshold.
In order to determine the imaginary part 
via relation similar to (\ref{impi}),
we introduce a slightly artificial 
cross section $\sigma_{tot}(e^+e^-\to Z^A\to h(I=1))$  
of hadron production mediated by $Z$ boson coupled 
to the axial-vector current. Then,
\begin{equation}
\mbox{Im}\Pi_5(s)= \frac{1}{12\pi}R_5^{(I=1)}(s)\,,
\label{im5}
\end{equation} 
where
\begin{equation}
R_5^{(I=1)}(s)=\frac{\sigma_{tot}^{(e^+e^-\to Z^A\to h(I=1))}(s)}{
\sigma^{(e^+e^-\to Z^A\to\mu^+\mu^-)}(s)}\,.
\label{sigmaAI1}
\end{equation}

Both $\Pi(q^2)$ and $\Pi_5(q^2)$ can be calculated at 
large $|q^2|$  from the same 2-point quark-loop diagrams 
shown in Fig.~\ref{fig:2loop}. In the chiral limit, the only difference
is in two extra $\gamma_5$ matrices 
present in  $\Pi^5_{\mu\nu}$.
Hence, in perturbative QCD
\begin{equation}
\Pi_5(q^2)=\Pi(q^2)\,,
\label{eqP}
\end{equation}  
at any order in $\alpha_s$. 
This equation is trivial
for the leading-order loop diagrams (Fig.~\ref{fig:2loop}a). 
Using Dirac algebra, it is easy to check that the 
extra $\gamma_5$ matrices cancel
each other ($\gamma_5^2=1$) in the absence of masses in the propagators. 
Furthermore, each gluon line inserted in the loop
brings two more $\gamma$-matrices which do not influence
that cancellation.

\begin{figure}
\begin{center}
\includegraphics*[width=8cm]{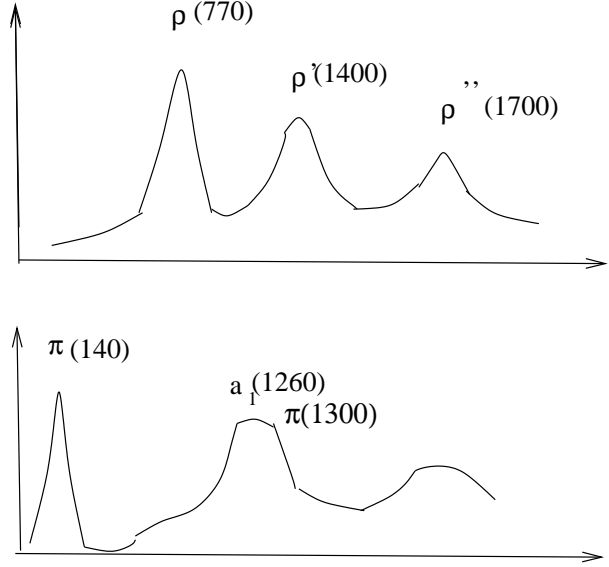}
\caption{{\em A schematic pattern of the resonances  
produced by the vector (above) and axial-vector (below) 
quark currents with $I=1$.}}
\label{fig:spectra}
\end{center}
\end{figure}

From (\ref{eqP}) follows the equation of 
two dispersion relations (\ref{disprel})
and (\ref{dispa}). To keep the dispersion integrals convergent, 
we differentiate them $n$ times at a certain $q^2<0$. The result is:
\begin{equation}
\int\limits_{s_{min}}^\infty ds\frac{R_5(s)}{(s-q^2)^n}=
\int\limits_{s_{min}}^\infty ds\frac{R(s)}{(s-q^2)^n}\,.
\label{eqdisp}
\end{equation}
Note that even in the presence of quark masses 
the corrections to (\ref{eqP}) and to (\ref{eqdisp}) are very small,
$O(m^2_{u,d})$.
If $q^2$ is not very large, $|q^2|\sim 1$ GeV$^2$, the integrals
in (\ref{eqdisp})
are dominated by the contributions of the low-mass hadronic states 
to the corresponding $e^+e^-\to h$ cross sections. In the case 
of vector current, the states are $\rho$-meson and its 
radial excitations. 
Hence, $R(s)$ represents a resonance curve 
with the peaks located at $s=m_\rho^2$, $m_{\rho'}^2$ etc. 
(see Fig.~\ref{fig:spectra}). The validity of (\ref{eqdisp})
at arbitrary (but large) $q^2<0$ implies that also $R_5(s)\simeq R(s)$.
We then expect the spectrum of resonances  
generated by the axial-vector current to resemble, in gross features,
the $\rho$-spectrum. However, experimental 
data \cite{PDG} reveal a completely different picture.
The lowest resonances 
in the axial-vector channel are the 
pion with $J^P=0^-$ , the axial meson 
$a_1(1260)$ and the radial excitation $\pi(1300)$. 
Summarizing, there is a clear evidence, based on the observed properties
of hadrons,  that chiral symmetry in QCD is violated much stronger
than it is expected from $L_{QCD}$.  

An additional source of the chiral symmetry violation
is provided by the quark condensate. 
Decomposing the quark and antiquark fields in (\ref{qqbar}) in the left-handed 
and right-handed components, 
\begin{equation}
\langle 0| \bar{\psi}_{q\,i}\psi^i_q |0\rangle=
\langle 0| (\bar{\psi}_{qR}+\bar{\psi}_{qL})
(\psi_{q R} +\psi_{qL})|0\rangle=
\langle 0|\bar{\psi}_{qR}\psi_{qL}+\bar{\psi}_{qL}\psi_{qR})) 
|0\rangle\neq 0\,,
\label{qqbar1}
\end{equation} 
we realize that condensate causes vacuum transitions 
between quarks of different chiralities.
Hence, in QCD one encounters  a {\em spontaneous broken} 
chiral symmetry,
a specific situation when the interaction (in this case $L_{QCD}$) 
obeys the symmetry (up to the small $O(m_{u,d})$ corrections), 
whereas the lowest-energy state (QCD vacuum) violates it \footnote{
That is quite similar to the electroweak sector of
the Standard Model \cite{Aitchison}, where the electroweak gauge symmetry 
is spontaneously broken by the nonvanishing vacuum average of the 
Higgs field.}. 
We conclude that  in order to correctly reproduce 
the properties of hadrons and hadronic amplitudes 
(e.g., correlation functions), one has to take into account 
the vacuum fields, in particular, the quark condensate.

\subsection{Condensate contributions to correlation functions}
\begin{figure}
\begin{center}
\includegraphics*[width=12cm]{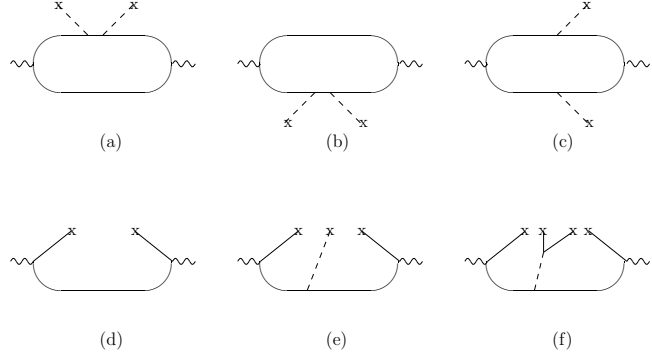}
\caption{{\em Diagrams used to calculate condensate contributions 
to the correlation functions of two quark currents. Lines
with crosses denote vacuum fields.}}
\label{fig:cond}
\end{center}
\end{figure}
Let us return to the correlation functions
(\ref{corrV}) and (\ref{corrA}), 
restoring nonzero $u$ and $d$ quark masses, that is, working with full
QCD. The vacuum quark-gluon fields 
generate  new contributions to  $\Pi$ or $\Pi_5$.
In addition to the perturbative loop diagrams in Fig.~\ref{fig:2loop}, 
there are  diagrams shown in Fig.~\ref{fig:cond}, 
where gluons, quarks and antiquarks penetrate to long
distances, being absorbed and emitted by vacuum fluctuations.   
The vacuum fields have  
characteristic momenta of $O(\Lambda_{QCD})$. 
Therefore, if the momentum scale in the correlator is large, $Q=\sqrt{-q^2}\gg \Lambda_{QCD}$, 
one can approximate the vacuum state by a set of constant  fields. 
In other words, the
virtual quark propagating at short distances/times between 
the points $x$ and $0$, cannot ``resolve'' the long-distance fluctuations of 
vacuum fields and percepts them in a form of  
averaged static fields. This approximation makes 
the calculation of diagrams in Fig.~\ref{fig:cond}  straightforward.
In addition to Feynman rules of perturbative QCD
for virtual quarks and gluons,
one has to form all possible combinations (\ref{vacuum})
of vacuum fields 
and replace them by the corresponding  condensate densities,
so that the 4-momenta of the crossed lines on these diagrams 
are neglected. 
For example, the product of quark and antiquark vacuum fields
in Fig.~\ref{fig:cond}~d has to be replaced by $\langle \bar{q}q\rangle$.     
More details on these calculations can be found, e.g., in the review \cite{CK}.
The result for the vector-current correlator 
has the following schematic form:
\begin{equation}
\Pi(q^2)= \Pi^{pert}(q^2)+
\sum\limits_{d=3,4,...} C_d(q^2)\langle 0|O_d |0\rangle\,,
\label{OPE}
\end{equation}
where the first term on r.h.s. corresponds to the perturbative diagrams
in Fig.~\ref{fig:2loop}, whereas the sum contains 
the contributions of vacuum condensates with dimensions $d$ 
obtained from the diagrams in Fig.~\ref{fig:cond}.
To compensate the growing dimension of the operators $O_d$, the
coefficients $C_d$ contain increasing powers of $1/Q$.
Thus, the condensate contributions  die away at $Q\to\infty$ and do not 
alter the 
perturbative asymptotics of the correlator 
given by  $\Pi^{pert}(q^2)$. Furthermore, as soon as we work at
relatively large $Q\gg \Lambda_{QCD}$, it is possible to retain
only a few first terms in the sum , that is, neglect diagrams
with more than 3-4 vacuum fields emitted from the virtual quarks 
in the correlation function.

For the axial-vector correlation function $\Pi_5(q^2)$  
one obtains an expression similar to (\ref{OPE}) with the same
perturbative part (up to very small corrections of $O(m_{u,d}^2)$),
but with different coefficients $C_d$ at certain condensate terms.
The most important deviation from the vector-current case
is in the value and sign of $C_6$, i.e., in the 4-quark condensate terms.
Thus, addition of condensate effects leads to an explicit violation 
of (\ref{eqP}). It is then not surprising 
that  hadron resonances contributing to $R(s)$ and $R_5(s)$ 
in (\ref{eqdisp}) are different. 

Furthermore, the method of 
correlation functions allows 
to reproduce  an important relation  for the pion mass,
explaining the smallness of $m_\pi$.
Note that from the point of view of naive quark model
$\pi$ and $\rho$ mesons differ only by orientations of quark spins.
Why is then $m_\pi \ll m_\rho $ and, moreover, $m_\pi<\Lambda_{QCD}$?
We consider
the correlation function similar to (\ref{corrA}), but for simplicity,
containing charged axial currents:
\begin{equation}
\Pi^5_{\mu\nu}=i\int d^4x e^{iqx}\langle 0 | 
T\{j_{\mu 5}(x)j^\dagger_{\nu 5}(0)\}|0\rangle\,=
-g_{\mu\nu}\Pi_5'(q^2)+ q_\mu q_\nu\Pi_5(q^2)\,,
\label{corrA1}
\end{equation}
where $j_{\mu5}= \bar{u}\gamma_\mu\gamma_5 d$.
This current is the part of the Standard Model weak current (\ref{jweak}),
responsible for $u\to d$ transition , e.g. the $\pi \to \mu \nu_\mu$
decay. The hadronic matrix element which determines this decay,
\begin{equation}
\langle 0| \bar{u} \gamma_\mu\gamma_5 d | \pi^+(q)\rangle = iq_\mu f_\pi \,, 
\label{fpi}
\end{equation}
is parameterized via the pion decay constant $f_\pi$ which plays
an essential role in our analysis. 
It is convenient to 
multiply (\ref{corrA1}) by $q_\mu q_\nu /q^2$, forming 
a combination of invariant amplitudes:
$$
-\frac{q^\mu q^\nu}{q^2}\Pi^5_{\mu\nu}=\Pi'_5(q^2)-q^2\Pi_5(q^2)\equiv \tilde{\Pi}_5(q^2)\,.
$$
Note that $\tilde{\Pi}_5=0$ at $m_{u,d}=0$ and, in particular, 
the perturbative part of $\tilde{\Pi}_5$ vanishes as $O(m_{u,d}^2)$.  
Concerning nonperturbative part, the only
first-order in $m_{u,d}$ contribution is given by the quark condensate diagram in Fig.~\ref{fig:cond}d
\begin{equation}
\tilde{\Pi}_5(q^2)=-\frac{(m_u+m_d)(\langle\bar{u}u\rangle +\langle\bar{d}d\rangle)}{q^2}+O(m_{u,d}^2)\,.
\label{pi5tilde}
\end{equation}
To proceed,  we use for $\tilde{\Pi}_5$ the dispersion relation of the type
(\ref{dispa}). To obtain 
the imaginary part, one has to return to the unitarity relation 
(\ref{tunit}), identifying $|i\rangle$ with $|0\rangle$ and $T_{ii}$ with the 
correlation function. The result is:
\begin{equation}
2\mbox{Im}\tilde{\Pi}_5(s)= -\frac{q^\mu q^\nu}{q^2}\sum\limits_{h_n}\langle 0
|j_{\mu5}|h_n\rangle \langle h_n|j^\dagger_{\nu5}(q)|0\rangle\,.
\label{unitPi}
\end{equation} 
Importantly, only pseudoscalar states contribute to the above sum,
because the matrix elements for the axial-vector mesons $a_1$ 
and its excitations vanish, being
proportional to the transverse polarization vectors of these mesons:
$$q^\mu\langle 0 |j_{\mu5}|a_1\rangle\sim q^\mu\epsilon_\mu^{a_1} =0 .$$ Using 
(\ref{unitPi}) and the definition (\ref{fpi})
we obtain the following expression for the dispersion relation:
\begin{eqnarray}
\tilde{\Pi}_5(q^2)&=& 
\frac1{2\pi}\int\limits_{m_\pi^2}^\infty
\frac{ds}{s-q^2-i\delta}
\left\{-\frac{q^\mu q^\nu}{q^2}(f_\pi q_\mu)( f_\pi q_\nu)
\int \frac{d^3 p_\pi (2\pi)}{2E_\pi}\delta^{(4)}(p_\pi-q)
\right\}_{q^2=s}+ ....\nonumber\\
&=&\int\limits_{m_\pi^2}^\infty
\frac{ds}{s-q^2-i\delta}\left\{-f_\pi^2 s\delta(s-m_\pi^2)\right\}+ ...\,,
 \label{disppi}
\end{eqnarray}
where the one-pion state contribution is shown explicitly 
(with $p_\pi^2=m_\pi^2$) including the phase space proportional to
$\delta$-function. Ellipses denote excited pions and
multiparticle states with the same quantum numbers.
Integrating out  $\delta$-function and substituting  (\ref{pi5tilde})
in l.h.s. we obtain
\begin{equation}
-\frac{(m_u+m_d)(\langle\bar{u}u \rangle + \langle \bar{d}d\rangle)}{q^2}+O(m_{u,d}^2) =
-\frac{f_\pi^2m_\pi^2}{m_\pi^2-q^2}-\sum\limits_{\pi'}\frac{f_{\pi'}^2m_{\pi'}^2}{m_{\pi'}^2-q^2}\,,
\label{finalpi}
\end{equation}
where the sum over higher states is also shown schematically.
To fulfil  this equation at large $q^2$ one has to demand
that $m_\pi^2\sim m_u+m_d$. Simultaneously, the decay constants 
of excited states have to obey: $f_{\pi'}\sim m_u+m_d$, otherwise the 
$q^2$ asymptotics of l.h.s and r.h.s.  in (\ref{finalpi})
is different at $O(m_{u,d})$ .
We then reproduce the well known Gell-Mann-Oakes-Renner relation
\begin{equation}
-(m_u+m_d) (\langle\bar{u}u\rangle +\langle \bar{d}d \rangle)+
O(m_{u,d}^2)= f_\pi^2m_\pi^2\,.
\label{GOR}
\end{equation}
This relation reflects the special nature of pion in QCD. 
The anomalously small pion mass is not accidental and is closely
related to the spontaneous chiral symmetry breaking via condensate.
If  a symmetry in quantum field theory is 
broken spontaneously, there should be massless states (Nambu-Goldstone
particles), one per each degree of freedom of broken symmetry. 
The three pions, $\pi^+,\pi^-,\pi^0$  play a role of massless 
Nambu-Goldstone particles in QCD. 
In other words, due to the specific structure of QCD vacuum fields, the
amount of energy needed to produce a pion state tends to zero.
The fact that pions still have small nonvanishing masses is due to the 
explicit violation of chiral symmetry via $u,d$ quark masses.

How large is the quark condensate density? Using $m_{\pi^{\pm}}\simeq 140
$ MeV, $f_\pi=131 $ MeV \cite{PDG} and taking the $u,d$ quark mass values 
from (\ref{qmasses}), renormalizing them at $Q=1$ GeV,  
one obtains from (\ref{GOR}), typically, 
\begin{equation}
\langle \bar{q}q \rangle(\mu=1 GeV) \simeq (-240\pm 10~ \mbox{MeV})^3\,,
\label{qqcondens}
\end{equation}
where we assume isospin symmetry for the condensates
$\langle \bar{q}q \rangle\simeq \langle \bar{u}u \rangle\simeq \langle \bar{d}d \rangle$. 
Not surprisingly, the estimated value is in the ballpark of $\Lambda_{QCD}$ !
Being not a measurable physical quantity, the condensate
density is a scale-dependent parameter. Since r.h.s. of (\ref{GOR})
is determined by hadronic parameters $f_\pi, m_\pi$, 
which are both scale-independent, the running of the 
quark condensate should compensate  
the running of the quark mass given in (\ref{runmass}), that is:
\begin{equation}
\langle \bar{q}q \rangle(Q)=\langle \bar{q}q \rangle(Q_0)
\left(\frac{\alpha_s(Q)}{\alpha_s(Q_0)} \right)^{-\gamma_0/\beta_0}\,.
\label{condrun}
\end{equation}

\subsection{Gluon condensate}

The gluon condensate density is another important characteristics
of nonperturbative QCD. This parameter cannot be easily 
estimated from correlation functions with light quarks,
because the latter are dominated by quark condensates.
A very useful object, sensitive to the gluon
condensate is the correlation function of $c$-quark currents:
\begin{equation}
\Pi_{\mu\nu}^c= 
i\int d^4x e^{iqx}\langle 0 | T\{ j_\mu^c(x) j^c_\nu(0)\}|0\rangle
= (-g_{\mu\nu}q^2+q_\mu q_\nu)\Pi^c(q^2)\,,
\end{equation}
where $j^c_\mu=\bar{c}\gamma_\mu c $ is the $c$-quark part of 
the quark e.m. current $j^{em}_\mu$. Following the same 
derivation as in Lecture 3, we write down the dispersion relation
for $\Pi^c(q^2)$ relating Im$\Pi^c(s)$  with the ratio $R_c(s)$ defined as:
\begin{equation}
R_c=\frac{\sigma(e^+e^-\to charm)}{\sigma(e^+e^-\to \mu^+\mu^-)}\,,
\label{Rc}
\end{equation}
where the cross section $\sigma(e^+e^-\to charm)$ includes 
hadronic states with $\bar{c}c$ content produced in $e^+e^-$:
charmonium resonances ($J/\psi,\psi',...$), pairs of charmed hadrons etc.  
In addition to the perturbative $c$-quark loop diagrams in 
 Fig~\ref{fig:2loop}, one has to include
also the contribution of the diagrams shown in Fig~\ref{fig:cond}a,b,c, 
with $c$ quarks emitting vacuum gluons. (Remember that nonperturbative quark 
condensate for the heavy $c$ quark is absent.)  The resulting relation 
has the following form:
\begin{equation}
\int\limits_{m_\psi^2}^\infty \frac{ds}{s-q^2}R_c(s)=\Pi^{pert}(q^2,m_c^2,\alpha_s)
+\frac{\alpha_s}{\pi} \langle G G\rangle f_c(q^2,m_c^2)\,,
\label{Rcdisp}
\end{equation}
where $m_\psi$ is the mass of the lowest $J/\psi$ state in this channel.
The function $\Pi^{pert}$ evaluated from the massive $c$ -quark loop
diagrams has  
a more complicated form than in the massless case. The function $f_c$ is the 
calculable short-distance part of the diagrams in Fig. ~\ref{fig:cond}a,b,c.
To achieve a better convergence at $s\to \infty$, the dispersion relation
is usually differentiated $n$ times. Note that in this case the point $q^2=0$
is also accessible: $c$ quarks are still highly virtual at $q^2=0$
because  the long-distance region starts at $q^2\sim 4m_c^2$. 
The set of power moments obtained from (\ref{Rcdisp}) 
with $O(\alpha_s^2)$ accuracy has a form
\begin{equation}
\int\limits_{m_\psi^2}^\infty \frac{ds}{s^{n+1}}R^c(s)=\frac{3Q_c^2}{(4m_c^2)^n}
r_n\left[1+\alpha_s(m_c)a_n+(\alpha_s(m_c))^2a_n'
+b_n\frac{\langle \frac{\alpha_s}{\pi}GG\rangle}{(4m_c^2)^2}\right]\,,
\label{moments}
\end{equation}
with calculable coefficients $r_n,a_n,a_n',b_n$. The natural scale 
for $\alpha_s$ in the perturbative loops is in this case the 
virtuality $Q\sim m_c$. The moments (\ref{moments}) are used to extract 
the gluon condensate density  and, simultaneously the $c$ quark mass, 
employing the experimental data on $R_c(s)$ on l.h.s. The estimate
of the condensate density obtained first in \cite{SVZ} is: 
\begin{equation}
\frac{\alpha_s}{\pi}\langle GG\rangle= (330 ~\mbox{MeV})^4 \pm 50\%,
\label{GGvalue}
\end{equation}  
again within the range of $\Lambda_{QCD}$.
The value of the gluon condensate density is usually 
given multiplied by $\alpha_s$  for convenience, because 
this product is scale-independent.

\section{RELATING QUARKS AND HADRONS: QCD SUM RULES}

\subsection{Introducing the method}

The relation (\ref{moments}) obtained first in \cite{NOSVVZ} 
is a well known example of a {\em QCD sum rule}.
The method  developed by Shifman, Vainshtein and Zakharov \cite{SVZ} 
employs  quark-current correlation functions 
calculating them  in the spacelike region, 
including perturbative and condensate contributions. 
Consider for example the correlation function (\ref{corrV}), with
the result of QCD calculation having the form (\ref{OPE}). 
Note that in the latter expression 
short- and long-distance effects are separated. The perturbative part 
$\Pi^{pert}(q^2)$ and the coefficients $C_d(q^2)$ 
take into account short-distance quark-gluon interactions
with characteristic momenta larger than the scale $Q$. 
Both $\Pi^{pert}$ and $C_d$ are {\em process-dependent}, 
i.e., depend on the choice of the currents.
On the other hand, the condensate densities absorb, in an averaged
way, the long-distance interactions with momenta less than $Q$
and are process-independent. 
The universality of condensates 
allows to calculate correlation functions
in different channels  without introducing new inputs, in an
almost model-independent way \footnote {The 
expansions similar to (\ref{OPE}) represent another example 
of OPE in QCD (see Lecture 3.3), when a product of two currents
is expanded in a set of local operators $O_d$. The perturbative part
in this case is interpreted as a unit operator with zero dimension.}.
To obtain the sum rule,  the QCD result for the correlation
function is matched, via dispersion relation, to  the sum 
over hadronic contributions (the integral over hadronic cross section).

A twofold use of the sum rule relations is possible.
Firstly, using experimental data
one saturates the hadronic sum in the dispersion integral and 
extracts the universal QCD parameters: quark
masses, $\alpha_s$, condensate densities.
One example is the $c\bar{c}$  sum rule (\ref{moments}) discussed above. 
Secondly,  having fixed QCD parameters, one calculates, with a certain 
accuracy, the masses and decay amplitudes 
of the lowest hadrons  entering the dispersion integral.
The relation (\ref{GOR}) obtained from the sum rule 
(\ref{finalpi}) can serve as an example. Note that (\ref{finalpi}) is unique, 
because all states in the hadronic sum except the pion and all QCD terms 
except the quark condensate are absent at $O(m_{u,d})$.

\subsection{The $\rho$-meson decay constant}
 
To demonstrate in more detail how the method works
\footnote{reviews can be found e.g., in \cite{CK},\cite{SRrev}.}, 
let me outline the calculation of the $\rho$-meson decay
constant from the correlation
function (\ref{corrV}). 
The dispersion relation for $\Pi_{\mu\nu}$ can be written in the 
following form:
\begin{eqnarray}
\Pi_{\mu\nu}(q) = \frac{1}{2\pi}\int\limits_{s_{min}}^\infty\frac{ds}{s-q^2-i\delta} 
\left\{
\sum\limits_{h_n}\langle 0 |j_{\mu}|h_n\rangle \langle h_n|j_{\nu}(q)|0\rangle
\right\}
\nonumber\\
= \int\limits_{s_{min}}^\infty\frac{ds}{s-q^2-i\delta} \left\{
\left[\frac{f_\rho^2}2\delta(s-m_\rho^2) +\rho^h(s)
\right](-g_{\mu\nu}q^2+q_\mu q_\nu)\right\}_{q^2=s}\,,
\label{disprhoo}
\end{eqnarray}
where the $\rho$-meson contribution ($h_n=\rho^0$) is isolated   
from the sum and the  hadronic matrix element is 
substituted: 
\begin{equation}
\langle 0 | j_\mu |\rho^0\rangle 
=\frac{f_\rho}{\sqrt{2}}m_\rho \epsilon_\mu^{(\rho)},
\end{equation}
determined by the $\rho$ decay constant $f_\rho$.
The integrand $\rho^h(s)$ ({\em spectral density}) includes 
the sum over excited  and multihadron states. 
We take into account the experimental fact that the $\rho$ resonance 
strongly dominates in the low-energy region $2m_\pi<\sqrt{s}<1$ GeV,
so that $\rho^h(s)$ practically starts from some threshold value
$s_0\sim 1$ GeV. 
Switching to the invariant amplitude $\Pi(q^2)$ and
using  the result (\ref{OPE})
of QCD calculation, one obtains from (\ref{disprhoo}) the desired relation,
a prototype of QCD sum rule:
\begin{equation}
\frac{f_\rho^2}{2(m_\rho^2-q^2)}+
\int\limits_{s_0}^\infty\frac{ds~ \rho^h(s)}{s-q^2-i\delta}=
\Pi^{pert}(q^2)+
\sum\limits_{d=3,4,...} C^d(q^2)\langle 0|O_d |0\rangle\,, 
\label{sumrule1}
\end{equation} 
which is valid at sufficiently large $|q^2|$.
To proceed, one applies to both sides of this equation the
{\em Borel transformation} defined as: 
\begin{equation}
\hat{ B}_{M^2}\Pi(q^2)=\lim_{\stackrel{-q^2,n \to \infty}{-q^2/n=M^2}}
\frac{(-q^2)^{(n+1)}}{n!}\left( \frac{d}{dq^2}\right)^n \Pi(q^2)
\equiv \Pi(M^2)~.
\label{Borel2}
\end{equation}
This transformation deserves a clarifying comment. 
Differentiating $\Pi(q^2)$ many times in $q^2$ 
means, one is effectively approaching the long-distance region. Indeed,
with an infinite amount of derivatives the 
function $\Pi(q^2)$ is defined at any $q^2$, including $q^2>0$. 
The $q^2\to -\infty$ limit works in the opposite direction:
one penetrates into the deep spacelike asymptotics.
Combining  the two transformations in (\ref{Borel2})
at $M^2=-q^2/n$, one fixes the virtuality scale at $O(M)$.  
Using the school-textbook definition of the exponent:
$e^x=\lim_{n\to \infty}(1+x/n)^n$, it is an easy exercise
to prove that  
\begin{equation}
\hat{B}_{M^2}\left \{\frac1{m_h^2\!-\!q^2}\right\}= e^{-m_h^2/M^2}.
\label{Borel}
\end{equation}
As a result, Borel transformation exponentially suppresses the 
integral over $\rho^h(s)$ in (\ref{sumrule1}) with respect
to the $\rho$-meson term. Furthermore,
after applying $\hat{B}$  to r.h.s. of (\ref{sumrule1}) the  
coefficients $C_d(M)$ contain  powers of $1/M^2$.
Hence, at large
$M^2$ it is possible to retain only a few low-dimension 
condensates in the sum, e.g., 
at $M^2\sim 1$ GeV$^2$  a reasonable approximation is to 
neglect all operators with $d> 6$. 
The explicit form of the QCD sum rule  (\ref{sumrule1})
after Borel transformation is \cite{SVZ}: 
\begin{eqnarray}
&\frac{f_\rho^2}{2}e^{-m_{\rho}^2/M^2}+ \int\limits_{s_{0}}^\infty ds~ \rho^h(s)e^{-s/M^2}
\nonumber
\\
&= \frac{M^2}2 \left[\frac1{4\pi^2} \left( 1+\frac{\alpha_s(M)}{\pi}\right) 
+\frac{(m_u+m_d)\langle\bar{q}q\rangle}{M^4}  
+\frac{1}{12}\frac{\langle \frac{\alpha_s}{\pi} G^a_{\mu\nu}G^{a\mu\nu}
\rangle}{M^4} 
-\frac{112\pi}{81} 
\frac{\alpha_s\langle \bar{q} q \rangle ^2 }{M^6}\right]\,.
\label{SVZrho}
\end{eqnarray}
In obtaining the above relation, the four-quark vacuum densities are 
factorized into a product of quark condensates. The quark-gluon
condensate has very small coefficient and is
neglected. The running coupling $\alpha_s$
is taken at the scale $M$ i.e., at the characteristic virtuality 
of the loop diagrams after Borel transformation. 

Importantly, there exists a SVZ 
region \cite{SVZ} of intermediate $M^2$ where the $\rho$ meson 
contribution alone saturates the l.h.s. of the sum rule (\ref{SVZrho}).
\begin{figure}
\begin{center}
\includegraphics*[width=13cm]{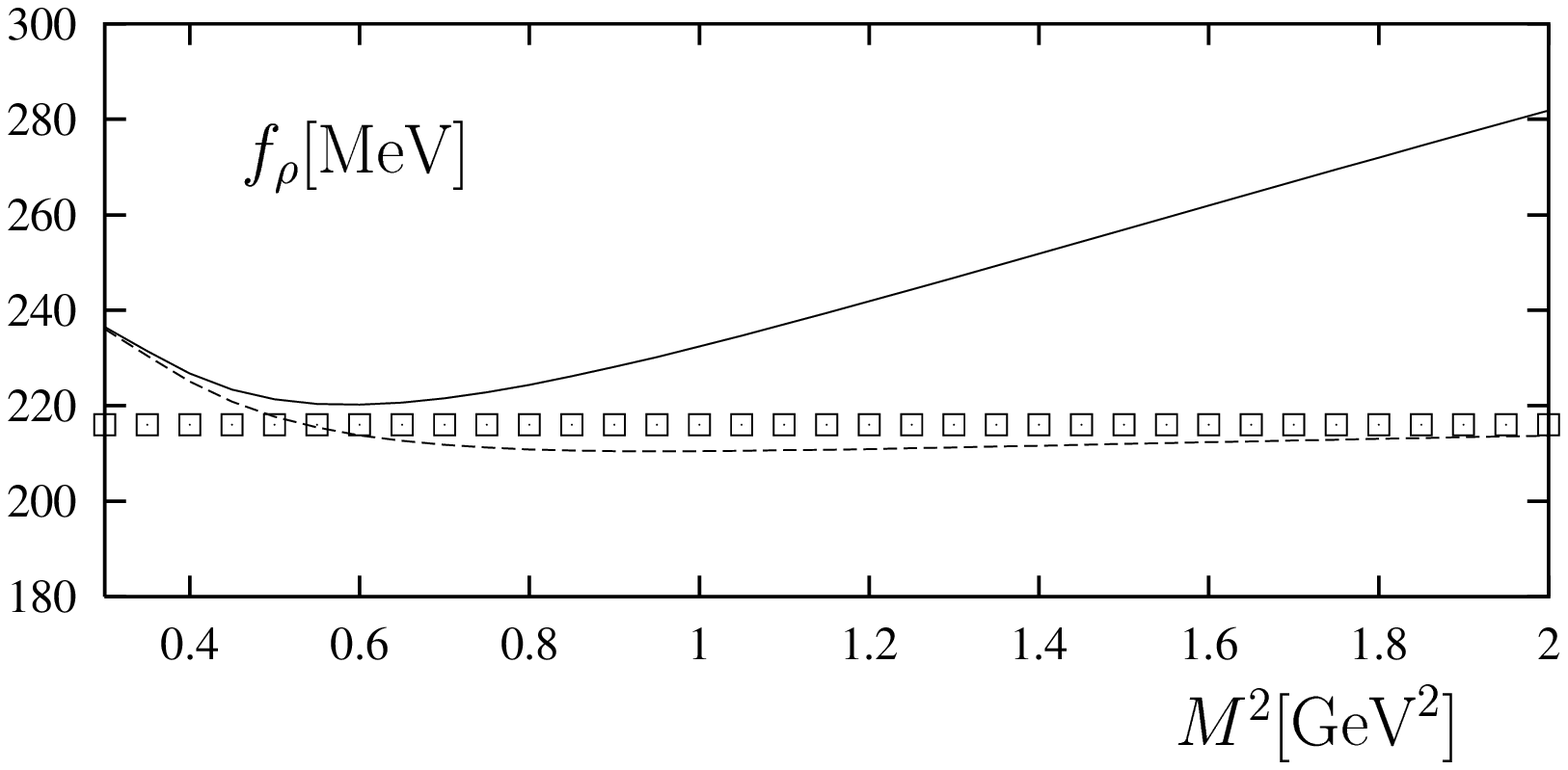}
\caption{{\em 
The $\rho$ meson decay constant calculated from the sum rule 
(\ref{SVZrho}) neglecting all excited and continuum states (solid), 
as a function of the Borel parameter, in
comparison with the experimental value (boxes). The dashed curve 
corresponds to an improved calculation , where the sum over excited
and continuum states is estimated using quark-hadron duality
with a threshold $s_0^\rho=1.7$ GeV$^2$. }}
\label{fig:frho}
\end{center}
\end{figure}
To illustrate this statement numerically, in Fig.~\ref{fig:frho} the 
experimentally measured $f_\rho$ (obtained from the $\rho^0\to e^+e^-$
width) is compared with the same hadronic parameter 
calculated from the sum rule (\ref{SVZrho})
where all contributions of excited and continuum states are neglected. 
One indeed observes a good agreement in the region $M^2\sim $ 1 GeV$^2$.

An important step to improve the sum rule (\ref{SVZrho})
is to use the {\em quark-hadron duality} 
approximation. The perturbative contribution to the correlation
function (the sum of Fig.~\ref{fig:2loop} diagrams) is represented
in the form of a dispersion integral splitted into two parts:
\begin{equation}
\Pi^{pert}(q^2)= \int\limits_0^{s_0}\frac{ds~ \rho^{pert}(s)}{s-q^2-i\delta}+
\int\limits_{s_0}^\infty \frac{ds~ \rho^{pert}(s)}{s-q^2-i\delta}\,.
\label{QCDdual}
\end{equation} 
The integral over the spectral function $\rho^h(s)$ 
in (\ref{sumrule1})
is approximated by the second integral over the perturbative spectral density
$\rho^{pert}(s)$ in (\ref{QCDdual}). 
The latter integral is then subtracted from both parts 
of (\ref{sumrule1}).  Correspondingly (\ref{SVZrho}) is
modified: the l.h.s. contains only the $\rho$ term,
and, on the r.h.s., the perturbative contribution has to be 
multiplied by  a factor $(1-e^{-s_0/M^2})$. The  numerical result obtained from  the duality-improved  sum rule (\ref{SVZrho}) is also shown in Fig.~\ref{fig:frho}.
The agreement between the sum rule prediction 
and experiment is impressive:
\begin{equation}
f_\rho^{(QCDSR)}= 213 ~\mbox{MeV}\pm (10-15)\%
\label{frho1}
\end{equation}
whereas $f_{\rho}^{exp}= 216\pm 5 ~\mbox{MeV}$ \cite{PDG}.  
The estimated theoretical uncertainty quoted in (\ref{frho1}) is typical
for QCD sum rules, reflecting the approximate nature of this method.
\begin{figure}
\begin{center}
\hspace{-0.5cm}
\includegraphics*[width=12cm]{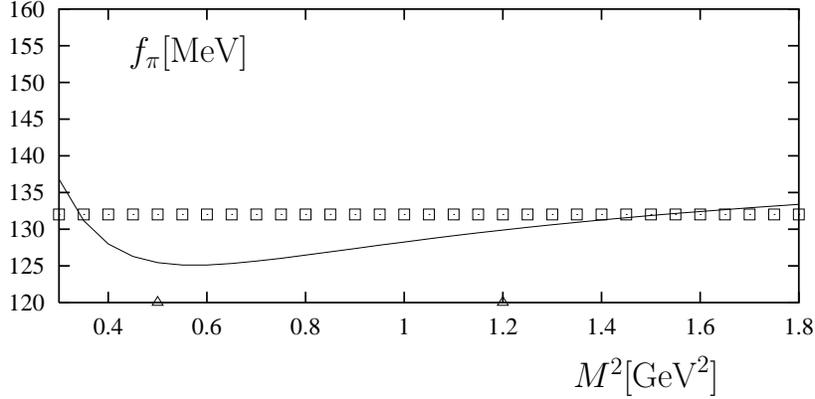}
\caption{{\em 
The $\pi$ meson decay constant calculated from QCD sum rule 
\cite{SVZ} for the correlation function of axial-vector currents
( solid line), in comparison with the experimental value (boxes). 
The quark-hadron duality threshold is fitted 
simultaneously as $s_0^\rho=0.7$ GeV$^2$.
The 
uncertainty of about 10-15\% has to be added to the theoretical
curve.}}
\label{fig:fpi}
\end{center}
\end{figure}
So far we only reproduced $f_\rho$, using the experimental
value of $m_\rho$. In principle, it is possible to    
go one step further and estimate also the $\rho$-mass from
the sum rule . Furthermore, a very similar sum-rule 
analysis for the correlation function of axial currents (\ref{corrA1})
(the invariant amplitude $\Pi_5$) successfully 
reproduces the value of $f_\pi$ (see Fig.~\ref{fig:fpi}).

\subsection{Baryons}

It is possible to extend the
method of QCD sum rules  to baryons
\cite{barIoffe,Doschetal}. The idea is to construct special 
quark currents with  baryon quantum numbers which can 
serve as a source of baryon production/annihilation from/to
QCD vacuum. In reality, such currents 
hardly exist \footnote{In
models of Grand Unification predicting proton decay 
via intermediate superheavy particles, 
the currents we are discussing are realized 
effectively in a form of localized 3-quark operators 
annihilating the proton.}, but they are 
allowed in QCD if colour-neutrality is obeyed.
A well known example is the Ioffe current with 
the nucleon quantum numbers (i.e., spin 1/2): 
\begin{equation}
J^N(x)= \epsilon_{abc} (u^{a T }(x) {\hat{C}} \gamma_\mu u^b(x) )
 \gamma_5 \gamma^\mu d^c(x)\,, 
\label{jN}
\end{equation}
where $a,b,c$ are color indices, $\hat{C}=\gamma_2\gamma_0$ 
is the charge conjugation matrix
and, for definiteness, the proton flavour content is chosen. 
As a next step, one constructs a correlation function
\begin{equation}
\Pi_N(q) = i\int d^4x e^{iqx}\langle 0| T\{J^N(x)\bar{J}^N(0)\}|0\rangle\,. 
\label{corrN}
\end{equation}
\begin{figure}
\begin{center}
\includegraphics*[width=12cm]{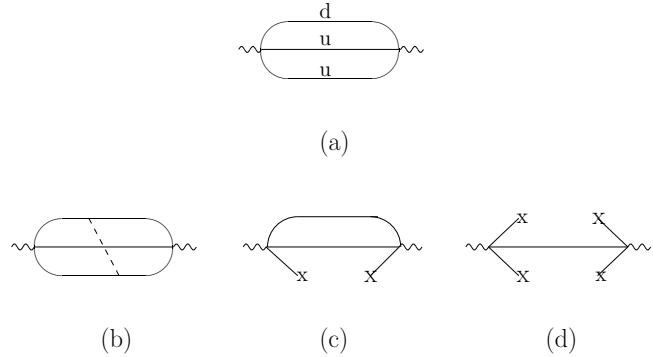}
\caption{{\em 
Some of the diagrams contributing to the correlation
function of two baryon currents: (a) the lowest-order tree-quark
two-loop diagram; (b) the $O(\alpha_s)$ correction, (c),(d) -quark
condensate diagrams.}}
\label{fig:baryons}
\end{center}
\end{figure}
The corresponding diagrams are shown in Fig.~\ref{fig:baryons}, 
including perturbative 
loops and vacuum condensates. They look quite different
from the quark-antiquark loops, but differences concern
the short-distance parts of the diagrams. The universality of 
condensates allows one to calculate $\Pi_N(q)$
without introducing new input parameters.
The hadronic contribution contains a total sum over states
produced and annihilated by the current $J^N$, starting from the lowest
possible state, the nucleon:
\begin{equation}
\Pi_N(q)= \frac{\langle 0| J^N|N \rangle \langle 0| \bar{J}^N|N \rangle}{m_N^2-q^2}+ \{\mbox{excited resonances,~multiparticle states}\}\,.
\label{nucleonSR}
\end{equation}
The derivation of the QCD sum rule is done along the same
lines as in the previous subsection. Omitting the details, 
let me only mention that  from this sum rule
an approximate formula for the nucleon mass is obtained, 
\begin{equation}
m_N\simeq [-(2.0)(2\pi)^2 \langle 0| \bar{q}q |0\rangle
(\mu=1\mbox{GeV})]^{1/3}\,,
\end{equation} 
relating it to the quark-condensate density. In fact, the quark masses 
$m_{u,d}$ themselves generate very small corrections to this 
relation and are neglected. 
Thus, QCD sum rules provide an answer to the question that was 
raised in Lecture 1:  almost $99\%$  of the baryonic mass in the Universe
are due to vacuum condensates.

\subsection{Quark mass determination}

As already mentioned, the sum rule (\ref{moments}) 
can be used to extract the $c$-quark mass. Here, the role of Borel
transformation is played by a simple differentiation at $q^2=0$,
which turns out to be more useful.  At low $n$, 
the moments (\ref{moments})  are especially convenient for 
$m_c$ determination because the gluon condensate effects are 
small. Replacing $c\to b$, $\psi \to \Upsilon$, etc., one 
obtains analogous sum rule relations for $b$ quark, 
where gluon condensate is much less important, being suppressed
by $m_b^{-4}$.
Recent analysis \cite{KS} of these sum rules yields
$m_c(m_c)= 1.304 \pm 0.027$ GeV, $m_b(m_b)= 4.209 \pm 0.05$ GeV. 
Another subset of charmonium sum rules (higher moments at fixed large
$q^2<0$) was recently employed in \cite{IoffeZ}, with 
a prediction for $m_c$
in  agreement with the above. 

The heavy-quark mass determination
using sum rules is also done in a different way, employing 
the large $n$  moments (\ref{moments}) which
are less sensitive to the cross-section above the
open flavour threshold. These moments, however, demand a careful treatment 
of Coulomb interactions between heavy quark and antiquark
in the perturbative diagrams. Remember that one-gluon exchange 
yields Coulomb potential. Close to the threshold of heavy quark-pair
production, $\sqrt{s}\simeq 2m_Q$ ( $Q=b,c$) this part of 
quark-antiquark interaction becomes important, and  at large $n$ 
the near-threshold region dominates
in the perturbative coefficients $a_n,a_n'$ in (\ref{moments}). 
A systematic treatment of this problem is 
possible in {\em nonrelativistic QCD (NRQCD)}, a specially designed 
effective theory obtained from QCD in the infinite heavy quark mass
limit (for a review  see, e.g. \cite{Hoang}).

Equally well, QCD sum rules allow to estimate
the masses of light $u,d,s$ quarks.
To give only one typical example from the vast literature
let me refer to the recent analysis  
\cite{JOP}, based on the correlation function 
for derivatives of the $s$-flavoured vector current 
$j_{s\,\mu}=\bar{s}\gamma_\mu q$, ($q=u,d$):
\begin{equation}
\Pi^s(q)=i\int d^4xe^{iqx}\langle 0 | T\{\partial_\mu j^{\mu}_s(x)\partial_\nu 
j^{\dagger\nu}_s(0)\}|0\rangle\,.
\end{equation}
The QCD answer for $\Pi^s$ is proportional to 
$(m_s-m_q)^2\simeq m_s^2$ (due to $\partial_\mu
j^\mu_s=(m_s-m_q)\bar{s}q$), making this correlator very sensitive
to $m_s$. One calculates the usual set of diagrams 
shown in Figs.~\ref{fig:2loop} and \ref{fig:cond}, where
the quark lines  are now $s$ and $q$. Furthermore, 
the recent progress in the
multiloop QCD calculations allows to reach the $O(\alpha_s^3)$ accuracy 
in the perturbative part of $\Pi^s$ and to include 
also $O(\alpha_s)$ corrections to the condensate
contributions. The hadronic spectral density $\rho_K(s)$ for this correlator 
is saturated by $J^P=0^+$ states with $s$ -flavour, e.g. $K\pi$ 
states with $L=0$.  
The sum rule  has the form
\begin{equation}
\int ds\rho_K(s)e^{-s/M^2}= (m_s-m_{u,d})^2\left
[\Pi^{pert}(M)+ \frac{C^K_4}{M^4} +\frac{C^K_6}{M^6}\right]\,,
\label{strange}
\end{equation} 
Data
of kaon $S$-wave scattering on $\pi,\eta,\eta'$ were used to reproduce 
$\rho^K(s)$. The resulting prediction \cite{JOP} for the mass is 
\begin{equation}
m_s(2 ~\mbox{GeV})=99 \pm 16 ~\mbox{MeV}.
\label{msvalue}
\end{equation}

The ratios of the light ($u,d,s$) quark masses
can be predicted in QCD from the relations for pions and kaons, 
similar to (\ref{GOR}). 
A systematic derivation is done employing  
{\em chiral perturbation theory}, an effective theory obtained
from QCD in the a low-energy limit, 
using instead of quarks and gluons, the pion and kaon degrees of freedom
(for a review see \cite{Leutwyler}). The result is \cite{Leutwyler1}:  
\begin{eqnarray}
\frac{m_u}{m_d}=0.553 \pm 0.043,~~~ 
\frac{m_s}{m_d}=18.9\pm 0.8,~~~ 
\frac{2m_s}{m_u+m_d}=24.4\pm 1.5 \,.
\label{rel}
\end{eqnarray}
From the above ratios and the value (\ref{msvalue})
one obtains $m_{u}(2 ~\mbox{GeV})=2.9 \pm 0.6 $ MeV
and $m_{d}(2 ~\mbox{GeV})=5.2 \pm 0.9 $ MeV.

We see that QCD sum rules are extremely useful
for the quark mass determination. The $m_q$ values extracted 
from sum rules are included, together with the lattice
determinations and results of other methods, in  the world-average intervals
in \cite{PDG} presented in (\ref{qmasses}).

\subsection{ Calculation of the $B$ meson decay constant}

In $B$-meson decays the CKM parameters of the Standard Model 
are inseparable from hadronic matrix elements. 
Hence, without QCD calculation of these matrix elements
with an estimated accuracy, it is impossible to use experimental data 
on $B$ decays for extracting the Standard Model 
parameters and for detecting/constraining new physics effects.
Currently, lattice QCD provides many hadronic parameters for $B$ physics,
with a continuously improving accuracy.
QCD sum rules represent another actively used working tool. 
With condensates and quark masses 
determined from a set of experimentally proven 
sum rules for light-quark and heavy quarkonium systems,
one has a real possibility to assess  the theoretical 
accuracy of the sum rule predictions by varying 
the input within allowed intervals. 

One of the most important applications 
of QCD sum rules is the determination of the $B$-meson
decay constant $f_B$ defined via the matrix element
\begin{equation}
m_b\langle 0 \mid\bar{q}i\gamma_5 b\mid B\rangle =m_B^2f_B~,
\label{fB}
\end{equation}
($ q=u,d,s$). 
Note that $f_B$ multiplied by $V_{ub}$ determines 
the width of leptonic $B$ decays, such as $B^- \to \tau^- \bar{\nu}_{\tau}$.
To calculate $f_B$ from QCD sum rules,
one usually employs the correlation function: 
\begin{equation}
\Pi_5^{(B)}(q)=i\int d^4x e^{iqx}\langle 0 | T\{
m_b\bar{q}(x)i\gamma_5 b(x), m_b\bar{b}(0)i\gamma_5 q(0)\}|0\rangle\,,
\label{Bcorr}
\end{equation}
so that the lowest $B$ meson term  in the hadronic sum for this 
correlation function contains  $f_B$:
\begin{equation}
\Pi_5^{(B)}(q)= \frac{m_B^4f_B^2}{m_B^2-q^2} + ....
\label{sumB}
\end{equation} 
To obtain the sum rule, one needs to calculate $\Pi_5^{(B)}(q)$ from the  
perturbative and condensate diagrams in Figs.~\ref{fig:2loop} and 
\ref{fig:cond}, with $\gamma_5$ vertices emitting and absorbing
$b$ and $q$ lines. One, therefore needs an input value
for $m_b$ and $m_s$ ($m_{u,d}$ can safely be neglected). These values
are taken from the analyses overviewed in the previous subsection.   
The sum rule has a form similar to (\ref{SVZrho}) for $f_\rho$,
Naturally, the expressions for the perturbative part and for the 
coefficients $C_d$ are completely different. Also the hierarchy of 
contributions in the heavy-light correlation function 
differs from the light-quark case. Now the quark condensate 
becomes very important being proportional to $m_b\langle \bar{q}q\rangle$. 
The recent updates of the sum rule for $f_B$ obtained in \cite{Jamin,PS} 
take into account the $O(\alpha_s^2)$ 
corrections to the heavy-light loop  calculated in \cite{ChetS}.
The numerical prediction of the sum rule, taking $m_b(m_b)= 4.21\pm 0.05$ GeV, is:
$ f_B=210 \pm 19 $ MeV and $ f_{B_s}= 244\pm 21 $ MeV \cite{Jamin},
in a good agreement with the most recent lattice QCD determinations.

\subsection{Light-cone sum rules and $B\to \pi$ form factor}

To complete our brief survey of 
QCD sum rules, let me introduce
one important version of this method, the light-cone sum rules (LCSR)
\cite{lcsr,cz} used to calculate 
various hadronic amplitudes relevant for exclusive processes.
In the following, we consider the application of LCSR 
to the $B\to \pi$ transition amplitude  (see Fig.~\ref{fig:Bpi}). 
The latter is determined  
by  the hadronic matrix element 
\begin{equation}
<\pi^+(p)|\bar{u} \gamma_\mu b |B(p+q)> =
2f^{+}_{B\pi}(q^2)p_\mu +\left[f^{+}_{B\pi}(q^2) +
f^{-}_{B\pi}(q^2)\right]q_\mu,
\label{Bpiformf}
\end{equation}
generated by the $b\to u$ weak current (\ref{jweak}).
Due to spin-parity conservation only the vector
part of the current contributes. There are two independent 4-momenta 
$p$ and $q$, and one independent 
invariant $q^2$, the momentum transfer squared. 
The initial and final mesons are on  shell, 
$p^2=m_\pi^2$ and $(p+q)^2=m_B^2$.  
It is quite obvious that
one needs two invariant functions of $q^2$, the {\em form factors} 
$f^+_{B\pi}(q^2)$ and $f^-_{B\pi}(q^2)$,  to parameterize this matrix
element. Only one form factor $f^+_{B\pi}$ is  interesting, the other one 
is kinematically suppressed in the measurable
$B\to \pi l \nu_l$ semileptonic decay rate ($l=e,\mu$).

To derive LCSR for $f^+_{B \pi}(q^2)$ one uses a
specific correlation function, which itself 
represents a  hadronic matrix element.
It is constructed from the product of the weak $\bar{u}\gamma_\mu b$ 
current and the current $ m_b\bar{b}i\gamma_5 d$ used 
to generate $B$ in (\ref{fB}). 
The currents are taken at two different 4-points and sandwiched
between vacuum and the one-pion state. 
The formal definition of this correlation function 
reads: 
\begin{eqnarray}
F_\mu(q,p)= i\int d^4x e^{iqx}
\langle\pi^+(p)\mid T\{\bar{u}(x)\gamma_\mu b(x),
m_b\bar{b}(0)i\gamma_5 d(0)\}\mid 0\rangle
\nonumber
\\
= F(q^2,(p+q)^2)p_\mu + \tilde{F}(q^2,(p+q)^2)q_\mu  ~\,,
\label{Bpicorr}
\end{eqnarray}
where it is sufficient to consider only one invariant
amplitude $F$. Note that we now have two kinematical invariants,
$q^2$ and $(p+q)^2$, still $p^2=m_\pi^2$.
\begin{figure}
\begin{center}
\includegraphics*[width=12cm]{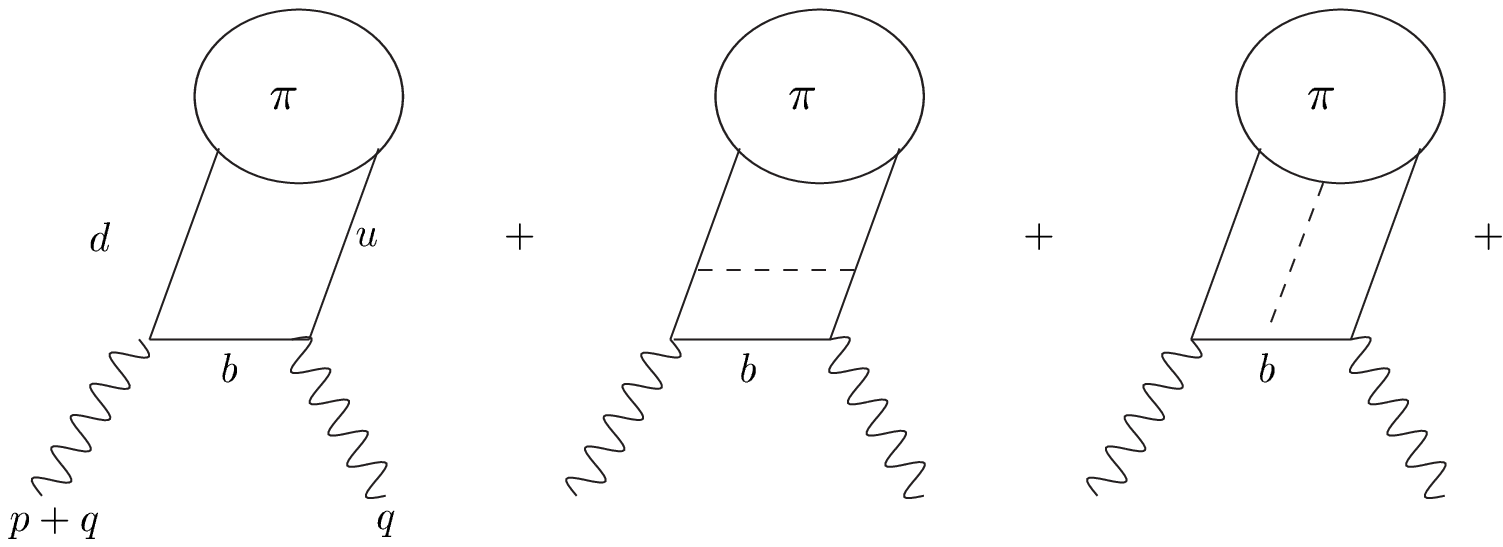}
\caption{{\em Diagrams contributing to the correlation
function (\ref{Bpicorr}), from left to right: the leading-order,
$O(\alpha_s)$ correction, soft gluon. The blob with $\pi$ denotes
the pion distribution amplitude.
}}
\label{fig:Bpidiag}
\end{center}
\end{figure}

Diagrammatically, the correlation function (\ref{Bpicorr})
is represented in Fig.~\ref{fig:Bpidiag}. At spacelike 
$(p+q)^2<0$ and $q^2<0$, very similar diagrams 
describe the process $\gamma^*\gamma^*\to \pi^0$ , the 
one-pion production by two virtual photons via e.m. currents
\footnote{This process
is experimentally accessible in $e^+e^- \to e^+e^-\pi^0$  two-photon
(double-tagged) collisions.}.
One only has to replace all quarks in the diagrams in  
Fig.~\ref{fig:Bpidiag} by either $u$ or $d$ quarks. 
Both objects: the heavy-light correlation function 
(\ref{Bpicorr}) and the $\gamma^*\gamma^*\to \pi^0$
amplitude contain one virtual quark propagating between vertices
and a quark-antiquark pair which is emitted at points $x$ and $0$ 
and converted into a real pion state. At large spacelike external 
momenta $|(p+q)^2|, |q^2| \gg
\Lambda_{QCD}^2$ the space-time interval $x^2\simeq 0$ 
approaches the light-cone. Hence, the virtual quark in both 
$\langle$two-currents $\to$ pion$\rangle$  amplitudes propagates 
at short distances allowing a perturbative QCD description.
The calculable short-distance parts are process-dependent.
In case of the correlation function (\ref{Bpicorr})
we have a virtual $b$ quark propagating between vertices 
of flavour-changing currents, 
whereas in the $\gamma^*\gamma^*\to \pi^0$ amplitude the light quark
propagates between e.m. vertices.
The long-distance part in both cases 
is, however, the same vacuum-to-pion matrix element of light quark
and antiquark emitted at points $x$ and $0$: 
\begin{equation}
\langle \pi(p) \mid \bar{q_1}(x)\Gamma_aq_2(0) \mid 0 \rangle\,,
\label{matrpi}
\end{equation}
where in one case $q_1=u$, $q_2=d$, $\pi=\pi^+$ 
and in the other case $q_1=u(d)$, $q_2=u(d)$, $\pi=\pi^0$.
Due to isospin symmetry, the difference
between these two configurations is indeed very small.
In order to treat the diagram with one extra gluon entering the pion, 
one has to introduce an additional quark-antiquark-gluon 
matrix elements of the type
\begin{equation}
\langle \pi(p)
\mid \bar{u}(x) g_sG^{\mu\nu}(y)\Gamma_b d(0) \mid 0 \rangle~,
\label{matrixglue}
\end{equation}
where $x^2\sim y^2\sim (x-y)^2 \to 0 $. In the above 
$ \Gamma_{a,b}$ denote certain combinations
of Dirac matrices.

Having separated short- and long-distances, 
one is able to calculate the correlation function (\ref{Bpicorr})
in a form of {\em light-cone OPE}, 
where the short-distance part (the virtual $b$ quark propagator
plus gluon corrections) is multiplied by a long-distance part,
the universal matrix elements such as
(\ref{matrpi}), (\ref{matrixglue}). 
The latter can be parameterized
in terms of {\em light-cone distribution amplitudes} 
of the pion \cite{BL},
The most important of them 
is defined by
\begin{equation}
\langle\pi(p)|\bar{u}(x)\gamma_\mu\gamma_5d(0)|0\rangle=
-ip_\mu f_\pi\int_0^1du\,e^{iupx}\varphi_\pi (u,\mu),
\label{pionwf}
\end{equation}
where $\mu $ is a characteristic  momentum scale,
determined by the average $x^2$ in the correlation function.  
In the above definition, $u$ and $1-u$ are the fractions of the pion momentum $p$
carried by the quark and antiquark, in the approximation
where one neglects the transverse momenta of the constituents with 
respect to the longitudinal momenta.
The leading-order answer for $F$ obtained
from the first diagram in Fig.~\ref{fig:Bpidiag} using (\ref{pionwf}) 
is quite  simple:
\begin{equation}
F( q^2,(p+q)^2)=
m_bf_\pi\int_0^1\frac{du ~\varphi_\pi(u,\mu) }{m_b^2-(q+up)^2}~.
\label{qcd}
\end{equation}
The pion DA $\varphi_\pi(u)$ plays here the same role
of nonperturbative input as the condensates in the conventional
QCD sum rules considered above.
Asymptotically, that is at $\mu\to\infty$, QCD
perturbation theory implies
$ \varphi_\pi(u,\infty) = 6u(1-u)$. However, at the physical scale
$\mu \sim m_b $, at which the OPE is applied to the
correlation function (\ref{Bpicorr}),
nonasymptotic effects also contribute, which we will not discuss
for brevity. 
Importantly, there is a power hierarchy of 
different contributions stemming from the diagrams in Fig.~\ref{fig:Bpidiag},
determined by the large scale in the correlation function.
This scale is given by the virtuality of $b$ quark, i.e. by the 
quantity which stands in  the denominator of the $b$ -quark 
propagator: $m_b^2-(q+up)^2= m_b^2-(p+q)^2u-q^2(1-u)$. 
Importantly, this quantity  remains large when $q^2$ 
is positive (timelike) but not very large, $q^2\ll m_b^2$, allowing one to 
penetrate into the lower part of the 
kinematical region $0<q^2<(m_B-m_\pi)^2$ of $B\to \pi l \nu_l$ decay. 
An example of a subleading contribution is the diagram with 
an additional gluon entering pion DA (the third one in Fig.~\ref{fig:Bpidiag}).
It has two powers of inverse scale and is suppressed with respect
to the leading-order diagram. 
Over recent years $O(\alpha_s)$ corrections and quark-antiquark-gluon
contributions have been calculated to improve the result (\ref{qcd}).

\begin{figure}
\begin{center}
\includegraphics*[width=13cm]{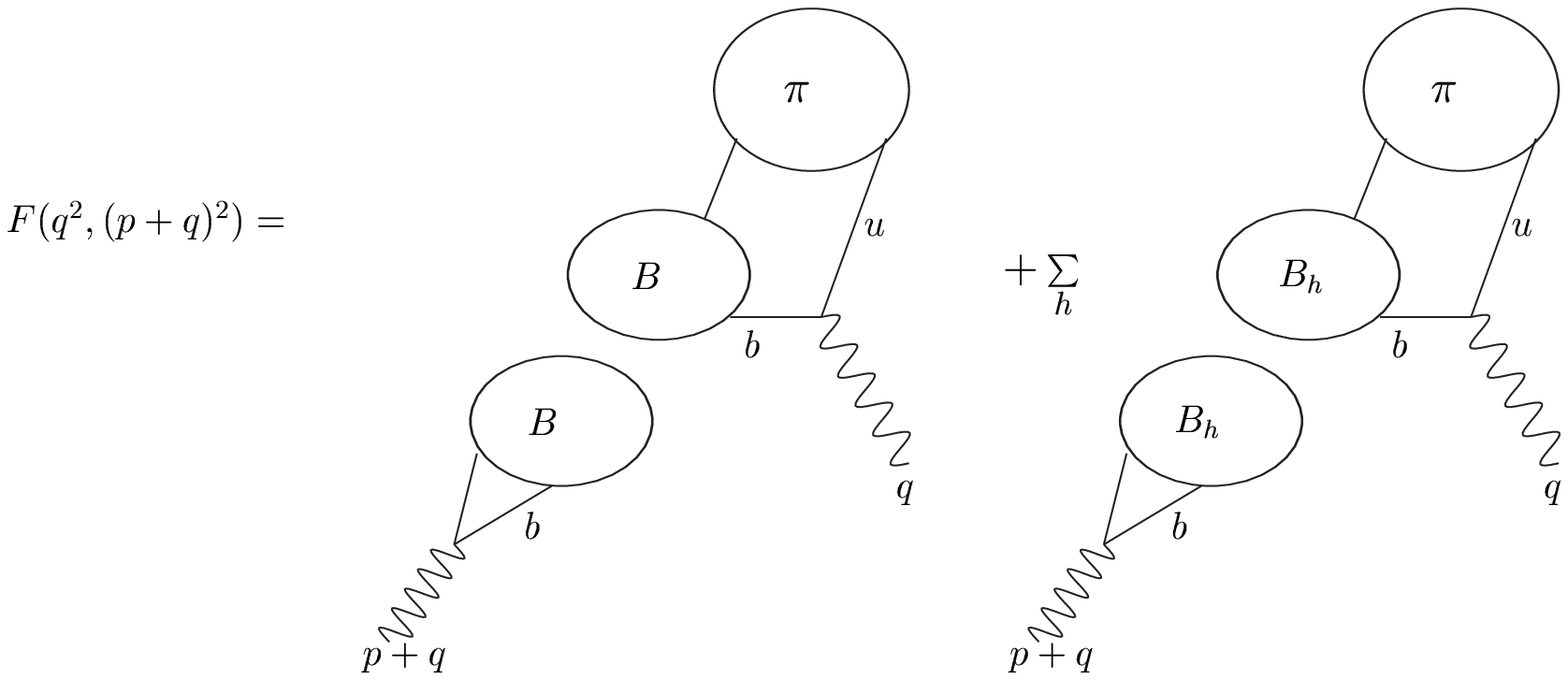}
\caption{{\em Diagramatic representation of the dispersion relation
for the correlation function (\ref{Bpicorr}.)}}
\label{fig:Bpidisp}
\end{center}
\end{figure}

Having evaluated $F$ as a function of $q^2$ and $(p+q)^2$,
we are still half-way from the final sum rule.
The next important step is writing down a dispersion relation in the variable
$(p+q)^2$, the external momentum of the current 
with $B$ meson quantum numbers. The diagrammatical representation 
of the dispersion relation is shown in Fig.\ref{fig:Bpidisp}.
It contains the same set of hadronic states as in the sum rule
for $f_B$, starting from $B$ meson ground state.  
Using (\ref{fB}) and (\ref{Bpiformf}) one 
obtains the hadronic representation we need:
\begin{equation}
F(q^2,(p+q)^2)= \frac{2m_B^2f_Bf_{B\pi}^+(p^2)}{m_b(m_B^2-(p+q)^2)}+\ldots~.
\label{disp}
\end{equation}
The ellipses in the above denote
the contributions from the excited $B$ and from continuum states.
Equating the QCD result (\ref{qcd}) in the region
of validity $(p+q)^2<0$ with
the dispersion relation (\ref{disp}) one obtains
a raw sum rule relation for $f_Bf^+_B(p^2)$.
The rest of the calculation follows the usual QCD sum rule
procedure: Borel transformation  in $(p+q)^2$ and subtraction
of the contribution from higher states invoking
quark-hadron duality. 
One finally arrives at an expression for the
desired form factor, the dominant term of which obtained 
directly from (\ref{qcd}) is given by
\begin{equation}
f^+_{B\pi}( q^2)= \frac{f_\pi m_b^2}{2f_Bm_B^2}\int_
\Delta^1\frac{du}{u}(\varphi_\pi(u,\mu_b) + \ldots)
exp \left(\frac{m_B^2}{M^2}
-\frac{m_b^2-q^2(1-u)}{uM^2}\right) ~.
\label{formf}
\end{equation}
Here, $M$ is the Borel mass parameter, and
the scale $\mu_b$ reflects the characteristic
virtuality of the correlation function, $\mu_b^2 = m_B^2-m_b^2$.
The integration limit $\Delta = (m_b^2-p^2)/(s_0-p^2) $
depends on the effective threshold $s_0^B$ above which the
contribution from higher states to the dispersion relation
(\ref{disp}) is canceled against the corresponding piece
in the QCD representation (\ref{qcd}). 
The parameters $f_B, s_0^B$ are usually taken from the sum rule
for $f_B$ considered in the previous subsection. 
The most recent predictions for 
$f^+_{B\pi}( q^2)$ obtained from LCSR \cite{BpiLCSR} 
were used to extract $|V_{ub}|$ from the 
measurements of the exclusive semileptonic width at $B$ factories \cite{Vub},
using the formula for the decay rate:
\begin{equation}
\frac{d\Gamma(B\to \pi l \bar{\nu})}{dq^2}=
\frac{G^2|V_{ub}|^2}{24\pi^3}(E_\pi^2-m_\pi^2)^{3/2}[f^+_{B\pi}(q^2)]^2\,.
\label{Bpiwidth}
\end{equation}
where $E_\pi$ is the pion energy in the $B$ meson rest frame.

\section{CONCLUSIONS}

QCD has a thirty-years history 
and embraces many approaches, some of them developed quite
independently from the others. Due to self-interactions 
of gluons this theory has 
an extremely rich dynamics, combining asymptotic
freedom at short distances with the self-emerging 
energy scale $\Lambda_{QCD}$ and confinement at long
distances. Accordingly, QCD has
two different phases: the perturbative one
responsible for quark-gluon processes at large momentum transfers 
and the nonperturbative one where the only observable states
are hadrons formed by confined
quarks and gluons. Yet there is no complete analytical
solution for the hadronic phase of QCD.
One has to rely on approximations:
either numerical (QCD on the lattice) or analytical (QCD sum rules).
In addition, several effective theories corresponding
to different limits of QCD and exploiting the rich symmetry
pattern of the theory are successfully used.
In the overview of QCD given in these lectures 
I tried to emphasize the importance of the hadronic aspects of QCD, 
where the most nontrivial phenomena and challenging problems  
are accumulated.

\section*{ACKNOWLEDGEMENTS}
I would like to thank the organizers of the School 
for inviting me to give these lectures and for 
an enjoyable meeting at Tsakhkadzor. I am grateful to   
the discussion leaders for their help, especially  
to A.~Pivovarov for useful remarks on the manuscript. 
My special thanks are to the students of the School for their 
stimulating interest and for many questions
and comments. 

This work was partially supported by 
the German Ministry for Education and Research (BMBF).

\end{document}